\documentclass[twocolumn]{aastex63}
\usepackage[utf8]{inputenc}
\usepackage{graphicx}
\usepackage{color}
\usepackage{amsmath,amssymb}
\usepackage{gensymb}
\usepackage{times}
\usepackage{nicefrac}
\usepackage{amsmath}
\usepackage{xcolor}
\usepackage[normalem]{ulem} % delete this after all 
\newcommand{\gaia}{\textsl{Gaia}}

\usepackage{savesym}
\savesymbol{tablenum}
\usepackage{siunitx}
\restoresymbol{SIX}{tablenum}
\DeclareSIUnit\year{yr}
\DeclareSIUnit\parsec{pc}
\DeclareSIUnit\msun{M_\odot}
\DeclareSIUnit\Rsun{R_\odot}

\newcommand{\kms}{\unit{\km\per\s}}
\newcommand{\kpc}{\unit{\kilo\parsec}}

\DeclareSIUnit\mstar{M_\star}

\newcommand{\usurfdens}{\unit{\msun.\parsec^{-2}}}
\newcommand{\uvoldens}{\unit{\msun.\parsec^{-3}}}

\newcommand{\abun}[2]{\ensuremath{{[\mathrm{#1}/\mathrm{#2}]}}}

\newcommand{\Teff}{\ensuremath{T_{\mathrm{eff}}}}
\newcommand{\logg}{\ensuremath{\log g}}

\newcommand{\zmax}{\ensuremath{z_{\textrm{max}}}}

\newcommand{\freqzero}{\ensuremath{\Omega_0}}

\newcommand{\rz}{\ensuremath{r_z}}
\newcommand{\rzp}{\ensuremath{\tilde{r}_z}}

\newcommand{\thzp}{\ensuremath{\tilde{\theta}_z}}

\newcommand{\dd}{\mathrm{d}}
\newcommand{\deriv}[2]{\frac{\mathrm{d}{#1}}{\mathrm{d}{#2}}}
\newcommand{\pderiv}[2]{\frac{\partial {#1}}{\partial {#2}}}

\sloppypar

\begin{document}
\title{Orbital Torus Imaging:\\
Acceleration, density, and dark matter in the Galactic disk measured with element abundance gradients}

\newcommand{\affcca}{
    Center for Computational Astrophysics, Flatiron Institute, \\
    162 Fifth Ave, New York, NY 10010, USA
}

\newcommand{\affcolumbia}{
    Department of Astronomy, Columbia University, \\
    550 West 120th Street, New York, NY 10027, USA
}

\newcommand{\affuw}{
    Department of Astronomy, Box 351580, University of Washington,\\ Seattle, WA 98195
}

\newcommand{\affccpp}{
    Center for Cosmology and Particle Physics, Department of Physics, New York University, \\
    726 Broadway, New York, NY 10003, USA
}
\newcommand{\affmpia}{
    Max-Planck-Institut f\"ur Astronomie, \\
    K\"onigstuhl 17, D-69117 Heidelberg, Germany
}

\newcommand{\affking}{
    Department of Physics, Engineering Physics, and Astronomy, \\
    Queen’s University, Kingston, Ontario, Canada
}

\newcommand{\affsurr}{
    School of Mathematics \& Physics, University of Surrey, Guildford, GU2 7XH, UK
}

\author[0000-0003-1856-2151]{Danny Horta}
\affiliation{\affcca}

\author[0000-0003-0872-7098]{Adrian~M.~Price-Whelan}
\affiliation{\affcca}

\author[0000-0003-2866-9403]{David~W.~Hogg}
\affiliation{\affcca}
\affiliation{\affmpia}
\affiliation{\affccpp}

\author[0000-0001-6244-6727]{Kathryn~V.~Johnston}
\affiliation{\affcolumbia}

\author{Lawrence Widrow}
\affiliation{\affking}

\author[0000-0002-1264-2006]{Julianne J. Dalcanton}
\affiliation{\affcca}
\affiliation{\affuw}

\author[0000-0001-5082-6693]{Melissa~K.~Ness}
\affiliation{\affcca}
\affiliation{\affcolumbia}

\author[0000-0001-8917-1532]{Jason A. S. Hunt}
\affiliation{\affsurr}
\affiliation{\affcca}

\correspondingauthor{Danny Horta}
\email{dhortadarrington@flatironinstitute.org}

\begin{abstract}\noindent
Under the assumption of a simple and time-invariant gravitational potential, many Galactic dynamics techniques infer the Milky Way's mass and dark matter distribution from stellar kinematic observations. 
These methods typically rely on parameterized potential models of the Galaxy and must take into account non-trivial survey selection effects, because they make use of the density of stars in phase space.
Large-scale spectroscopic surveys now supply information beyond kinematics in the form of precise stellar label measurements (especially element abundances).
These element abundances are known to correlate with orbital actions or other dynamical invariants.
Here, we use the Orbital Torus Imaging (OTI) framework that uses abundance gradients in phase space to map orbits.
In many cases these gradients can be measured without detailed knowledge of the selection function.
We use stellar surface abundances from the \textsl{APOGEE} survey combined with kinematic data from the \textsl{Gaia} mission. 
Our method reveals the vertical ($z$-direction) orbit structure in the Galaxy and enables empirical measurements of the vertical acceleration field and orbital frequencies in the disk. 
From these measurements, we infer the total surface mass density, $\Sigma$, and midplane volume density, $\rho_0$, as a function of Galactocentric radius and height. Around the Sun, we find $\Sigma_{\odot}(z=1.1~\kpc) = 72^{+6}_{-9}~\usurfdens$ and $\rho_{\odot}(z=0) = 0.081^{+0.015}_{-0.009}~\uvoldens$ using the most constraining abundance ratio, [Mg/Fe].
This corresponds to a dark matter contribution in surface density of $\Sigma_{\odot,\mathrm{DM}}(z=1.1~\kpc) =24\pm4~\usurfdens$, and in total volume mass density of $\rho_{\odot,\mathrm{DM}}(z=0) = 0.011\pm0.002~\uvoldens$. Moreover, using these mass density values we estimate the scale length of the low-$\alpha$ disc to be $h_R = 2.24\pm0.06~\kpc$.

\end{abstract}

\section{Introduction}

Galaxies are dynamic and evolving structures driven by the interactions and assembly of their constituent stars, gas, and dark matter \citep{White1991}. The resulting observed structure of galaxies provides important information about the properties and behavior of dark matter, and the processes involved in galaxy formation. Unfortunately, we only observe galaxies at (effectively) a single snapshot in time. To turn this snapshot of stellar and gas kinematics into inferences about galaxy formation processes and dark matter, the study of galactic dynamics relies on theoretical frameworks and methods built on statistical physics \citep{Binney2008}.
Over the last century, this endeavor has led to important measurements of the structure of mass and dark matter in our galaxy, the Milky Way.

A key quantity that has been used historically to summarize the mass distribution in the Milky Way is the total surface mass density near the sun, $\Sigma_\odot$ (\citealp[e.g.,][]{Oort1960,Kuijken1989,Kuijken1991,Creze1998,Holmberg2000}).
The total surface mass density is typically measured from stellar kinematics using methods that rely on strong assumptions about the stellar distribution function (DF) and form of the gravitational potential (\citealp[e.g.,][]{Kuijken1991,Flynn2006,McMillan2011,Bovy2013,Zhang2013}). In all cases, measurements of the total surface mass density together
with estimates of the surface density from stars and gas provides a means to infer the local contribution of dark matter to the mass distribution in the galaxy (\citealp[e.g.,][]{McKee2015, Bland2016}).

One prominent approach is ``Jeans modeling'' (\citealp[e.g.,][and references therein]{Jeans1919,Binney2008}), which employs a set of equations (the Jeans equations) that relate spatial derivatives of second-order velocity moments to the gravitational potential. 
More general approaches instead attempt to model the DF explicitly (\citealp[e.g.,][]{Kuijken1991,Bovy2013,Piffl2014, Widmark2022,Li2021,Li2023,Cheng2023}). 
However, these models become computationally expensive for large data sets or are unable to incorporate flexible model forms. 
Moreover, these methods rely on an accurate accounting of the survey's selection function, making them challenging to apply precisely.

The advent of large-scale stellar surveys --- especially the \textsl{Gaia} mission
(\citealp{Gaia2016,Gaia2022}) --- has revolutionised Galactic astrophysics. 
We are now able to measure the phase-space coordinates of over a billion stars throughout the Galaxy.
Owing to large spectroscopic surveys, it is also now possible to obtain high-quality measurements of stellar parameters/labels (temperatures, surface gravities, metallicities, and other element abundances) for over a million stars in the Milky Way (e.g., \textsl{APOGEE}: \citealt{Majewski2017}; \textsl{GALAH}: \citealt{Martell2017}; \textsl{LAMOST}: \citealt{Cui2012}; amongst others). 
This unprecedented amount of high-quality information is supplying the necessary means to further develop dynamical inference frameworks. 
It also provides a window of opportunity to study the relationship between stellar labels, that should be approximately invariant over long timescales, with the kinematics and orbits of stars in the Galaxy. 

Recent results have suggested that combining spectroscopic and astrometric data for large samples can provide precise measurements of the Galactic mass distribution (\citealp[e.g.,][]{Sanders2015, Price2021, Binney2023}). 
One example of such techniques is Orbital Torus Imaging \citep[OTI;][]{Price2021}, which utilises chemical abundance information to directly map the orbit structure of stars in the Galactic disk. 
It uses the idea that, in a well-mixed (steady-state) population, abundance moments (like mean abundances) can depend only on actions, or dynamical invariants, or integrals of the motion; they cannot depend on orbital phases.
OTI was originally developed and used in the context of modeling the vertical kinematics of stars (i.e., position $z$ and velocity $v_{z}$ perpendicular to the Galactic disk). 
While this framework is still new, it is a promising avenue toward developing models of our Galaxy that relaxes many assumptions that have been necessary in past modeling efforts.
In particular, provided that the stars are selected in an abundance-neutral way, the method does not require much (or any) knowledge of the survey selection function.

In the first of a series of papers focusing on this topic, here we set out to combine stellar labels (namely, mean element abundances ratios) obtained from spectroscopic measurements using the Apache Point Observatory Galactic Evolution Experiment \citep[\textsl{APOGEE}][]{Majewski2017} survey, with kinematic observations from the latest \textsl{Gaia} data release 3 \citep{Gaia2022}, in order to
measure the surface mass density and vertical acceleration field across the disk of the Milky Way. 
We do this by constructing a model under the Orbital Torus Imaging framework, building on the method described in (Price-Whelan et al, in prep). 

In Section~\ref{sec_data} we describe the data and sample used. In Section~\ref{sec_abungrad} we show and discuss the abundance gradients in phase space. In Section~\ref{sec_modelling} we describe the method. Section~\ref{sec_results} contains our results, which are then discussed in Section~\ref{sec_discussion}. We finalise by providing our concluding thoughts in Section~\ref{sec_conclusions}.

\section{Data}
\label{sec_data}

\begin{figure*}
    \centering
    \includegraphics[width=\textwidth]{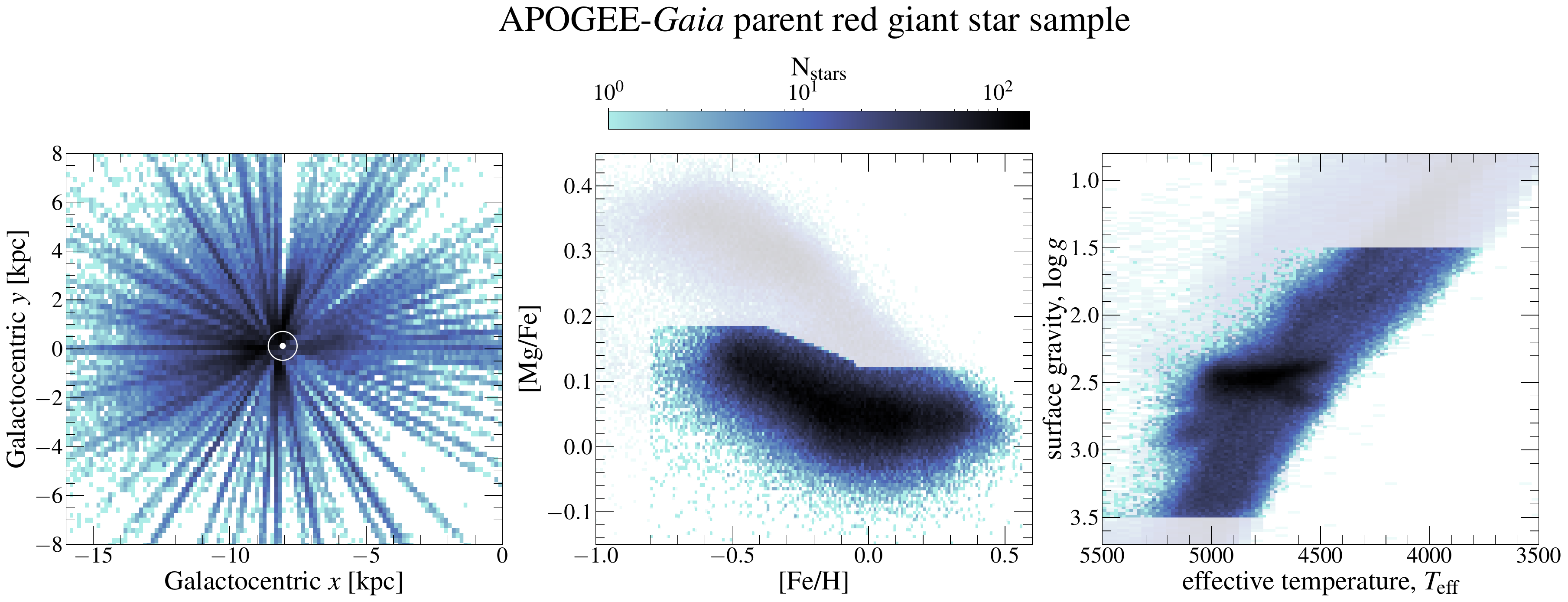}
    \caption{Our parent sample of red giant branch stars in the low-$\alpha$ disk sequence used in this work. The left panel shows a 2D histogram of the positions of stars in our sample projected onto the Galactic plane (i.e. in Galactocentric $x,y$ coordinates); The position of the Sun is marked with the $\odot$ symbol. The middle panel shows the selection used to define the low-$\alpha$ sequence using a cut in element abundance space, i.e., the \abun{Mg}{Fe} vs. \abun{Fe}{H} plane. This chemical cut was performed in order to restrict our sample to stars that are likely part of the Galaxy's thin disk. The right panel shows the spectroscopic stellar parameters, effective temperature \Teff\ and surface gravity \logg. In the middle and right panels, the full \textsl{APOGEE}--\gaia\ sample is shown as the faint background histogram.}
    \label{fig_data}
\end{figure*}

We use a cross-matched catalogue of the latest spectroscopic data from the \textsl{APOGEE} survey \citep[DR17][]{Majewski2017, SDSSDR17} and astrometric data from the \textsl{Gaia} mission data release 3 \citep[\textsl{Gaia} DR3][]{Gaia2022}. The \textsl{APOGEE} data are based on observations collected by two high-resolution, multi-fibre spectrographs \citep{Wilson2019} attached to the 2.5m Sloan telescope at Apache Point Observatory \citep{Gunn2006} and the du Pont 2.5~m telescope at Las Campanas Observatory \citep{BowenVaughan1973}, respectively. Element abundances are derived using the ASPCAP pipeline \citep{Perez2015} based on the FERRE code \citep[][]{Prieto2006} and the line lists from \citet{Cunha2017} and \citet[][]{Smith2021}. The spectra themselves were
reduced by a customized pipeline \citep{Nidever2015}. For details on
target selection criteria, see \citet{Zasowski2013} for \textsl{APOGEE}, \citet{Zasowski2017} for \textsl{APOGEE}-2, \citet[][]{Beaton2021} for \textsl{APOGEE} north, and \citet[][]{Santana2021} for \textsl{APOGEE} south. 

The \textsl{Gaia} mission/survey (\citealp[]{Gaia2016}) delivers detailed sky positions, proper motion, and parallax measurements for $\sim2$ billion stars, limited only by their apparent magnitudes (\textsl{Gaia} $G\lesssim20.7$). Here we use only astrometric (parallax and proper motion) measurements released in \textsl{Gaia} DR3 (\citealp[][]{Gaia2022}).

Together, the \textsl{APOGEE} and \textsl{Gaia} data supply full 6D phase space information from which kinematic and orbital properties can be derived. 
The \textsl{APOGEE} data additionally provide detailed chemical abundance measurements for up to $\sim20$ different elemental species spanning the $\alpha$, odd-Z, iron-peak, and $s$-process nucleosynthetic channels.

\subsection{Parent Sample}
The parent sample used in this work is comprised of stars that satisfy the following selection criteria:
\begin{description}
    \item[Red Giant Branch stars] \hfill \\ 
        \textsl{APOGEE}-determined atmospheric parameters, effective temperature and surface gravity $3500<\Teff<5500$~K and $1.5 < \logg < 3.5$,
    \item[High signal-to-Noise spectra] \hfill \\ 
        \textsl{APOGEE} spectral S/N $> 50$,
    \item[High-quality derived spectral parameters] \hfill \\ 
        \textsl{APOGEE} \texttt{STARFLAG} bits \textit{not} set to 0, 1, 3, 16, 17, 19, 21, 22 and \textsl{APOGEE} \texttt{ASPCAPFLAG} bits \textit{not} set to 23,
    \item[No star clusters] \hfill \\ 
        Stars that are not within the \textsl{APOGEE} globular cluster value added catalogue (Schiavon et al., in prep) or the catalogue from \citet{Horta2020},
    \item[Low-$\alpha$ disk] \hfill \\ 
        Stars that fall in the ``low-$\alpha$'' disk sequence (see Figure~\ref{fig_data}), and thus are likely part of the Milky Way's thin disk.
\end{description}

Our selection yields a total of 172,656 stars in our \textsl{APOGEE}-\textsl{Gaia} parent data sample; These stars are shown in Figure~\ref{fig_data} compared to the full \textsl{APOGEE}-\textsl{Gaia} cross-matched catalog in spatial location within the Galaxy (left panel), element abundance measurements (\abun{Mg}{Fe} vs. \abun{Fe}{H}; middle panel) and spectroscopic parameters (right panel).
We use the 6D phase space information\footnote{The positions, proper motions, and distances are taken from \textsl{Gaia} DR3 data, whilst the radial velocities are taken from \textsl{APOGEE} DR17. We use the photo-astrometric StarHorse distances \citep{Anders2022}.} to compute Galactocentric Cartesian coordinates assuming a solar velocity $\boldsymbol{v}_\odot = (8.4, 251.8, 0)~\kms$ and distance to the Galactic center $R_{0} = 8.275~\kpc$ \citep{Gravity2022}. 
Note that we zero-out the $z$-component of the solar velocity, $v_{z,\odot}$, and set the solar height above the galactic midplane to $z_\odot = 0$ as we measure these quantities below.

Lastly, for our modelling we only use stars within $|z| < 2~\kpc$ from the solar position, and have vertical velocities with respect to the Sun smaller than $|v_z| < 80~\kms$. This removes likely halo contaminants and/or stars not part of the Milky Way disk.
We additionally select stars that have small eccentricities (i.e. on near-circular orbits) by enforcing $|v_{R}|<50~\kms$
and $|\Delta R| < 1~\kpc$, where $R$ is the present-day cylindrical radius of each star, and $\Delta R=R - R_{g}$, where $R_{g} = L_{z}$/$v_{\mathrm{circ}}$ is the approximate guiding-center radius of each star computed  using the rotation curve from \citet{Eilers2019}. We chose a cut of $|v_{R}|<50~\kms$ as after inspecting the $\Delta R$--$v_R$ plane, where we find that the approximate maximum radial velocity a star with $|\Delta R| < 1~\kpc$ could reach is $v_R\sim50~\kms$.
Our final working sample is comprised of 94,685 stars.

\section{Element abundance gradients in vertical kinematics}
\label{sec_abungrad}

\begin{figure*}[t!]
    \centering
    \includegraphics[width=\textwidth]{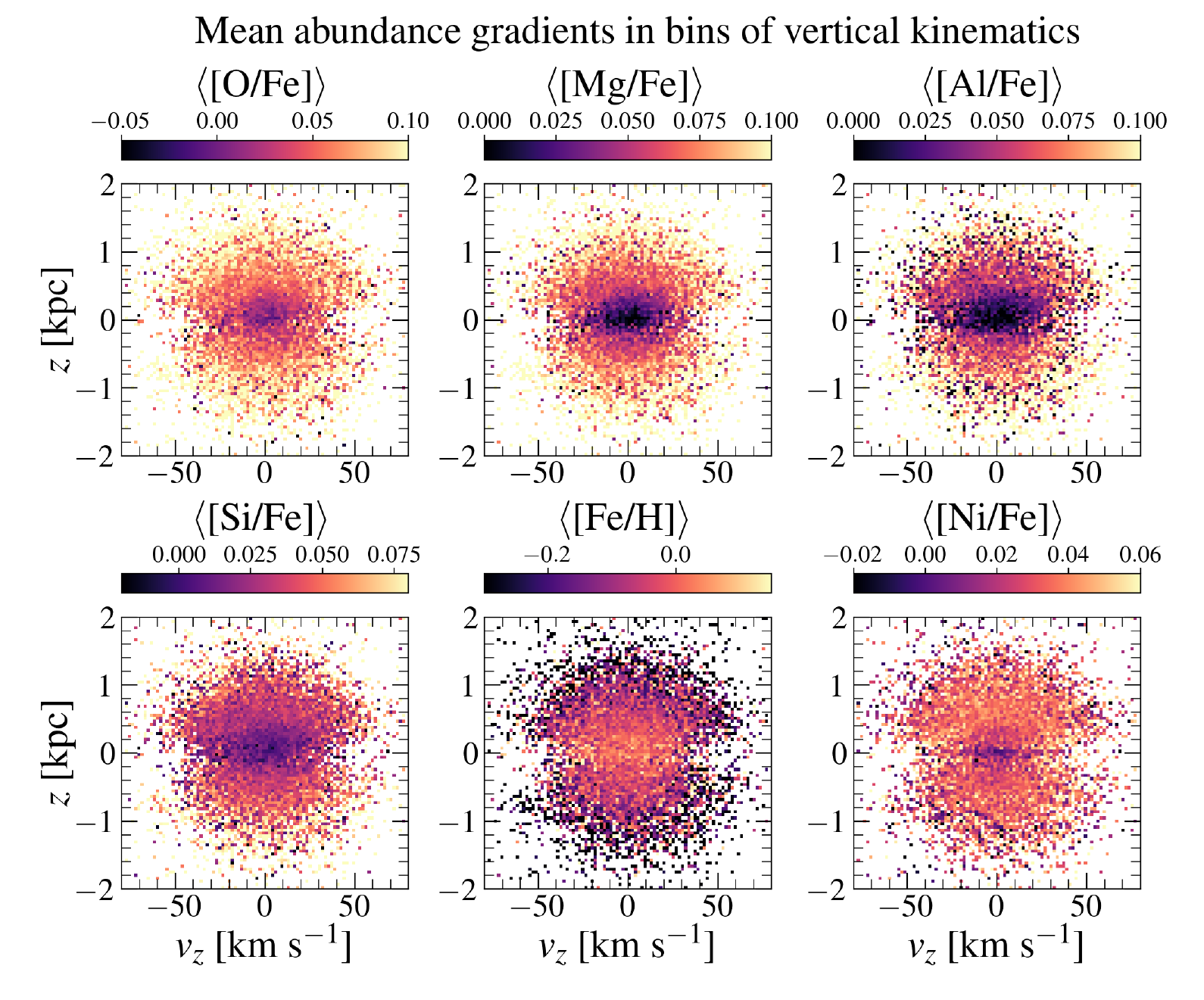}
    \caption{Each panel shows the mean element abundance, $\langle\mathrm{[X/Y]}\rangle$, of stars in bins of vertical phase space coordinates, $z, v_z$, for the abundance ratios stated in the panel titles.
    All stars are red giant branch stars in the low-$\alpha$ disk sequence (see Section~\ref{sec_data}) around the Sun (i.e., $|R - R_{\odot}| < 0.5~\kpc$) with close-to-circular orbits ($|R - R_g|<1~\kpc$ and $|v_{R}|<50~\kms$). 
    The typical spread in abundance values in each pixel (as estimated with the mean absolute deviation) is $\lesssim0.02$ dex, except for [Fe/H] where it is $\lesssim 0.08$ dex. 
    We only show the abundances that have the smallest scatter in our low-$\alpha$ sample and/or are best determined by ASPCAP/\textsl{APOGEE}.}
    \label{fig_abun}
\end{figure*}

The stars in the Milky Way disk have a well-known dependence on position, velocity, and chemical abundance within the Galaxy.
For example, there is a known interrelation between Galactic position, iron abundance [Fe/H], and age for stars in the Galactic disk \citep[e.g.,][]{Frankel2018}. 
The Milky Way disk has also been shown to be homogeneous in terms of its star formation environment for most elements (\citealp[e.g.,][]{Ness2019,Ness2022,Horta2022}), and to show a continuity in mono-abundance populations (\citealp[e.g.,][]{Bovy2016,Mackereth2017}). 
These element abundance gradients also exist with respect to the vertical kinematics of stars: the mean abundance (in any abundance ratio) of stars in the Galactic disk typically depends on both the vertical height $z$ and vertical velocity $v_{z}$ of the stars \citep[e.g.,][]{Price2021}. 
We leverage these gradients by developing a model in the Orbital Torus Imaging framework to fit mono-abundance contours in the vertical phase space ($z,v_z$) for different element abundance ratios.

In this section, we first visualize the available abundance data as a function of vertical phase space coordinates (Section~\ref{sec_abungrad_2}) and show how these abundance gradients contain information about the orbital structure of the Galactic disk (Section~\ref{sec_ti}). 
We then show that some element abundance ratios have a stronger dependence on orbital properties, which should make them more informative for our analysis (Section~\ref{abun_shape}). 
Finally, we explore the dependence of the vertical abundance distribution as a function of spatial location in the disk for our most promising abundance ratio, \abun{Mg}{Fe} (Section~\ref{sec_abunshape}).

\subsection{Exploring different element abundances}
\label{sec_abungrad_2}

Figure~\ref{fig_abun} shows mean element abundances, $\langle\mathrm{[X/Y]}\rangle$, binned by vertical phase space coordinates, ($z,v_z$), for a subset of our parent sample positioned around the Sun that have low eccentricities. Each pixel has a size of $\delta z = 0.04~\kpc$ and $\delta v_z = 1.76~\kms$. Specifically, we show six element abundance ratios provided by the \textsl{APOGEE} survey that represent the $\alpha$ elements (O, Mg, Si), odd-Z elements (Al), and iron peak elements (Fe and Ni). 
We provide equivalent plots for all elements measured by \textsl{APOGEE} in Appendix~\ref{app_gradients}. 
The six elements shown in Figure~\ref{fig_abun} are, in theory, the most reliable in the \textsl{APOGEE} survey \citep{Jonsson2020}. 
With these six elements, we are sensitive to differences in contributions from different supernova (SN) types (SN II and SN Ia) for stars with different orbital properties.

Upon inspection of $\langle\mathrm{[X/Y]}\rangle$ in ($z,v_z$), it is clear is that there are gradients with respect to both height $z$ and vertical velocity $v_z$. For elements synthesised predominantly in the explosions of SN II (like $\alpha$ and odd-Z), the gradient in $\langle\mathrm{[X/Y]}\rangle$ is positive with increasing $z,v_z$. Conversely, for heavier elements like [Fe/H], synthesized in both SN II and SN Ia, the gradient in $\langle\mathrm{[X/Y]}\rangle$ is in the opposite direction, growing in magnitude with decreasing $z,v_z$. We also note that the range, [min$_{\langle\mathrm{[X/Y]}\rangle}$,max$_{\langle\mathrm{[X/Y]}\rangle}$], in the gradient for each element abundance is different. 
We find that \abun{Fe}{H} has the largest range, spanning from $\sim -0.3-0.1$~dex, and \abun{Ni}{Fe} has the smallest range, spanning just $\approx 0.06$~dex --- the fact that such a small gradient is detectable speaks to the high quality element abundance measurements produced by the \textsl{APOGEE} survey.

Another interesting property that is apparent in the element abundance gradients in $z$,$v_z$ space is the shape of the mono-abundance contours. Figure~\ref{fig_abun} reveals that both the shape and element abundance gradient are more pronounced/steeper for element abundances synthesised via SN II (for example, Mg and Si) as compared with other elements (e.g., Ni). These small yet noticeable differences are likely caused by different elements being synthesised at different rates, leading to stellar populations manifesting different mono-abundance contour shapes in phase space. Despite such differences, the fact that we observe mono-abundance $z,v_z$ contours for all elements is proof that this chemical-dynamical data is rich in information, and in principle should retain vital clues about the structure of the Galactic disk.

\subsection{Mapping orbits with mono-abundance contours}
\label{sec_ti}

As seen in Figure~\ref{fig_abun}, there are clear gradients of $\langle\mathrm{[X/Y]}\rangle$ in $z,v_z$ space. Stars at small heights $z$ and small vertical velocities $v_z$ have (on average) different mean element abundance ratios than stars at larger $z$ or $v_z$. 
In fact, these $\langle\mathrm{[X/Y]}\rangle$ gradients are related with respect to phase space position in $z,v_z$. 
Stars that have higher vertical velocities will reach higher maximum heights above the disk, $\zmax$, as they orbit, and vice versa. 
As stars orbit in this phase plane, their trajectories roughly follow elliptical shapes. 
Those elliptical trajectories are the shapes we see in Figure~\ref{fig_abun} for pixels of constant $\langle\mathrm{[X/Y]}\rangle$.

\begin{figure}[t!]
    \centering
    \includegraphics[width=\columnwidth]{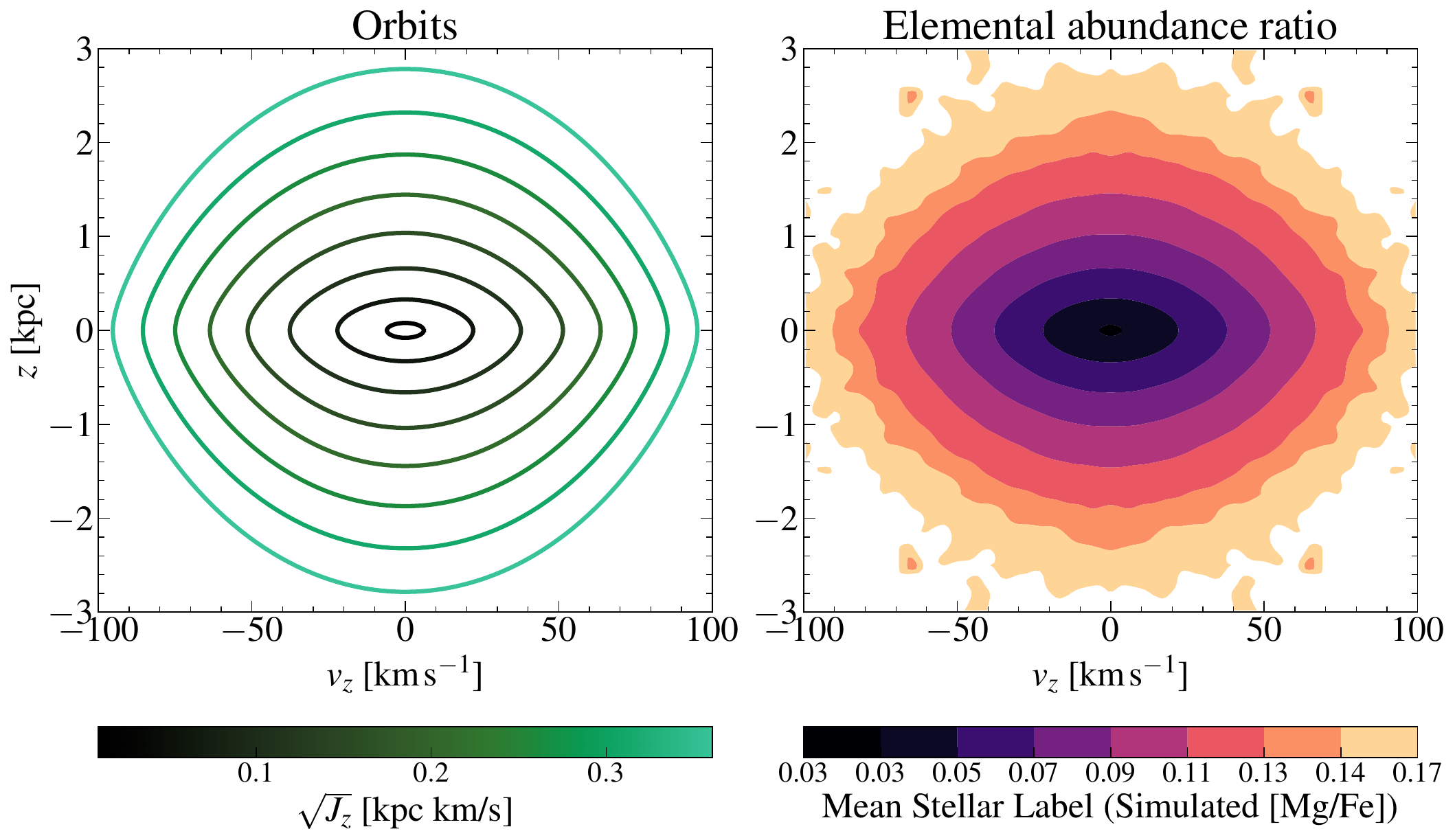}
    \caption{This figure demonstrates, with simulated data, that contours of constant element abundance ratio (right panel) nearly delineate orbits in $z,v_z$ space (left panel). 
    \textit{Left}: This panel shows eight orbits with different values of the vertical action $J_z$ computed in a simple Milky Way mass model.
    For all of these orbits, the value of the other actions are set such that the orbits have zero radial action, $J_R=0$, and a constant azimuthal action set to the solar value, $J_\phi = J_{\phi, \odot}$.
    Orbits with lower $J_z$ values are more elliptical and reach smaller maximum heights from the Galactic plane, \zmax. 
We note that this is our only usage of a parameterized Galactic mass model in this work, and this is only for demonstration purposes.}
    \textit{Right}: We paint element abundances onto the simulated stars with a linear dependence on vertical action $J_z$ with a slope similar to that of \abun{Mg}{Fe} (see Figure~\ref{fig_abun}).
    The filled contours then show curves of constant element abundance ratio for a simulated population of orbits in the vertical phase space.
    See Price-Whelan et al. (in prep) for more details.
    
    \label{fig_contours}
\end{figure}

Figure~\ref{fig_contours} illustrates that orbital trajectories in $z,v_z$ space are identifiable using curves of $\langle\mathrm{[X/Y]}\rangle$ gradients seen in Figure~\ref{fig_abun}. 
The left panel shows eight orbital trajectories in $z,v_z$ with different values of the vertical action, $J_z$, obtained from using a toy model of the Galactic potential from Price-Whelan et al., (in prep).
We note that this is our only usage of a parameterized Galactic mass model in this work, and this is only for demonstration purposes.
The orbits of stars with lower $J_z$ values trace out nearly elliptical trajectories that reach smaller heights above the Galactic plane, \zmax, than stars with larger $J_z$ values. 
The right panel shows the same vertical phase plane but now shows contours of constant $\langle\mathrm{[X/Y]}\rangle$, computed by painting on chemical abundance values to stars based on a linear relation between $\langle\mathrm{[Mg/Fe]}\rangle$ and the vertical action $J_z$ (see Figure~\ref{fig_jz-z}). Under assumptions enumerated below (see Section~\ref{model_assump}), contours of constant $\langle\mathrm{[X/Y]}\rangle$ in $z,v_z$ delineate orbits. In what follows, we use this link between element abundances and orbits to model the mass distribution across the Galactic disk. 

\subsection{Identifying informative element abundance gradients}
\label{abun_shape}

\begin{figure*}[t!]
\centering
\includegraphics[width=\textwidth]{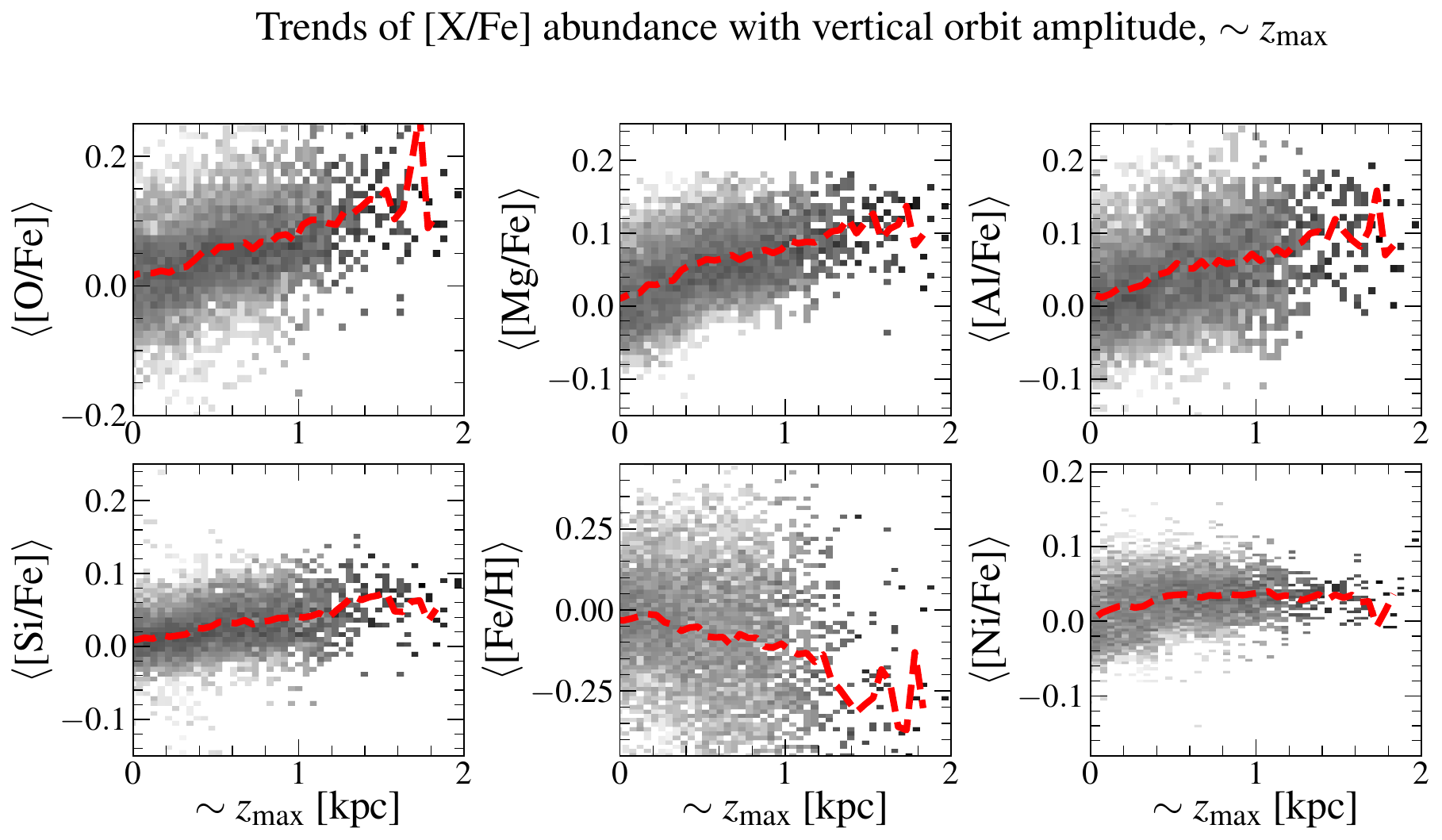}
\caption{Column-normalized histograms of element abundance values as a function of approximate maximum vertical height, $\sim z_{\mathrm{max}}$, for stars with orbits similar to the Sun (i.e., $R_g - R_{g, \odot} < 0.5~\kpc$) with low vertical velocities, $|v_z| < 10~\kms$, from our parent sample. Each panel shows a different element abundance ratio indicated on the vertical axis. The dashed (red) line shows the running mean as a function of $\sim z_{\mathrm{max}}$. 
A steeper gradient in the mean abundance of a given panel indicates that a given abundance ratio will lead to better constraints on the orbit structure. The abundance ratio with the steepest and smoothest trend between element abundance and vertical kinematics is \abun{Mg}{Fe}.}
\label{fig_jz-z}
\end{figure*}

\begin{figure*}[t!]
    \centering
    \includegraphics[width=\textwidth]{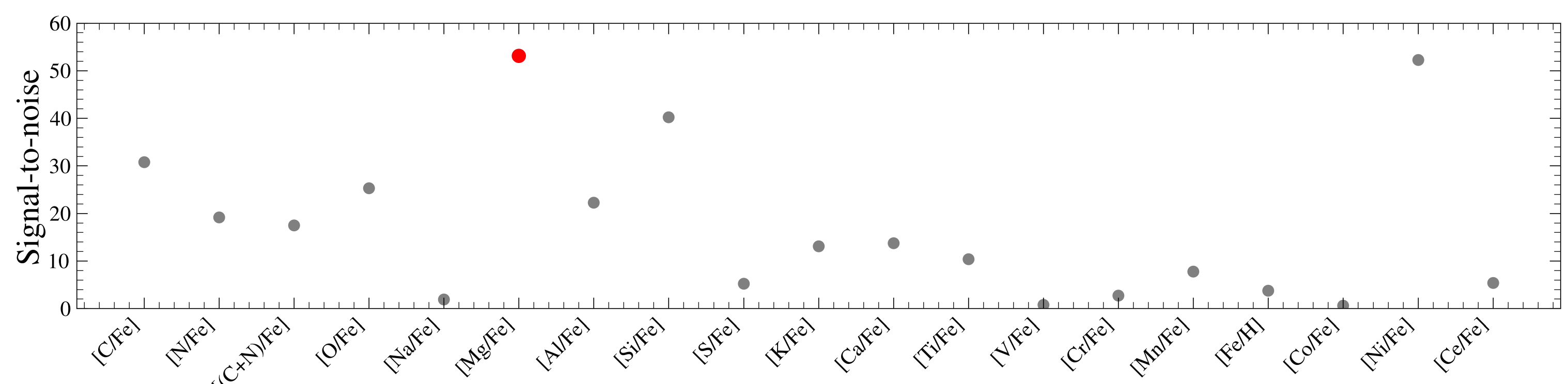}
    \caption{Signal-to-noise measurement, computed by taking the ratio of the slope, $\Delta\langle\mathrm{[X/Y]}\rangle/\Delta\sim z_{\mathrm{max}}$, with the average variance around the median line, $\sigma_{\langle\mathrm{[X/Y]}\rangle}^{2}$, for all element abundance ratios supplied by APOGEE. The higher the signal-to-noise ratio, the more constraining power an element abundance ratio has for modelling the Milky Way disc in $z,v_z$ space. Quantitatively, [Mg/Fe] is the element abundance ratio with the highest constraining power. 
    }
    \label{fig_xferat}
\end{figure*}

The shapes of contours of constant $\langle\mathrm{[X/Y]}\rangle$ in Figure~\ref{fig_abun} nearly correspond to orbits in the vertical phase space (Figure~\ref{fig_contours}). 
However, our ability to identify contours of constant $\langle\mathrm{[X/Y]}\rangle$ will depend on the steepness and scatter of the gradients of $\langle\mathrm{[X/Y]}\rangle$ vs. vertical action $J_z$.
Thus, in order to assess which element abundance ratio we expect to give the best constraining power, we investigate the trends between our set of well-measured \textsl{APOGEE} abundances as a function of a proxy for the maximum $z$ height a star reaches, $\sim\zmax$. 
To compute \zmax, which is related to the vertical action $J_z$, one has to assume a form for the Galactic mass distribution, which we are trying to avoid.
We instead here sub-select our data to have $|v_z| < 10~\kms$ so that those stars' instantaneous vertical positions, $z$, can serve as a proxy for \zmax --- i.e. $z_{|v_z| < 10~\kms} \sim \zmax$.
Figure~\ref{fig_jz-z} shows the resulting dependence of each element abundance ratio as a function of our potential-independent proxy for the maximum $z$ height, $\sim\zmax$.
The pixels show the column-normalized densities of stars in this space for stars with orbits similar to the Sun, $R_g - R_{g, \odot} < 0.5~\kpc$, where $R_g$ is the guiding-centre radius.
All of our selected element abundance ratios show a trend in their mean value (red dashed line) below $z \lesssim 1~\kpc$. 
Above this height, $z \gtrsim 1~\kpc$, the clearest trends are shown by the elements in the $\alpha$ group, with [Mg/Fe] being the clearest. 

To quantify the power an element abundance ratio has for inferring the orbital structure of the Milky Way disk, in Figure~\ref{fig_xferat} we show the signal-to-noise value for every element abundance ratio supplied by APOGEE. This value is computed by taking the ratio between the slope of $\langle\mathrm{[X/Y]}\rangle$ with $\sim z_{\mathrm{max}}$ with the average variance around the median line, $\sigma^{2}_{\langle\mathrm{[X/Y]}\rangle}$. Figure~\ref{fig_xferat} confirms our qualitative analysis from Figure~\ref{fig_jz-z}, and shows that [Mg/Fe] is the element abundance ratio with the most constraining power.
Given this test, we will use [Mg/Fe] as our representative element abundance ratio for our model fitting (Section~\ref{sec_modelling}).

\subsection{Changes in vertical mono-abundance contours across the Galactic disk}
\label{sec_abunshape}

\begin{figure*}
    \centering
    \includegraphics[width=\textwidth]{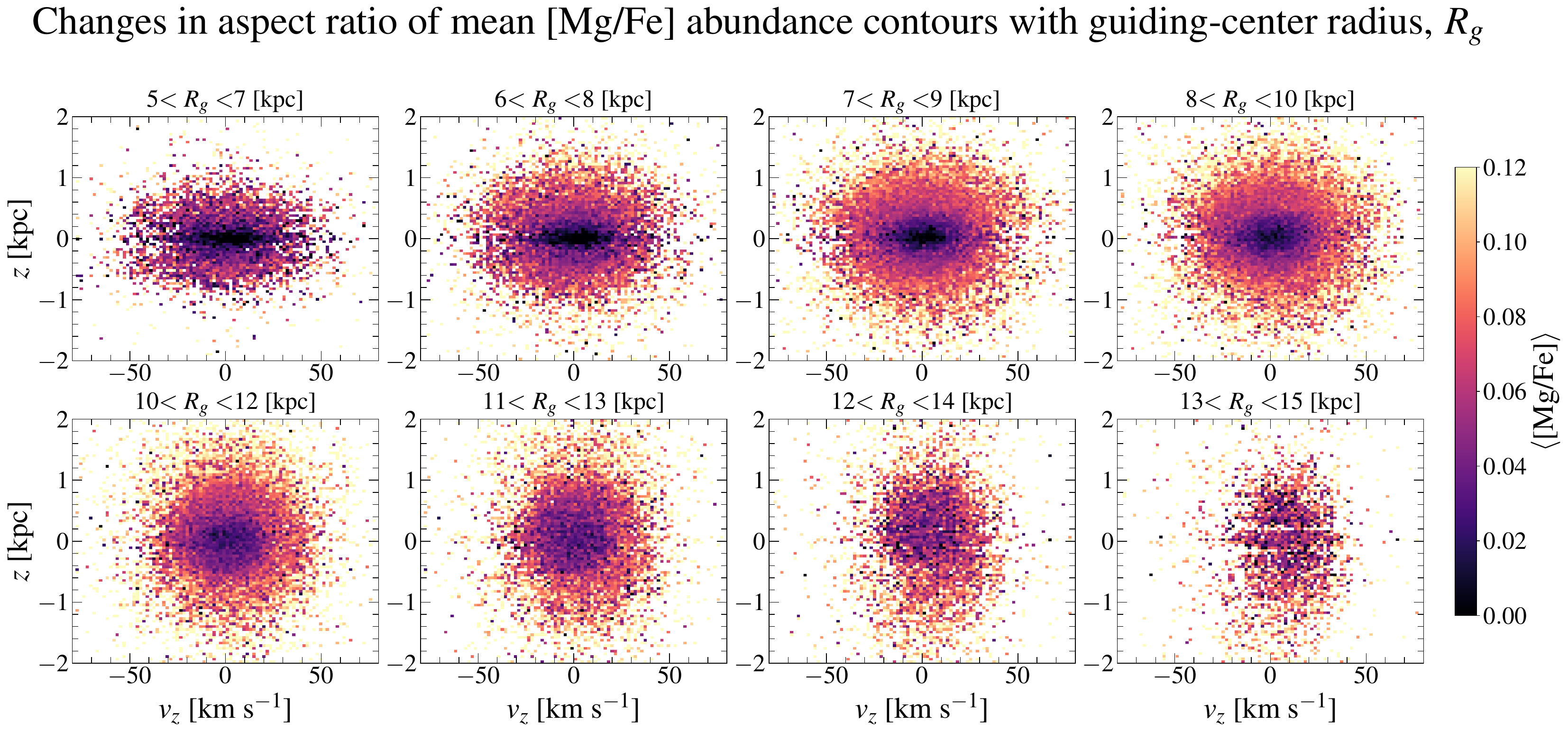}
    \caption{The same as Figure~\ref{fig_abun} but now showing only $\langle$\abun{Mg}{Fe}$\rangle$ and binned into eight overlapping bins of guiding-center radius, $R_g$.
    Each panel shows a $2~\kpc$-wide bin of $R_g$, increasing from the top left (inner galaxy) to the bottom right (outer galaxy), and the Sun would appear in the top right panel.
    The aspect ratio of each panel is kept constant.
    Stars orbiting in the inner galaxy have more oblate contours of constant abundance as compared to stars further out, which is especially apparent in for the low-\abun{Mg}{Fe} contour values.
    For example, compare the darker colors in the two panels in the far left column.
    This is a result of the surface mass density, as stars in the inner galaxy feel a stronger gravitational pull and will therefore reach smaller vertical heights for a given vertical velocity.}    \label{fig_spatialbins}
\end{figure*}

Figure~\ref{fig_abun} shows element abundance trends with vertical kinematics for stars around the Sun.
However, our sample contains stars with a wide range of orbits in the Galactic disk, owing to the exquisite volume and precision of data from \textsl{Gaia}.
In Figure~\ref{fig_spatialbins}, we therefore examine how the $\langle\mathrm{[X/Y]}\rangle$ contours in vertical kinematics for \abun{Mg}{Fe} vary as a function of orbital location across the Galactic disk.
We illustrate this using the same visualization of $\langle\mathrm{[Mg/Fe]}\rangle$ in $z,v_z$, but now divide our sample into bins of similar guiding-center radius, $R_g$, from top left to bottom right going from orbits in the inner to outer Galactic disk. Only stars with $5 < R_g < 15~\kpc$ are shown, binned in 2 kpc-wide bins for illustration purposes.

Figure~\ref{fig_spatialbins} shows that the morphology of constant $\langle\mathrm{[Mg/Fe]}\rangle$ contours varies as a function of guiding-center radius. 
Stars in the inner Galaxy (top left) show a more oblate mono-abundance contour than stars in the outer Galaxy (bottom right). 
This ``squashing'' is around $z$ for different $v_z$, and becomes less pronounced as one increases in $R_g$, turning from an oblate shape to a more circular and then prolate shape. 
As we will see in Section~\ref{sec_results}, the change in mono-abundance contour shapes is related to the surface mass density in the disk. 
Towards the inner disk, where the surface mass density within a given column is higher, stars in the Galactic disk feel a stronger gravitational influence as they oscillate above and below the Galactic midplane. 
This leads to stars of a given element abundance ratio not being able to reach larger vertical heights for a given vertical velocity value, which leads to flattened mono-abundance contours in $z,v_z$ space. 
Conversely, towards the outer regions of the Galaxy, where the surface mass density in the Galactic disk is lower, and the influence from the gravitational pull of the dark halo is stronger, stars are able to reach higher vertical heights given a particular vertical velocity magnitude. This leads to a more prolate and then diamond-shaped profile for stars with a given element abundance ratio.

The fact that we observe such changes in the morphology of constant $\langle\mathrm{[Mg/Fe]}\rangle$ in $z,v_z$ highlights that the disk is rich in chemical-dynamical structure that is useful for inferring properties of the Galaxy. 
From the contours of constant $\langle\mathrm{[Mg/Fe]}\rangle$, we can already see, conceptually, how these data can be used to infer mass density: 
The aspect ratio of an ellipse in these panels has units of frequency (i.e. $v_z / z$), and frequency is related to density through Poisson's equation, 
\begin{equation}
    4\pi G\rho(z) \approx \Omega_{z}^{2}.
\end{equation}

In the following section, we describe a method for empirically modelling the mono-abundance $z,v_z$ contours, following the method from Price-Whelan et al. (in prep.).

\section{Orbital Torus Imaging: Modeling the orbital structure and mass of the disk with element abundances}
\label{sec_modelling}

Orbital Torus Imaging (OTI) is a framework for exploiting the idea that, in a steady-state galaxy, the distribution of intrinsic stellar invariants of a tracer population can only depend on dynamical invariants and not on time or orbital phase.
In a 2D slice of phase-space, for example the vertical kinematics $\boldsymbol{w} = (z, v_z)$ for a small region of the Galactic disk, an implication of this is that contours of constant mean element abundance should correspond to orbital trajectories.
With orbital trajectories, we can measure properties of the acceleration field and mass distirbution of the Galaxy.
A full description of our methodology is described in a companion paper (Price-Whelan et al., in prep.), but here we briefly summarize the framework we use to infer the disk properties from the combined vertical kinematics and element abundance data.
Our model for fitting $\langle\mathrm{[X/Y]}\rangle$ works only in $z,v_z$, but we will explore how the properties of the disk vary as a function of radius by binning the data.
Our framework is summarized in the following steps:
\begin{enumerate}
    \item We bin the the parent sample data into 25 equally-sized (2 kpc-wide) but overlapping bins of guiding-center radius, $R_g$, ranging from $5 < R_g < 13~\kpc$. For each bin of $R_g$, we compute $\langle\mathrm{[X/Y]}\rangle$ and the median absolute deviation (MAD, defined as $\mathrm{median}(|X_{i}-X|)$, where $X$ is $\langle\mathrm{[X/Y]}\rangle$), for each of our element abundance ratios in small bins ($\delta_z=0.04~\kpc,\delta_{v_{z}}=1.76~\mathrm{km}~\mathrm{s}^{-1}$) of $z, v_z$.
    \item We fit a flexible, empirical model of the element abundance trends to the pixelized $\langle\abun{X}{Y}\rangle$ values in $z,v_z$ space. Using this model, we predict the values of $\langle\abun{X}{Y}\rangle$ in every $\boldsymbol{w}=(z,v_z)$ pixel, $p(\langle\abun{X}{Y}\rangle\,|\,\boldsymbol{w}$), using the method described in Price-Whelan et al. (in prep.) \footnote{The accompanying software tool can be found at \url{https://github.com/adrn/TorusImaging}.}. We use the empirical scatter of the abundances in each pixel (the MAD values) scaled by $1/\sqrt{N}$, where $N$ is the number of stars in a pixel, as our uncertainty when fitting this model. 
    \item We optimize the following likelihood function,
    \begin{equation}
    \label{log-like}
        \ln \mathcal{L}(\theta|\langle\abun{X}{Y}\rangle(\boldsymbol{w})) = \sum\limits_{i=1}^{N}\frac{Q_{i}^{2}}{2},
    \end{equation}
    where $Q$ is 
    \begin{equation}
        Q = \frac{(\langle\abun{X}{Y}\rangle(\boldsymbol{w})_{\mathrm{data}} -\langle\mathrm{[X/Y]}\rangle(\boldsymbol{w})_{\mathrm{model}})}{\xi},
    \end{equation}
    for each pixel in $\boldsymbol{w}=(z,v_z)$ space. Here, $\xi=1.5\times\mathrm{MAD}(\boldsymbol{w})_{ \mathrm{data}}/\sqrt{N_{\mathrm{stars}}(\boldsymbol{w})}$, and $\theta$ is the set of model parameters we aim to optimize (described in detail at the end of this Section).
    \item We use the best-fit model and interpret the level sets of the fitted function as orbits, allowing us to estimate the vertical acceleration $a_z(z)$, surface mass density $\Sigma(z)$, and midplane volume density $\rho_{\odot}(z=0~\kpc)$ for each bin of guiding radius and for each element abundance ratio. To compute the uncertainties on these measured quantities, we bootstrap sample with replacement, instead of using the standard error estimate.
\end{enumerate}

In the following, we briefly outline the Orbital Torus Imaging approach and assumptions that go into this method and discuss how we map orbits from element abundance gradients (Section~\ref{model_assump}), before presenting the method for determining the vertical acceleration and surface mass density from our model (Section~\ref{method_acc}).

\subsection{Modeling approach and underlying assumptions}
\label{model_assump}

The core idea of Orbital Torus Imaging (OTI) is that any stellar label (e.g., element abundance ratios) that correlates intrinsically with the orbital properties of stars in the Galaxy can be used in place of dynamical quantities to ``image'' the orbits. The element abundance distribution should be constant along an orbit, and stars at different phases along an orbit should have the same element abundance distribution. Stellar labels can therefore be used to trace the orbits by finding curves of constant label value (see \citealt{Price2021} for more details).
Once the shapes of orbits are traced using this method, we can use standard dynamical arguments to extract quantities of interest like the acceleration, surface mass density, and midplane mass density.
The advantage of this approach is that our model is very flexible: we never parameterize the global form of the gravitational potential and instead only locally (in guiding radius) parameterize the shapes of orbital trajectories in the vertical phase space.
Our model form and methodology is described in detail in Price-Whelan et al. (in prep), which is based on ideas presented previously \citep{Price2021}.

Our current implementation of OTI relies on a set of assumptions described below:
\begin{description}
    \item[Axisymmetric and separable] The gravitational potential that the stars orbit in is axisymmetric, smooth, and separable in cylindrical radius $R$ and vertical position $z$: $\Phi(R,z) = \Phi(R)+\Phi(z)$.
    This allows us to work with just one position and velocity component (namely, $z$ and $v_{z}$), but this could equally be applied to radial kinematics $R$ and $v_{R}$.
    In future work, we will explore models that simultaneously model the radial and vertical phase space.
    \item[Near circular orbits] For the previous assumption to be a valid simplification, we also require that stars have negligible eccentricity (or close to zero radial action $J_R=0$) and have the same $z$-component of the angular momentum $L_z$ (or azimuthal action $J_\phi$). In order to get close to this, we have restricted our disk sample to $|R-R_{g}|< 1~\kpc$ in any bin of guiding radius $R_g$, and have ensured the stars have low radial velocities (or radial action), $|v_R|<50~\kms$.
    \item[Phase-mixed] The stellar distribution function in vertical kinematics, $f(z, v_z)$, in any bin of guiding radius is phase mixed. This assumption is the key assumption; it is what ensures that contours of constant element abundance will correspond to curves of constant orbital actions (or other invariants).
    \item[Negligible kinematic measurement uncertainties] When fitting our model below, we always bin our stellar data into small bins of vertical phase-space coordinates ($\delta z,\delta v_z$). We assume that most measurements of vertical position $z$ and velocity $v_z$ for stars in our sample are more precise than our adopted bin sizes, so we ignore measurement uncertainties on the kinematic quantities.
\end{description}

For a slice of phase space at fixed values of the other actions (namely, $J_{R}$ and $J_{\phi}$), under our assumptions, contours of constant phase space $\langle\mathrm{[X/Y]}\rangle$ delineate orbital trajectories \citep{Price2021}. This concept is powerful and motivates a path towards empirically mapping the orbital structure in vertical kinematics directly from the observed element abundance ratios in a slice of phase space.

Our method works by modeling the shapes of level sets of $\langle\mathrm{[X/Y]}\rangle$ in $z,v_z$ space.
We do this by parameterizing the shapes of the contours as a low-order Fourier distortion away from ellipses of constant axis ratio (i.e. constant frequency).
Very near the Galactic midplane, where the density distribution is approximately uniform, we expect the orbits (and therefore contours of constant $\langle\mathrm{[X/Y]}\rangle$) to be very nearly elliptical.
For orbits that reach larger heights above the plane, we expect their frequencies to be smaller (i.e. they have longer oscillation periods).
For orbits that reach many scale heights above the midplane (i.e. $\zmax \gtrsim 1~\kpc$), the influence of the Milky Way's dark matter halo will further distort the shapes of the orbits to be more diamond-shaped and less elliptical.

We therefore parameterize the shapes of these level sets using a mixture of $m=2$ and $m=4$ Fourier distortions away from a constant axis ratio ellipse.
That is, we define an elliptical (euclidian) distance $\rzp$ and angle $\thzp$ in the vertical phase space,
\begin{align}
    \rzp &= \sqrt{\Delta z^2\, \freqzero + \Delta v_z^2\,\freqzero^{-1}} \\
    \thzp &= \tan^{-1}\left(\frac{\Delta z}{\Delta v_z}\freqzero\right)
\end{align}
where 
\begin{align}
    \Delta z &= z - z_0\\
    \Delta v_z &= v_z - v_{z,0}
\end{align}
and $z_0, v_{z, 0}$ are parameters that set the solar position and velocity with respect to the midplane, and $\freqzero$ is a parameter that sets the frequency of an orbit with zero $\rzp$.
We parameterize the level sets of the element abundances as curves of constant \textit{distorted} distance $\rz$, defined as
\begin{equation}
    \rz = \rzp \, \left[1 + \sum_m^{\{2, 4\}} e_m(\rzp) \, \cos(m\thzp) \right] \quad .
    \label{eq:rz}
\end{equation}
In the above expression, the functions $e_2(\rzp)$ and $e_4(\rzp)$ are yet-to-be specified amplitude functions that set the oblateness of the contours and diamond-shaped distortion due to the dark matter halo.
We require that these functions go to zero as $\rzp \rightarrow 0$ so that \freqzero\ can be interpreted as the asymptotic midplane orbital frequency.
As we expect the orbits to get more prolate with increasing $J_z$, we require that the $m=2$ coefficient function ($e_2$) is monotonic and strictly positive.
We also expect the orbits to become more ``diamond-shaped'' at large $J_z$ and not more square, so we require the $m=4$ coefficient function to be monotonic and negative.
To enforce these things within a fitting procedure, we represent the functions $e_2(\rzp)$ and $e_4(\rzp)$ as monotonic quadratic splines.
We fix the knot locations of the splines to be equally spaced in $\rzp^{2}$, and treat the function derivative values at the fixed knot locations as free parameters.

With the above, we have a formalism for labeling (and foliating) curves in the vertical phase space using the distorted elliptical distance $\rz$ (Equation~\ref{eq:rz}).
To fit $\langle\abun{X}{Y}\rangle$ as a function of $z,v_z$, we then must also model the dependence of $\langle\abun{X}{Y}\rangle$ as a function of this distorted radius, i.e. $\langle\abun{X}{Y}\rangle(\rz)$.
We represent this function also using a monotonic quadratic spline. The knot locations of the splines are also set to be equally spaced in $\rzp^{2}$.

Our model is a generative model of $\langle\abun{X}{Y}\rangle$ at any location in $z, v_z$.
We fit the model parameters by maximizing the log-likelihood of the data (Equation~\ref{log-like}) --- the value of the $\langle\mathrm{[X/Y]}\rangle$ at locations of $z,v_z$ space --- by treating the MAD value in each pixel of $z, v_z$ as our uncertainty, and ignoring any uncertainties in the phase space coordinates.
Our implementation is built on the \texttt{JAX} framework \citep{jax2018github} such that we can easily auto-differentiate through our model, allowing us to efficiently use a gradient-based optimizer (BFGS; \citealt{Flet87}) to fit our model parameters.
For a given element abundance ratio, for each bin of guiding radius, our model has eighteen free parameters: the midplane frequency \freqzero, the solar position and velocity $z_0$ and $v_{z, 0}$, the spline function derivative values (10) for the $e_2$ and $e_4$ coefficient functions, and the spline function values (5) for the abundance function $\langle\abun{X}{Y}\rangle(\rz)$.

\subsection{Measuring the vertical acceleration field and surface mass density}
\label{method_acc}

Any function of the phase-space coordinates, $F$, that is time-invariant and constant along an orbit (i.e. a constant of the motion), $F(z(t), v_z(t)) = \textrm{const.}$, provides a means to measure the acceleration:
\begin{align}
    \deriv{F}{t} &= \pderiv{F}{t} + \pderiv{F}{z}\,\deriv{z}{t} + \pderiv{F}{v_z}\,\deriv{v_z}{t} \\
    0 &= \pderiv{F}{z}\,v_z + \pderiv{F}{v_z}\,a_z \\
    a_z &= -v_z \, \pderiv{F}{z} \, \left(\pderiv{F}{v_z}\right)^{-1} \quad .
\end{align}
For a given setting of our parameters (see Section~\ref{model_assump}), we can compute the vertical acceleration inferred by our model using the chain rule applied to our orbit label function, the distorted elliptical radius $r_z$, which, by construction, is a constant of the motion.
As we are interested in evaluating the acceleration as a function of $z$ and not $v_z$, we take the limit of this expression as $v_z \rightarrow 0$ and $\thzp \rightarrow \frac{\pi}{2}$:
\begin{align}
    a_z(z) &= \lim_{v_z \rightarrow 0} a_z(z, v_z) \\ 
    &= \lim_{v_z \rightarrow 0} \left[-v_z \, \pderiv{r_z}{z} \, \left(\pderiv{r_z}{v_z}\right)^{-1} \right] \quad .
\end{align}
This already reveals a subtlety in our parameterization and modeling method: in a gravitating system, the acceleration should not be allowed to be a function of the velocity $v_z$, but our model permits this; we return to this point in the discussion (Section~\ref{sec_discussion}).
If we substitute our expression for $r_z$ (Equation~\ref{eq:rz}) into the above limit, we find that
\begin{multline}
    a_z(z) = - \freqzero^2 \, z \, \\
        \times \frac{\left[
            1 + \sum_m (-1)^{m/2} \,
                \left(e_m + \rzp \, \pderiv{e_m}{\rzp}\right)
        \right]}{\left[
            1 + \sum_m (-1)^{m/2} \,
                \left(e_m\,(1 - m^2) + \rzp \, \pderiv{e_m}{\rzp}\right)
        \right]} 
        \label{eq:az}
\end{multline}
for $m=\{2, 4\}$, and the whole expression is evaluated at $\rzp = |\Delta z|\,\sqrt{\freqzero}$ (see Price-Whelan et al., in prep., for a full derivation).
We note that the factor $-\freqzero^2 \, z$ is equivalent to the acceleration in a simple harmonic oscillator potential, but for our parameterization, this is made anharmonic through the Fourier coefficient functions $e_2$ and $e_4$. 

At this point, it is worth asking ``\textit{How reliably can we empirically measure $a_z$?}'' In a companion paper, Price-Whelan et al., (in prep), we focus on the method and test the Orbital Torus Imaging framework described here in a number of different scenarios to assess the accuracy of the $a_z$ measurement. Specifically, Price-Whelan et al., (in prep) test: $i$) a simple harmonic oscillator; $ii$) a quasi-isothermal disk DF simulation; $iii$) a quasi-isothermal disk DF simulation with a selection function imprinted resembling that of the APOGEE survey; $iv$) a perturbed N-body disk simulation. In all cases, the empirical measurement of the $a_z$ matches well the true value for measurements below the range we explore in this work ($|z|<1.2~\kpc$, see Section~\ref{sec_results}).

Once the vertical acceleration is determined, we then compute the volume and surface mass density using Poisson's equation.
We assume that the total mass distribution is axisymmetric such that the volume density only depends on cylindrical radius and vertical position, $\rho = \rho(R, z)$.
In an axisymmetric and separable mass distribution, Poisson's equation is
\begin{align}
    4 \pi G \, \rho(R, z) &= -\frac{1}{R} \, \frac{\partial}{\partial R}\left(R \, a_R \right) - \deriv{a_z}{z} \\
    &= -\frac{2\,v_c}{R} \, \deriv{v_c}{R} - \deriv{a_z}{z} \label{eq:poisson}
\end{align}
where $a_R$ and $a_z$ are the radial and vertical acceleration, respectively, and the second line replaces the radial terms with an expression involving the circular velocity, $v_c(R)$.
We note that the midplane density, at any radius, $\rho_0(R)$, can be estimated using the fact that the $z$-derivative of Equation~\ref{eq:az} evaluated at $z=0$ is simply related to our frequency parameter $\freqzero$,
\begin{equation}
    \rho_0(R) = \rho(R, z=0) \approx \frac{\freqzero^2}{4\pi G} \label{eq:rho0}
\end{equation}
where we have neglected the term involving the radial gradient of the circular velocity. We tested the impact of including this radial dependence in our results, and found a maximum impact on the order of $\sim10\%$, which is well within our measurement uncertainties. 

Integrating the expression in Equation~\ref{eq:poisson} over $z$, we obtain an expression for the surface mass density
\begin{align}
    \Sigma(R, z) &= \int_{-|z|}^{|z|} \dd z' \rho(R, z') \\
    &= \frac{|a_z|}{2\pi G} - \frac{v_c}{\pi G \, R} \, \deriv{v_c}{R} \, |z| 
    \label{eq:surfdens}
\end{align}
where we emphasize that $\Sigma(z)$ is the \textit{total} mass density within a vertical column (stars, gas, and dark matter). In the following section, we utilize this method to estimate key quantities about the mass distribution of the Galactic disk. We note that with our current method, we do not measure the circular velocity curve $v_c(R)$, so we adopt a linear fit to the circular velocity measured in \citet{Eilers2019} over the radius range considered below ($6 < R < 13~\kpc$).

\subsection{Measuring the contribution of dark matter to the surface and volume mass density}

If knowledge of the baryonic contribution to the column mass density, $\Sigma_{\mathrm{bary}}$, is available, it is then straightforward to determine the contribution from dark matter to the total surface mass density, $\Sigma_{\mathrm{tot}} = \Sigma_{\mathrm{bary}} + \Sigma_{\mathrm{DM}}$. 
 Similarly, under the assumption that the density of dark matter, $\rho_{\mathrm{DM}}$, is constant below a given vertical height, $|z|$, it is also possible to directly infer $\rho_{\mathrm{DM}}$ using
\begin{equation}\label{eq_dm}
    \rho_{\mathrm{DM}} = \frac{\Sigma_{\mathrm{tot}}(z)-\Sigma_{\mathrm{bary}}(z)}{2z}.
\end{equation}

In the following Section, we will use estimates of $\Sigma_{\mathrm{bary}}$ from the literature (\citealp[e.g.,][]{McKee2015}) to compute $\Sigma_{\mathrm{DM}}$ and $\rho_{\mathrm{DM}}$ around the Sun, employing estimates of the total surface mass density computed using element abundances.

\subsection{Measuring the disc scale length}

With measurements of the total volume mass density at the midplane, $\rho(z=0)$, across radius, it is also possible to estimate the disk's exponential scale length, $h_R$, and the total volume mass density at the midplane at the solar radius, $\rho(z=0)_{\odot}$, via the following relation \citep{Kuijken1989}

\begin{equation}
\label{eq_hr}
    \rho(R,z=0) = \rho(z=0)_{\odot} \times \exp{\Bigg( -\frac{(R-R_{\odot})}{h_R}\Bigg)}.
\end{equation}

We will use this equation in Section~\ref{sec_scaleradius} to estimate $h_R$ and $\rho(z=0)_{\odot}$, respectively.

\section{Results}
\label{sec_results}

\subsection{Modelling the vertical element abundance trends across the Galactic disk}

\begin{figure*}
    \centering
    \includegraphics[width=\textwidth]{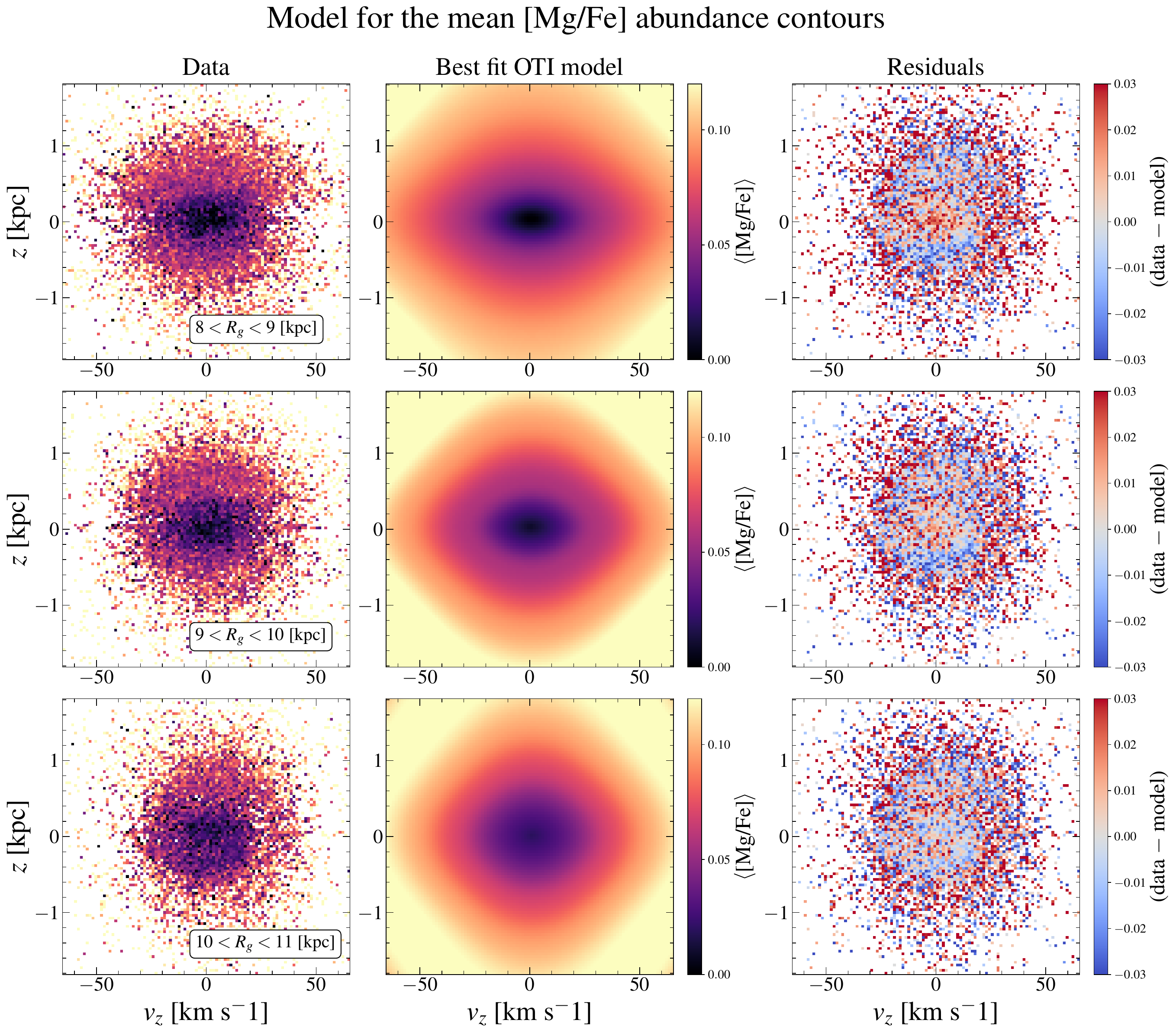}
    \caption{
    A demonstration of our modeling procedure (Section~\ref{sec_modelling}) that fits $\langle\mathrm{[Mg/Fe]}\rangle$ trends in vertical kinematics in three bins of guiding radius. 
    Each row of panels corresponds to a given guiding-center radius bin: From top to bottom, $8<R_g<9~\kpc$, $9<R_g<10~\kpc$, and $10<R_g<11~\kpc$. 
    The left panels show the mean \abun{Mg}{Fe} abundance in bins of $(z, v_z)$. 
    The middle panels show the optimized OTI models.
    The right panels show the residuals (i.e. the difference between the data and the optimized model, in units of \abun{Mg}{Fe} dex).
    The model captures the general trends of chemical gradients despite the obvious sampling limitations, but a clear spiral pattern is apparent in the residual plots. 
    This feature illustrates that there is a chemical phase-space spiral in the Galactic disk, which is likely related to the kinematic phase spiral \citep[see, e.g., ][]{Antoja2018}.
    }    \label{fig_rg}
\end{figure*}

As an initial demonstration of our method, we use the modeling framework described above (Section~\ref{sec_modelling} to fit the vertical element abundance gradient of $\langle\mathrm{[Mg/Fe]}\rangle$ in bins of guiding radius for our \textsl{APOGEE}--\textsl{Gaia} parent sample.
Figure~\ref{fig_rg} illustrates the result of optimizing our model applied to this data in three bins of guiding radius, $8 < R_g < 9~\kpc$, $9 < R_g < 10~\kpc$, and $10 < R_g < 11~\kpc$, where each row in the figure corresponds to one guiding radius bin. 
The panels in each row of Figure~\ref{fig_rg} show the data, our optimized model, and the residual, from left to right.
The data panel (left) shows $\langle\mathrm{[Mg/Fe]}\rangle$ in bins of $z,v_z$ coordinates, where each pixel has a size of $\delta z = 0.04~\kpc$ and $\delta v_z = 1.76~\kms$. 
The model panel (middle) shows our optimized model, fit to the data in the left column, evaluated for all pixel locations in the left panel. 
The residual panel (right) shows the difference between the data and the optimized model. 

Comparing the top row of Figure~\ref{fig_rg} to the bottom row (i.e. stars in $8 < R_g < 9~\kpc$ versus $10 < R_g < 11~\kpc$), we note that the optimized model prefers a more oblate contour shape at a given value of $\langle\mathrm{[Mg/Fe]}\rangle$ (e.g., $\langle \abun{Mg}{Fe} \rangle \approx 0.05$) for the inner radial bin.
Conversely, the outer radial bin (i.e. the bottom row) shows a more oblate/diamond-like contour shape far from the Galactic midplane.
The fact that the residuals are generally small suggests that the functional form used is flexible enough to capture the overall structure of the data, at least in the regions of vertical phase space that are well-represented by the data.
However, we note that the residuals do show a clear spiral pattern, which is most visible in the outer radial bin (i.e. the lower right panel).
This closely resembles the morphology of the \textsl{Gaia} ``phase spiral'' \citep{Antoja2018} in phase-space density, suggesting that this feature also manifests in chemical abundance space (see also \citealp[][, Frankel et al. in prep]{Gaia2023_spiral}). 
This feature is further evidence of the complex chemical-dynamical structure of the Galactic disk, but we reserve any study of the chemical $z,v_z$ spiral for future work. 

In the subsections that follow, we show summary properties of the mass distribution of the Milky Way disk as inferred from the most constraining element abundance ratio, [Mg/Fe], in all bins of guiding-center radius.

\subsection{Measuring the vertical acceleration field and surface mass density across the Galactic disk}
\label{results_mass}

In this Section, we measure the vertical acceleration, $a_z(z)$, and surface mass density, $\Sigma(z)$, as a function of height above the Galactic plane in bins of guiding-center radius across the disk for all six element abundance ratios considered above. 
In detail, we use the abundance ratios shown in Figures~\ref{fig_abun} and \ref{fig_jz-z} and use vertical kinematic data for stars in overlapping, 2~kpc-wide bins of guiding radius, $R_g$, spaced by $0.25~\kpc$ from $6$ to $12~\kpc$.
We evaluate the quantities ($a_z$ and $\Sigma$) as a function of vertical position between $z=0-1.2~\kpc$ in steps of $50~\unit{\parsec}$. 
We truncate at $z=1.2~\kpc$ as this is the approximate height to which our model represents well the data (see the residuals panels in Figure~\ref{fig_rg} and Price-Whelan et al., in prep). 
We use the bootstrap resamplings to estimate our measurement error on the resulting acceleration and surface density values; we  compute the 50$^{th}$ and [16$^{th}$,84$^{th}$] percentiles and use these as our representative median and uncertainty values for all measurables.

\begin{figure*}
    \centering
    \includegraphics[width=\textwidth]{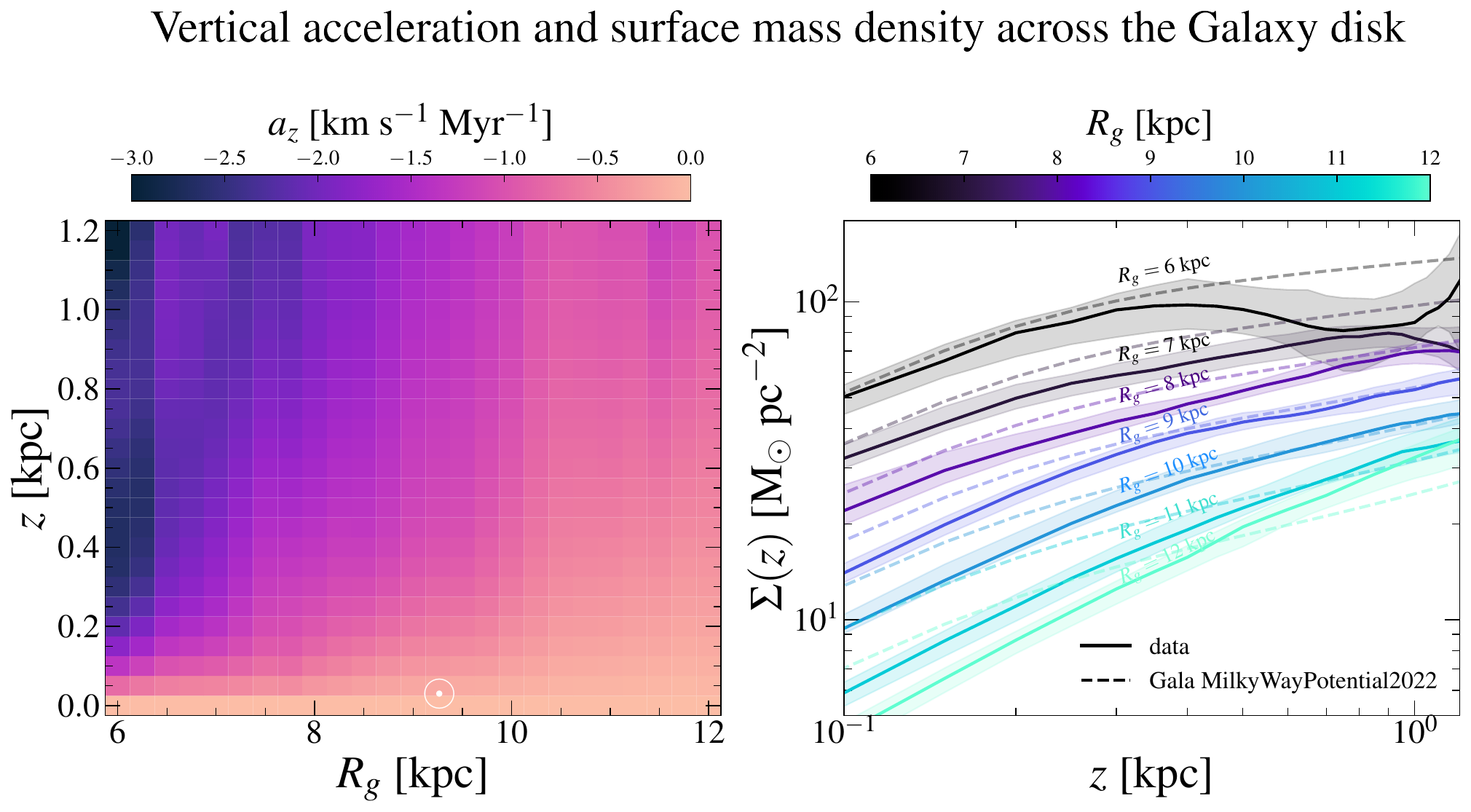}
    \caption{\textit{Left}: mean vertical acceleration estimate determined when fitting the $\langle$[Mg/Fe]$\rangle$ distribution in $z,v_z$ space for stars in parent sample binned into $R_g$ and approximate $z_{\mathrm{max}}$ ($z_{|v_{z}|<10~\kms}~\sim z_{\mathrm{max}}$). Each pixel is an independent measurement of the vertical acceleration. The $\odot$ symbol shows the position of the Sun in this diagram. \textit{Right}: Estimates of the median (solid line) and [16$^{th}$,84$^{th}$] percentiles (shaded) \textit{total} surface mass density as a function of vertical height $z$ for seven guiding-center radii ($R_g=[6,7,8,9,10,11,12]~\kpc$). Also shown here as dashed lines are the analytical values computed using the \texttt{MilkyWayPotential2022} in \texttt{Gala} \citep{Price2017}. Using orbital torus imaging, we empirically map the acceleration field and surface mass density across the Galaxy disk with element abundances.}    \label{fig_forcelaw}
\end{figure*}

The left panel of Figure~\ref{fig_forcelaw} shows our inferred mean vertical acceleration values when using our best element abundance ratio ([Mg/Fe]). From this, it is clear evident that $a_z$ varies with both $R_g$ and $z$, revealing beautiful trends across radius and height above the plane. 
Close to the Galactic midplane ($z\sim0.1~\kpc$), the vertical acceleration is almost zero for all guiding-centre radii, as is to be expected for any model. 

In a similar vein, the right panel of Figure~\ref{fig_forcelaw} shows estimates of the surface mass density given our measurements of $a_z$ and Equation~\ref{eq:surfdens} for a range of $z$ and $R_{g}$ (solid lines). 
As a comparison, we also show the $\Sigma(z)$ values (dashed lines) computed using the \texttt{MilkyWayPotential2022} in \texttt{gala} \citep{Price2017}.
We find that the inferred vertical surface mass density profile of the disk is generally consistent with this potential model at low $z$.
At $R_g = 6~\kpc$ and $z < 0.2~\kpc$, we estimate a surface mass density $\Sigma(z<0.2~\kpc)\sim80~\usurfdens$. 
For similar heights but larger guiding-center radius ($R_g =10~\kpc$), we find a much lower surface mass density estimate ($\Sigma(z<0.2~\kpc)\sim 20~\usurfdens$). 
Around the Sun ($R_g\sim9~\kpc$), we estimate a value of the total surface mass density of $\Sigma_{\odot}(z)\sim70~\usurfdens$ at $z=1.1~\kpc$.

In the following subsections, we set out to constrain this quantity in more detail and compare it to estimates from the literature. We will also estimate the contribution of dark matter to the surface mass density around the Sun by using our estimates from Figure~\ref{fig_forcelaw} and subtracting from it values of the stellar (and stellar remnants) surface mass density at around the Solar Neighborhood, $\Sigma_{\odot,\mathrm{bary}}(z)$. 

\subsection{The total and dark matter contribution to the surface mass density in the solar neighborhood}
\label{results_dm}

In this section, we focus on the surface mass density around the solar neighborhood and estimate the contribution to this value from dark matter. 
To do this, we select stars with precise abundance measurements within a narrow range of radius around the Solar neighbourhood (i.e., $|R-R_\odot| < 1~\kpc$ for $R_{\odot} = 2.75~\kpc$). This cut yields 37,729 stars. We then bootstrap sample with replacement 25 trials, and compute the 50$^{th}$ and [16$^{th}$,84$^{th}$] percentiles and use them as our representative median and 1$\sigma$ measurement and uncertainty.
Upon obtaining our model fits, we repeat the procedure from Section~\ref{method_acc}, and estimate the vertical acceleration as a function of vertical height, binning the data in 50 pc bins spanning from $z = 0~\kpc$ to $z = 1.2~\kpc$. 
As a last step, we convert the vertical acceleration into an estimate of the total surface mass density. 

A commonly referenced value in the literature is $\Sigma_{\odot}(z=1.1~\kpc)$, the total surface mass density around the Solar neighborhood at a height of $z=1.1~\kpc$. 
The reason for the choice of $1.1~\kpc$ is historic \citep{Kuijken1991}, and is hypothesized to be the height where the dark halo starts to dominate the surface mass density of the disk. 
For this reason, in this Section we will use this value to compare to previous measurements. 

\begin{figure}
    \centering
    \includegraphics[width=\columnwidth]{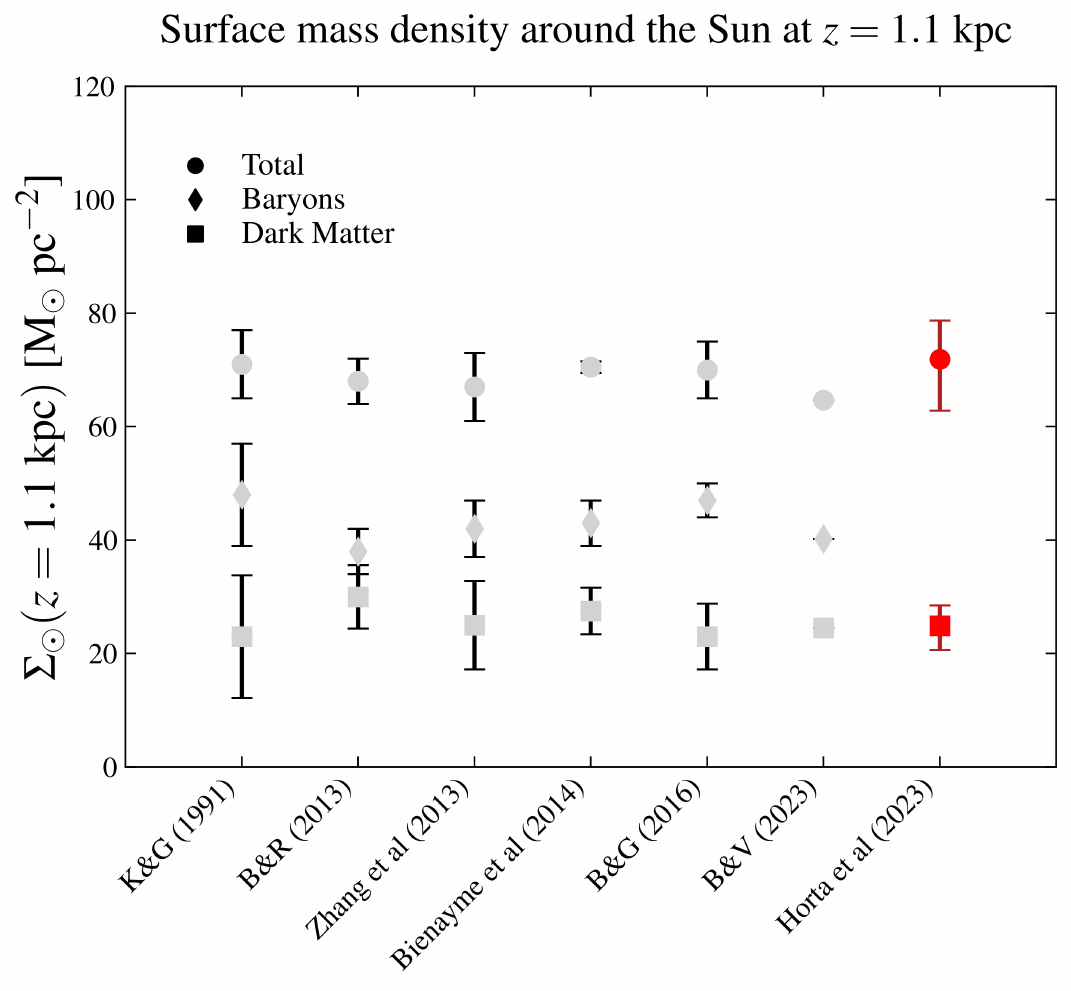}
    \caption{Estimated median and [16$^{th}$,84$^{th}$] percentile values of the \textit{total} (circles) and dark matter (squares) surface mass density at $z=1.1$ kpc for stars in the solar neighbourhood, computed by modelling 25 bootstrapped samples with replacement of the $\langle$[Mg/Fe]$\rangle$ distribution in $z,v_z$ space, using the most constraining element abundance ratio, [Mg/Fe]. For comparison, we also show previous estimates of this quantity from previous work (see Table~\ref{tab_sigmasun}), including the values of the surface mass density from baryonic material (diamonds). The values of the dark matter surface mass density are computed by subtracting the $\Sigma_{\odot,\mathrm{bary}}(z=1.1~\kpc)=47\pm3$ value from \citet{McKee2015} to our estimates of the total surface mass density. Our estimates of the total and dark matter $\Sigma_{\odot}(z=1.1~\kpc)$ for Mg agree well with previous estimates of $\Sigma_{\odot}(z=1.1~\kpc)$ around the solar neighbourhood computed using other methods. We note that previous works determined these values with smaller samples of stars, on the order of $N_{\mathrm{stars}}\lesssim10,000$, and the uncertainties are likely underestimated. Our measurement is computed using a sample of 37,729 stars; the uncertainties on our measurement will only become smaller with the advent of more data (e.g., \citealp[SDSS-V/MWM][]{Kollmeier2017}). 
    }    \label{fig_sigma1.1}
\end{figure}

Figure~\ref{fig_sigma1.1} shows $\Sigma_{\odot} (z=1.1~\kpc)$ for the different element abundance ratios used in this work. In this illustration we also show the estimate of $\Sigma_{\odot} (z=1.1~\kpc)$ from previous studies (\citealp[][]{Kuijken1991,Bovy2013,Zhang2013,Bienayme2014,Bland2016,Binney2023}), for comparison. We find that within the uncertainties, our estimates of $\Sigma_{\odot} (z=1.1~\kpc)$ for [Mg/Fe], $\Sigma_{\odot}(z=1.1~\kpc)=72^{+6}_{-9}~\usurfdens$, which is in agreement with previous works ($\sim 65-71\pm 10~\usurfdens$, see Table~\ref{tab_sigmasun}). 
In a similar vein, we also show the dark matter contribution to the solar neighbourhood surface mass density at $z=1.1$ kpc as squares, $\Sigma_{\odot,\mathrm{DM}} (z=1.1~\kpc)$. To compute this quantity, we have subtracted the estimate for the baryonic contribution (namely, $\Sigma_{\odot,\mathrm{bary}}(z=1.1~\kpc)=47\pm3~\usurfdens$, \citealp[][]{McKee2015}) to the surface mass density around the solar neighborhood at $z=1.1$ kpc from our measurements of $\Sigma_{\odot} (z=1.1~\kpc)$. We determine a value of $\Sigma_{\odot,\mathrm{DM}}(z=1.1~\kpc)=24\pm4~\usurfdens$. 
All our inferred surface mass density values are summarized in Table~\ref{tab_sigma}.

% Example table
\begin{table*}
	\centering
	\caption{From left to right, the work or element abundance ratio used to compute the the total surface mass density at the Solar neighbourhood and $z=1.1$ kpc, the stellar (and stellar remnants) surface mass density at the Solar neighbourhood and $z=1.1$ kpc, and the dark matter surface mass density at the Solar neighbourhood and $z=1.1$ kpc. Our estimates of $\Sigma_{\odot,\mathrm{DM}}, (z=1.1~\kpc)$ are computed assuming $\Sigma_{\odot,\mathrm{bary}}(z=1.1~\kpc) = 47\pm3~\usurfdens$ \citep{Bland2016}. For reference, we also list the estimates from previous studies (\citealp[][]{Kuijken1991,Bovy2013,Zhang2013,Bienayme2014,Bland2016,Binney2023}).}
	\label{tab_sigmasun}
	\begin{tabular}{lcccr} % four columns, alignment for each
		\hline
		 & $\Sigma_{\odot}(z=1.1~\kpc)~[\usurfdens]$& $\Sigma_{\odot,\mathrm{bary}}(z=1.1~\kpc)~[\usurfdens]$& $\Sigma_{\odot,\mathrm{DM}}( z=1.1~\kpc)~[\usurfdens]$\\
   \hline
    Kuijken$\&$Gilmore (1991) & $71\pm6$& $48\pm9$& $23\pm11$\\
  Bovy$\&$Rix (2013) & $68\pm4$& $38\pm4$& $30\pm6$\\
  Zhang et al. (2013) & $67\pm6$& $42\pm5$& $25\pm8$\\
  Bienaym\'e et al (2014) & $70.5\pm1.0$ & $43\pm4$& $27.5\pm4.0$\\
  Bland-Hawthorn $\&$ Gerhard (2016) & $70\pm5$ & $47\pm3$\footnote{This estimate is originally from \citet{McKee2015}.} & $23\pm6$\\
  Binney$\&$Vasiliev (2023) &  64.7 & 40.2 & 24.5 \\
    \hline
     Horta et al. (2023) (this work; \abun{Mg}{Fe}) & $72^{+6}_{-9}$ & --& $24\pm4$\\
     \hline
	\end{tabular}
 \label{tab_sigma}
\end{table*}

\subsection{The total volume density at the midplane across the Galaxy disk}
\label{sec_volumedens}
\begin{figure}
    \centering
    \includegraphics[width=\columnwidth]{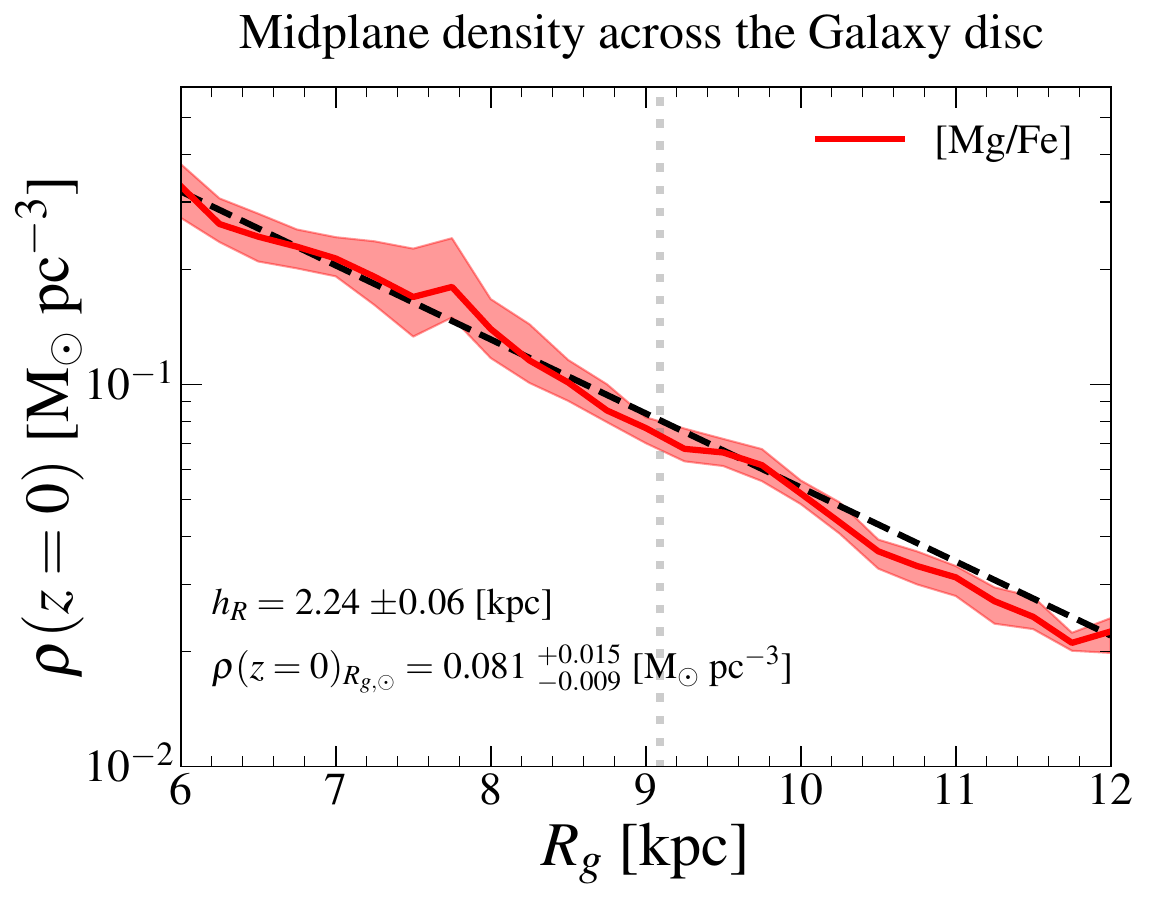}
    \caption{
    Total volume density at the midplane $\rho(z=0)$ computed using [Mg/Fe] as our element abundance tracer as a function of guiding-center radius. The solid line indicates the median of 25 bootstrap with resampling, whilst the shaded regions show the [16$^{th}$,84$^{th}$] percentile ranges. There is a declining density at the midplane with increasing guiding-center radius, which is best fitted by a scale radius of $h_R=2.24\pm0.06~\kpc$. The dashed line shows an exponential profile assuming a value for the scale radius of $h_R=2.24$. The vertical dotted line corresponds to the $R_g$ value of the Sun.}    \label{fig_rho0}
\end{figure}

Another important quantity in the Milky Way is the volume mass density throughout the Galactic disk.
Here, we use our inferred asymptotic midplane frequency values, $\freqzero$, to measure the total volume mass density at the Galactic midplane (Equation~\ref{eq:rho0}) for \abun{Mg}{Fe}.
Figure~\ref{fig_rho0} shows our resulting measurements of the total volume density at the midplane, $\rho (R, z=0)$ as a function of guiding-center radius, $R_g$. 
The median and [16$^{th}$,84$^{th}$] percentiles from 25 bootstrap with replacement samples are shown as the solid line and shaded regions. The general trend shows a declining value of $\rho(z=0)$ with increasing guiding-center radius, as expected. At $R_g=6$ kpc, we determine a value of $\rho(z=0) \sim 0.3~\uvoldens$. This value drops to $\rho(z=0) \sim 0.08~\uvoldens$ at $R_g=9$ kpc, and $\rho(z=0) \sim 0.02~\uvoldens$ at $R_g=12~\kpc$.

\subsection{The scale radius}
\label{sec_scaleradius}
Fig~\ref{fig_rho0} shows the total volume mass density estimates we obtain at the midplane ($z=0$) across guiding-center radii. Using these values and equation~\ref{eq_hr}, we optimize\footnote{To optimize equation~\ref{eq_hr} we used the \texttt{curve-fit} function in \texttt{scipy}.} the exponential model to find the best fitting parameters representing the scale radius, $h_R$, and the midplane total volume mass density at the solar radius, $\rho(z=0)_{\odot}$. We obtain a value (for the low-$\alpha$ disc) of $h_R=2.24\pm0.06~\kpc$, and $\rho(z=0)_{\odot}=0.081^{+0.015}_{-0.009}~\uvoldens$. The scale radius estimate we obtain is in good agreement with previous results ($h_R\sim2-3$; \citealp[e.g.,][]{Bovy2013}).

\subsection{The total and dark matter contribution to the total volume mass density in the solar neighbourhood}
\begin{figure}
    \centering
    \includegraphics[width=\columnwidth]{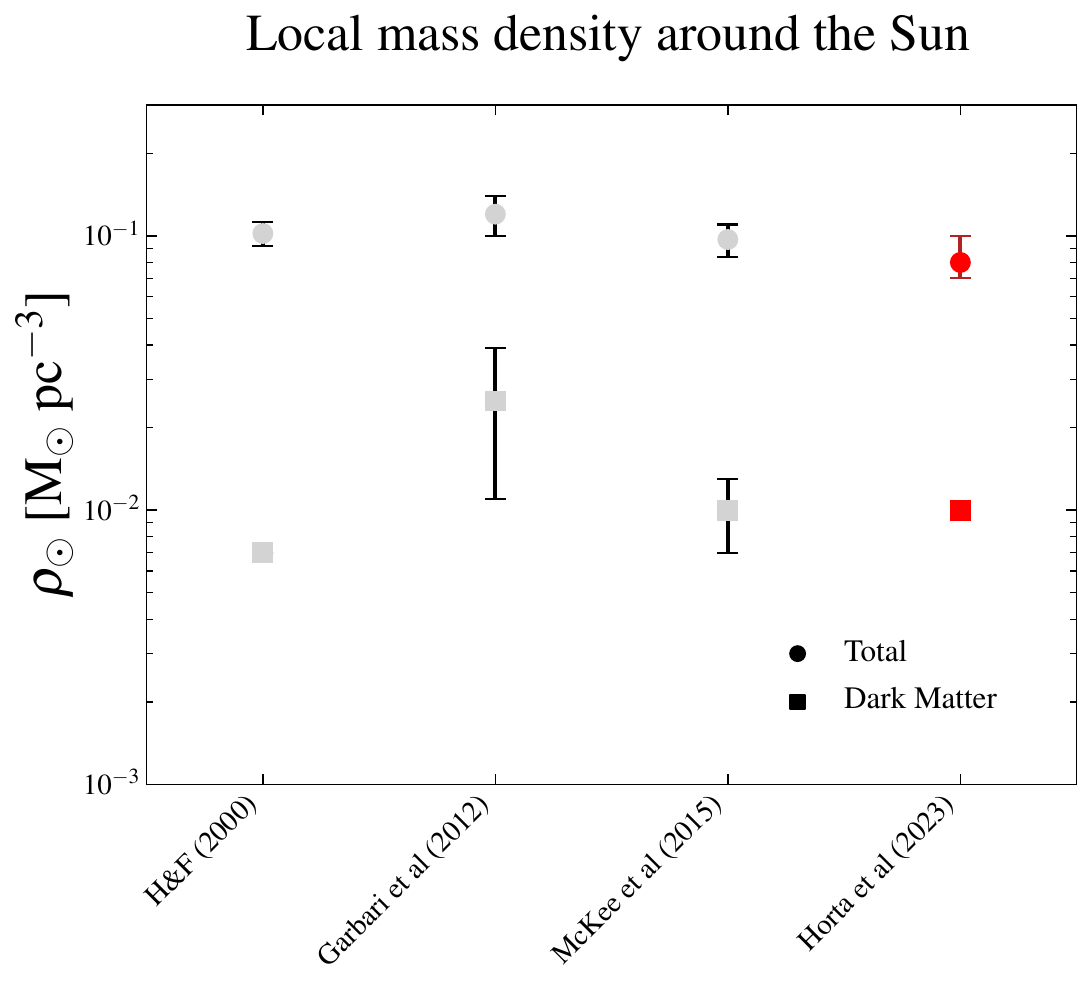}
    \caption{Estimated median and [16$^{th}$,84$^{th}$] percentile values of the total volume mass density in the Solar Neighbourhood, computed by modelling the 25 bootstrapped samples with replacement of the $\langle$[Mg/Fe]$\rangle$ distribution in $z,v_z$ space. For comparison, we also show estimates from \citet{Holmberg2000}, \citet{Garbari2012}, and \citet{McKee2015}. See \citet{DeSalas2021} for a more extensive review of $\rho_{\odot,\mathrm{DM}}$. Our estimates of $\rho_{\odot}$ and $\rho_{\odot,\mathrm{DM}}$ are in good agreement with the latest previous measurements.}    \label{fig_rho_sun}
\end{figure}

In Figure~\ref{fig_rho_sun}, we show our measurement of $\rho_{\odot}(z=0)$ in the solar neighbourhood. 
For reference, we also show the value from previous work (\citealp[][]{Holmberg2000,Garbari2012,McKee2015}{}) for both $\rho_{\odot}(z=0)$ and $\rho_{\odot,\mathrm{DM}} (z=0)$.
We estimate a value for the total midplane volume mass density at the Sun of $\rho_{\odot} (z=0) = 0.081^{+0.015}_{-0.009}~\uvoldens$. This is in good agreement with the value reported in recent studies ($\rho_{\odot}(z=0)\sim0.08-0.1~\uvoldens$; \citealp[e.g.,][]{Holmberg2000,Garbari2012,McKee2015}). 

In addition, using Equation~\ref{eq_dm}, we also compute the value of the local volume mass density contributed by dark matter to be $\rho_{\odot,\mathrm{DM}}(z=0)=0.011\pm0.002~\uvoldens$, using the value of $\Sigma_{\odot,\mathrm{bary}}(z=1.1~\kpc)~\usurfdens$ from \citet{McKee2015}. Our estimate of $\rho_{\odot,\mathrm{DM}}(z=0)$ is also in good agreement with the wealth of previous measurements \citep[][and references therein]{DeSalas2021}. Our results are summarised in Table~\ref{tab_rho}.

\begin{table}
	\centering
	\caption{Total volume mass density around the solar neighbourhood ($\rho_{\odot}$) at the midplane, $z=0$. Our value of $\rho_{\mathrm{DM},\odot}$ is computed using Equation~\ref{eq_dm} and assuming that the dark matter density is constant in a column around the Sun from $z=0$ to $z=1.1$ kpc, using the value of $\Sigma_{\mathrm{bary},\odot}$ from \citet{McKee2015}. For reference, we also show estimates from previous results for $\rho_{\odot}$ and $\rho_{\odot,\mathrm{DM}}$. See \citet{DeSalas2021} for a more extensive review on measurements of $\rho_{\odot,\mathrm{DM}}$.  }
	\label{tab_midplane}
	\begin{tabular}{lccr} % four columns, alignment for each
		\hline
		  & $\rho_{\odot}$ [M$_{\odot}$pc$^{-3}$] & $\rho_{\odot,\mathrm{DM}}$ [M$_{\odot}$pc$^{-3}$] \\
   \hline
   Holmberg $\&$ Flynn (2000) &$0.102\pm0.01$ & 0.007\\
    Garbari et al., (2012) &$0.12\pm0.02$ & $0.025\pm0.014$\\
    Mckee et al. (2015) &$0.097\pm0.013$ & $0.013\pm0.003$\\
  \hline
 Horta el al., (2023) \\
 (This work; $\mathrm{[Mg/Fe]}$)  & 0.081$^{+0.015}_{-0.009}$ & $0.011\pm0.002$\\
    \hline
\label{tab_rho}
	\end{tabular}
\end{table}

\subsection{The midplane position and vertical velocity at the Solar Neighborhood}
In addition to the main results in this paper, other quantities that we can directly measure from our fitting procedure are the midplane position, vertical velocity, and vertical frequency around the Solar neighbourhood. To do so, we take the same stars from Section~\ref{method_acc} and measure $z_{\odot}$, $v_{z,\odot}$, and $\Omega_{z,\odot}$ around the Sun. We perform this exercise using $\langle\mathrm{[Mg/Fe]}\rangle$.
Table~\ref{tab_midplane} shows the median and [$16^{th}$,$84^{th}$] percentile values from fitting the data using each element abundance ratio and bootstrapping with replacement 25 times. We find that our fits return measurements for $z_{\odot}$ and v$_{z,\odot}$ that are close to previous estimates (e.g., $z_{\odot}=29.5\pm1$ pc and $v_{z,\odot}=8.66\pm0.05$ $\kms$, see  \citealt{Schonrich2010,Bennett2019}). This leads to an estimated value for the Solar vertical frequency of $\Omega_{z,\odot}=0.30\pm0.01$ Myr$^{-1}$. Our measurements are in good agreement with previous estimates (see Table~\ref{tab_midplane}).

\begin{table}
	\centering
	\caption{From left to right, the work or element abundance ratio to compute $z_{\odot}$, $v_{z,\odot}$, and $\Omega_{z,\odot}$, the Solar midplane position, the Solar vertical velocity, and the Solar vertical frequency.}
	\label{tab_midplane}
	\begin{tabular}{lcccr} % four columns, alignment for each
		\hline
		  & $z_{\odot}$ [pc]&  $v_{z,\odot}$[km s$^{-1}$]& $\Omega_{z,\odot}$[Myr$^{-1}$]\\
   \hline
Bennett$\&$Bovy (2019) & 20.8$\pm$0.3& --& --\\
  Sch\"onrich et al. (2010) & --&7.25$\pm$0.9 &-- \\
  \hline
   Horta el al., (2023) \\(This work; $\mathrm{[Mg/Fe]}$) & $29.5\pm1.0$ & $8.66\pm0.05$& $0.30\pm0.01$\\
\hline
	\end{tabular}
\end{table}

\section{discussion}
\label{sec_discussion}
\subsection{First empirical estimates of the acceleration field and mass density across the disk from stellar labels}

We believe that this article presents the first dynamical measurement of the Milky Way disk's profile empirically from stellar labels. The ability to infer dynamical properties of the disk from kinematics via contours of constant density was proposed approximately four decades ago \citep{Kuijken1989}. However, until the advent of \textsl{Gaia}, insufficient numbers of stars have prohibited such analysis. In this work, we have inferred dynamical properties of the Milky Way disk by modelling the mono-abundance contours, which reveals the vertical orbital structure of stars across the Galaxy. 
We have shown that this unique synergy between stellar labels and kinematics enables measuring the vertical acceleration field, and therefore the surface mass density profile of the Galactic disk without the assumption of a functional form for the gravitational potential. 
Our method is also much less affected by spatial selection function effects when compared to previous techniques (\citealp[e.g.,][]{Kuijken1991,Bovy2013,Zhang2013,McMillan2013,Sanders2015,Li2021}), as we are much less sensitive to  spatial selection functions the survey data we use. 

When compared to previous estimates, we find that our measurements for our most constraining element, [Mg/Fe], agrees well with previous studies for estimates of the local surface mass density, $\Sigma_{\odot}(z=1.1~\kpc) = 72^{+6}_{-9}~\usurfdens$ and $\Sigma_{\odot,\mathrm{DM}}(z=1.1~\kpc) = 24\pm4~\usurfdens$ (\citealp[][and references therein]{Kuijken1991,Bovy2013,Zhang2013,Bienayme2014,Bland2016,Binney2023}), and for the local volume mass density, $\rho_{\odot} (z=0) = 0.081^{+0.015}_{-0.009}~\uvoldens$ and $\rho_{\odot,\mathrm{DM}}(z=0) = 0.011\pm0.002~\uvoldens$ (\citealp[][and references therein]{Holmberg2000,McKee2015,DeSalas2021}). This is also the case for measurements of the vertical midplane position and vertical velocity, $z_{\odot}=29.5\pm1$ pc and $v_{z,\odot}=8.66\pm0.05~\kms$ (\citealp[][]{Schonrich2010,Bennett2019}).

\subsection{The impact of survey selection effects on our measurements}
\label{selfunc}

In detail, the effective selection function for our \textsl{APOGEE}--\textsl{Gaia} sample depends on position, magnitude, and color of the stars. 
However, in our analysis below, we only model the element abundance trends as a function of orbital properties and not the absolute distribution of orbital properties represented in our sample (i.e. we model the element abundance \textit{conditioned} on the orbital actions).
Because of this, if the effective selection function factorizes in terms of its dependence on position and its dependence on element abundances, our analysis will not be sensitive to selection biases.
This is especially important for cross-matches to \textsl{APOGEE} because of the pencil-beam nature of the survey (see for example \citealt{Lane2022}).
We do expect the effective selection function of our sample to approximately factorize in this way because of the uniform selection criteria \textsl{APOGEE} used to define its core samples of targets, which only depends on $J-K_s$ color and $H$-band magnitude \citep{Zasowski2013, Zasowski2017, Beaton2021, Santana2021}.
However, even with simple selection criteria, stars at different evolutionary stages are represented differently as a function of sky position and distance.
This could introduce correlations in the selection function as the reported chemical abundances depend on stellar parameters in complex ways \citep{Eilers2019,Eilers2022}.
To mitigate this issue, we select stars in a narrow range of surface gravity $\logg$ along the red giant branch, $1.5 < \logg < 3.5$. Moreover, in the low-$\alpha$ disc, stars of constant mean element abundance and orbit (e.g., $R_g$) present similar age distributions \citep[][]{Ness2019}.

In summary, the advantage of OTI over other methods is that it does not require detailed knowledge of the positional selection function, provided that the sample selection does not strongly impact the chemical abundance ratios as a function of phase space selections. 
For the case of our \textsl{APOGEE} disk sample, we argue that this condition is sufficiently satisfied.

\subsection{Orbital torus imaging in the context of chemical-dynamical models of the Galaxy}
In this work we have built on the Orbital Torus Imaging (OTI) framework \citep{Price2021} to map the orbital structure of disk stars using their element abundance ratios. 
Our framework effectively models the conditional distribution over some stellar label or element abundance ratio, \abun{X}{Y}, at (conditioned on) a given location in phase-space, i.e. $p(\abun{X}{Y} \,|\, \boldsymbol{w})$ (where we use $\boldsymbol{w}$ to represent the phase-space coordinates; here, these would be $\boldsymbol{w} = (z, v_z)$).
Other techniques have constructed models that unify stellar labels with the distribution function of stars to create chemo-dynamical models of the Galaxy. 
One example of this is the dynamical modeling of ``mono-abundance populations'' (MAPs; \citealt{Bovy2016,Mackereth2017}), which constrains the stellar distribution function conditioned on element abundances, or $p(\boldsymbol{w} \,|\, \abun{X}{Y})$.
Another example is the ``extended distribution function'' (eDF) framework used \citep{Sanders2015, Binney2023}, where here the model uses and constrains the joint distribution over kinematics and element abundances, $p(\boldsymbol{w} , \abun{X}{Y})$.
In both of these cases, the phase-space coordinates appear on the \textit{left} of the conditional, in the distribution being modeled.
The power of the OTI framework is that we model the element abundance (or stellar label) distribution \emph{conditioned} on phase-space coordinates, which, as mentioned above, requires much less knowledge of the survey selection functions that supply our data.

\subsection{The impact of disequilibrium and phase spirals}

As a result of our method, in the residual plots of the 10$<$$R_g$$<$11 kpc (bottom row of Figure~\ref{fig_rg}, but also other $R_g$ bins not shown) we have detected a spiral pattern that matches the phase space spiral discovered by \citet{Antoja2018} (but also hinted at by \citealt{Widrow2012}). This result highlights that in addition to the over-arching smooth distribution, there is substructure in the chemical gradients of disk populations in $z,v_z$ space. The properties and origin of the phase space spiral are currently a debated topic in the field of Galactic dynamics using \textsl{Gaia} (e.g., \citealp{Binney2018,Li2020,Bland2021,Hunt2022,Widmark2022,Widmark2022a,Widmark2022b,Antoja2023,Frankel2023,Tremaine2023}) and numerical/cosmological simulations (\citealp{Laporte2019,Khoperskov2019,Hunt2021,Grand2022}). Although we have not explored the properties of this chemical phase spiral in this work, it would be interesting to model the shape, size, and amplitude of the chemical spiral across the Galaxy disk, and compare its properties to the density phase spiral (see Frankel et al. in prep for an example). We reserve this exploration for future work. 

In principle, the effect of disequilibrium in the Milky Way disc, and the phase spirals that arise from it, could bias our measurements of the midplane vertical position and velocity. However, we choose to ignore this effect as we argue that it will be small (our model is able to capture well the distribution of mean element abundance ratio in vertical kinematic space. If the effect of the chemical spiral was large, this would not be the case). In the future, it would be worth-while accounting for the effect of the chemical spiral in our method, using a model that can also varies with vertical phase, $\theta_{z}$.

\subsection{Future prospects of Orbital Torus Imaging}

In this paper we have used the Orbital Torus Imaging (OTI) framework to directly map orbits and mass in the Galactic disc using one element abundance ratio with the highest signal to noise gradient, \abun{Mg}{Fe} (see Figure~\ref{fig_xferat}). 
However, there are clear observed element abundance gradients in many other element abundance ratios as well (see Figure~\ref{fig_app_gradients}). 
In principle, all element abundance ratios should provide constraints on the orbit shapes and mass distribution, as done here. 
However, most element abundance ratios are not independent \citep[e.g.,][]{Ness2022}, and therefore cannot be naively combined in a likelihood sense.
In future work, it will be interesting to identify and use independent combinations of abundance ratios as our stellar labels, especially those originating from different nucleosynthetic channels (i.e., SN II, SN Ia, AGB winds, neutron-star mergers), which can be separated by studying the element abundance distributions of stellar populations \citep[e.g.,][]{Weinberg2022, Griffith2023}.

This work has made use of spectroscopic data from the APOGEE survey, but many other large spectroscopic surveys are operating or coming online soon (e.g., MWM/SDSS-V: \citealp{Kollmeier2017}; WEAVE: \citealp[][]{Dalton2012}; 4MOST: \citealp[][]{DeJong2012}). 
The next generation of surveys will deliver precise stellar label (element abundance) measurements for many millions of stars across large swaths of the Milky Way. 
This will extend the reach of the OTI framework to new parts of the Galaxy and enable even more precise constraints on the Galactic mass distribution.

\section{Conclusions}
\label{sec_conclusions}

In this paper we use the latest \textsl{APOGEE} and \textsl{Gaia} data to measure the bulk properties of the mass distribution around the Galactic disk.
We use a novel modeling framework (Orbital Torus Imaging; \citealt{Price2021}, Price-Whelan et al. in prep.) that leverages correlations between stellar element abundance measurements (provided by \textsl{APOGEE}) with stellar kinematics (provided by \textsl{Gaia}) to infer the orbit structure of stars in the vertical phase-space, $z, v_z$.
We never use a parameterized model of the global Galactic potential and instead partition the data into bins of guiding-center radius to perform our measurements with the vertical phase-space coordinates.
Our main results and findings are summarized below:
\begin{itemize}
    \item The stellar label (i.e., mean element abundances) gradients in the vertical ($z,v_z$) phase space can be well mapped by an empirical model that includes a low-order Fourier expansion distortion to an ellipsoidal radius (Figure~\ref{fig_rg}). From this mapping, it is possible to estimate the vertical acceleration field and surface mass density across the low-$\alpha$ disk (Figure~\ref{fig_forcelaw}).
    \item At the Solar radius and at a vertical height of $z=1.1$ kpc from the midplane, we estimate an average value for the total surface mass density of $\Sigma_{\odot}(z=1.1~\kpc)=72^{+6}_{-9}~\usurfdens$ using our the element abundance ratio with highest constraining power, \abun{Mg}{Fe} (Figure~\ref{fig_sigma1.1}). 
    Using the value from \citet{McKee2015} for the contribution from baryons, we estimate a contribution of dark matter of $\Sigma_{\odot,\mathrm{DM}}(z=1.1~\kpc)=24\pm4~\usurfdens$.
    \item We estimate the total volume density at the midplane as a function of orbital radius across the Galaxy (Figures~\ref{fig_rho0} and \ref{fig_rho_sun}). This value drops from $\rho(z=0) \sim 0.3~\uvoldens$ at $R_g=6~\kpc$ to $\rho(z=0) \sim 0.02~\uvoldens$ at $R_g=12~\kpc$. At the position of the Sun, we estimate a value of $\rho_{\odot}(z=0)=0.081^{+0.015}_{-0.009}~\uvoldens$ and $\rho_{\odot,\mathrm{DM}}(z=0)=0.011\pm0.002~\uvoldens$ using \abun{Mg}{Fe}. 
    \item We fit an exponential profile to the $\rho(z=0)$-$R_g$ relation (Fig~\ref{fig_rho0}), and obtain a scale length of the low-$\alpha$ disc of $h_R=2.24\pm0.06~\kpc$.
    \item Our model directly fits the vertical position and velocity at the Sun. Using \abun{Mg}{Fe} as our most reliable element abundance ratio, we find a midplane vertical position of $z_{\odot}=29\pm1~\unit{\parsec}$, a vertical velocity relative to the midplane of $v_{z,\odot}=8.66\pm0.05~\kms$, and a midplane vertical frequency $\Omega_{z,\odot}=0.30\pm0.01$ Myr$^{-1}$.   
    \item In the residual from our best model fits (right panels of Figure~\ref{fig_rg}), we find a clear imprint of the vertical phase-space spiral, but now detected in element abundances. 
    This ``chemical spiral'' matches in position and structure the recently-discovered \textsl{Gaia} phase space spiral \citep{Antoja2018}.
\end{itemize}

Orbital Torus Imaging (OTI) is a promising new method for inferring the orbital structure and mass distribution of the Galaxy that does not rely on parameterised functions of the Galactic potential and is much less affected by the spatial selection function. Instead, it relies on the ability to image orbital tori from gradients of stellar labels in phase space. In this work, we have used this technique to map the acceleration field and surface mass density of a region of the Galaxy disc in vertical phase space. To do so, we have had to assume the potential can be decomposed in $R$ and $z$. However, the next steps should involve combining the radial dependence in our model. Along those lines, insofar we have only been able to map the orbital structure of the Galaxy disc with a limited sample. The advent of the Galactic Genesis Milky Way Mapper survey (SDSS-V; \citealt{Kollmeier2017}) will deliver over $\gtrsim5$ million all-sky spectra with \textsl{APOGEE} precision, that will be perfect for measuring the structure of the Milky Way with OTI.

\section*{Acknowledgements}
We thank the referee for a constructive report that helped improve the paper. DH also warmly thanks Sue, Alex, and Debra for their constant support. The Flatiron Institute is funded by the Simons Foundation. This work has made use of data from the European Space Agency (ESA) mission
{\it Gaia} (\url{https://www.cosmos.esa.int/gaia}), processed by the {\it Gaia}
Data Processing and Analysis Consortium (DPAC,
\url{https://www.cosmos.esa.int/web/gaia/dpac/consortium}). Funding for the DPAC
has been provided by national institutions, in particular the institutions
participating in the {\it Gaia} Multilateral Agreement.
%%%%%%%%%%%%%%%%%%%%%%%%%%%%%%%%%%%%%%%%%%%%%%%%%%
\section*{Data Availability}
All \textsl{APOGEE} and \textsl{Gaia} data used in this study are publicly available and can be downloaded directly from \url{https://www.sdss4.org/dr17/} and \url{https://gea.esac.esa.int/archive/}, respectively. 

\software{
    Agama \citep{Vasiliev2019},
    Astropy \citep{astropy2013, astropy2018, astropy2022},
    gala \citep{gala},
    JAX \citep{jax2018github},
    JAXOpt \citep{jaxopt:2021},
    matplotlib \citep{Hunter:2007},
    numpy \citep{numpy},
    pyia \citep{pyia},
    scipy \citep{scipy}.
}

\clearpage

\appendix
\section{Abundance gradients with vertical kinematics}
\label{app_gradients}
\begin{figure*}
    \centering
    \includegraphics[width=1\textwidth]{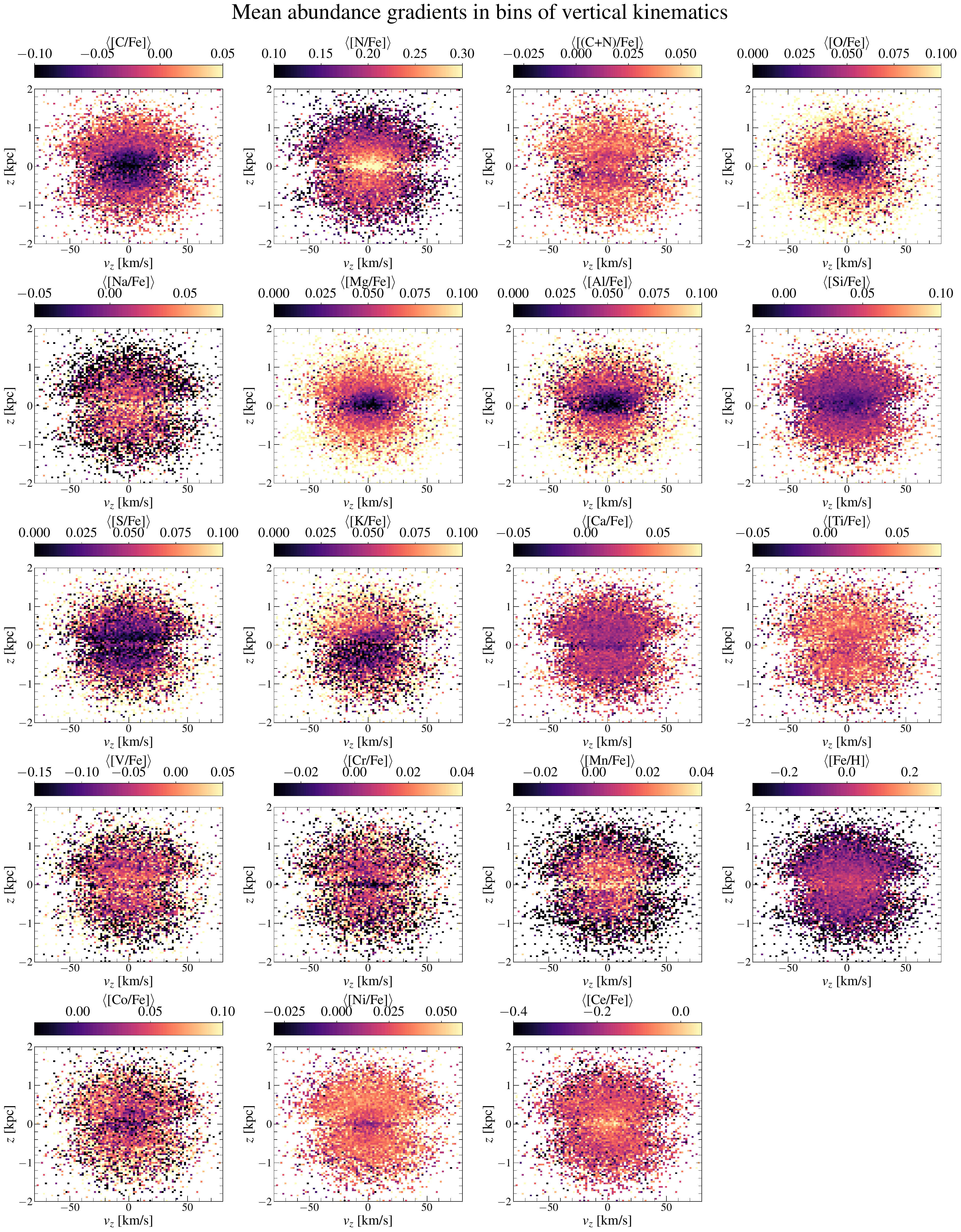}
    \caption{Abundance gradients in vertical kinematic space for all element abundances provided by the \textsl{APOGEE} survey. [(C+N)/Fe] is determined using the following relation: $\log_{10}(10^{(\mathrm{[C/Fe]}+\mathrm{[Fe/H]+8.39})}+10^{(\mathrm{[N/Fe]}+\mathrm{[Fe/H]+7.78})})-\log_{10}(10^{8.39}+10^{7.78})-\mathrm{[Fe/H]}$. We note that trends in [C/Fe], [N/Fe], and [O/Fe] could be affected by dredge-up processes in giant stars. However, this will not affect our results as we did not use these element abundance ratios in our study.}    \label{fig_app_gradients}
\end{figure*}

\bibliography{refs}

\begin{thebibliography}{}
\expandafter\ifx\csname natexlab\endcsname\relax\def\natexlab#1{#1}\fi
\providecommand{\url}[1]{\href{#1}{#1}}
\providecommand{\dodoi}[1]{doi:~\href{http://doi.org/#1}{\nolinkurl{#1}}}
\providecommand{\doeprint}[1]{\href{http://ascl.net/#1}{\nolinkurl{http://ascl.net/#1}}}
\providecommand{\doarXiv}[1]{\href{https://arxiv.org/abs/#1}{\nolinkurl{https://arxiv.org/abs/#1}}}

\bibitem[{{Abdurro'uf} {et~al.}(2022){Abdurro'uf}, {Accetta}, {Aerts}, {Silva
  Aguirre}, {Ahumada}, {Ajgaonkar}, {Filiz Ak}, {Alam}, {Allende Prieto},
  {Almeida}, {Anders}, {Anderson}, {Andrews}, {Anguiano}, {Aquino-Ort{\'\i}z},
  {Arag{\'o}n-Salamanca}, {Argudo-Fern{\'a}ndez}, {Ata}, {Aubert},
  {Avila-Reese}, {Badenes}, {Barb{\'a}}, {Barger}, {Barrera-Ballesteros},
  {Beaton}, {Beers}, {Belfiore}, {Bender}, {Bernardi}, {Bershady}, {Beutler},
  {Bidin}, {Bird}, {Bizyaev}, {Blanc}, {Blanton}, {Boardman}, {Bolton},
  {Boquien}, {Borissova}, {Bovy}, {Brandt}, {Brown}, {Brownstein}, {Brusa},
  {Buchner}, {Bundy}, {Burchett}, {Bureau}, {Burgasser}, {Cabang}, {Campbell},
  {Cappellari}, {Carlberg}, {Wanderley}, {Carrera}, {Cash}, {Chen}, {Chen},
  {Cherinka}, {Chiappini}, {Choi}, {Chojnowski}, {Chung}, {Clerc}, {Cohen},
  {Comerford}, {Comparat}, {da Costa}, {Covey}, {Crane}, {Cruz-Gonzalez},
  {Culhane}, {Cunha}, {Dai}, {Damke}, {Darling}, {Davidson}, {Davies},
  {Dawson}, {De Lee}, {Diamond-Stanic}, {Cano-D{\'\i}az}, {S{\'a}nchez},
  {Donor}, {Duckworth}, {Dwelly}, {Eisenstein}, {Elsworth}, {Emsellem},
  {Eracleous}, {Escoffier}, {Fan}, {Farr}, {Feng}, {Fern{\'a}ndez-Trincado},
  {Feuillet}, {Filipp}, {Fillingham}, {Frinchaboy}, {Fromenteau}, {Galbany},
  {Garc{\'\i}a}, {Garc{\'\i}a-Hern{\'a}ndez}, {Ge}, {Geisler}, {Gelfand},
  {G{\'e}ron}, {Gibson}, {Goddy}, {Godoy-Rivera}, {Grabowski}, {Green},
  {Greener}, {Grier}, {Griffith}, {Guo}, {Guy}, {Hadjara}, {Harding},
  {Hasselquist}, {Hayes}, {Hearty}, {Hern{\'a}ndez}, {Hill}, {Hogg},
  {Holtzman}, {Horta}, {Hsieh}, {Hsu}, {Hsu}, {Huber}, {Huertas-Company},
  {Hutchinson}, {Hwang}, {Ibarra-Medel}, {Chitham}, {Ilha}, {Imig}, {Jaekle},
  {Jayasinghe}, {Ji}, {Johnson}, {Jones}, {J{\"o}nsson}, {Katkov}, {Khalatyan},
  {Kinemuchi}, {Kisku}, {Knapen}, {Kneib}, {Kollmeier}, {Kong}, {Kounkel},
  {Kreckel}, {Krishnarao}, {Lacerna}, {Lane}, {Langgin}, {Lavender}, {Law},
  {Lazarz}, {Leung}, {Leung}, {Lewis}, {Li}, {Li}, {Lian}, {Liang}, {Lin},
  {Lin}, {Lin}, {Lintott}, {Long}, {Longa-Pe{\~n}a}, {L{\'o}pez-Cob{\'a}},
  {Lu}, {Lundgren}, {Luo}, {Mackereth}, {de la Macorra}, {Mahadevan},
  {Majewski}, {Manchado}, {Mandeville}, {Maraston}, {Margalef-Bentabol},
  {Masseron}, {Masters}, {Mathur}, {McDermid}, {Mckay}, {Merloni},
  {Merrifield}, {Meszaros}, {Miglio}, {Di Mille}, {Minniti}, {Minsley},
  {Monachesi}, {Moon}, {Mosser}, {Mulchaey}, {Muna}, {Mu{\~n}oz}, {Myers},
  {Myers}, {Nadathur}, {Nair}, {Nandra}, {Neumann}, {Newman}, {Nidever},
  {Nikakhtar}, {Nitschelm}, {O'Connell}, {Garma-Oehmichen}, {Luan Souza de
  Oliveira}, {Olney}, {Oravetz}, {Ortigoza-Urdaneta}, {Osorio}, {Otter},
  {Pace}, {Padilla}, {Pan}, {Pan}, {Parikh}, {Parker}, {Peirani}, {Pe{\~n}a
  Ram{\'\i}rez}, {Penny}, {Percival}, {Perez-Fournon}, {Pinsonneault},
  {Poidevin}, {Poovelil}, {Price-Whelan}, {B{\'a}rbara de Andrade Queiroz},
  {Raddick}, {Ray}, {Rembold}, {Riddle}, {Riffel}, {Riffel}, {Rix}, {Robin},
  {Rodr{\'\i}guez-Puebla}, {Roman-Lopes}, {Rom{\'a}n-Z{\'u}{\~n}iga}, {Rose},
  {Ross}, {Rossi}, {Rubin}, {Salvato}, {S{\'a}nchez}, {S{\'a}nchez-Gallego},
  {Sanderson}, {Santana Rojas}, {Sarceno}, {Sarmiento}, {Sayres}, {Sazonova},
  {Schaefer}, {Schiavon}, {Schlegel}, {Schneider}, {Schultheis}, {Schwope},
  {Serenelli}, {Serna}, {Shao}, {Shapiro}, {Sharma}, {Shen}, {Shetrone}, {Shu},
  {Simon}, {Skrutskie}, {Smethurst}, {Smith}, {Sobeck}, {Spoo}, {Sprague},
  {Stark}, {Stassun}, {Steinmetz}, {Stello}, {Stone-Martinez},
  {Storchi-Bergmann}, {Stringfellow}, {Stutz}, {Su}, {Taghizadeh-Popp},
  {Talbot}, {Tayar}, {Telles}, {Teske}, {Thakar}, {Theissen}, {Tkachenko},
  {Thomas}, {Tojeiro}, {Hernandez Toledo}, {Troup}, {Trump}, {Trussler},
  {Turner}, {Tuttle}, {Unda-Sanzana}, {V{\'a}zquez-Mata}, {Valentini},
  {Valenzuela}, {Vargas-Gonz{\'a}lez}, {Vargas-Maga{\~n}a}, {Alfaro},
  {Villanova}, {Vincenzo}, {Wake}, {Warfield}, {Washington}, {Weaver},
  {Weijmans}, {Weinberg}, {Weiss}, {Westfall}, {Wild}, {Wilde}, {Wilson},
  {Wilson}, {Wilson}, {Wolf}, {Wood-Vasey}, {Yan}, {Zamora}, {Zasowski},
  {Zhang}, {Zhao}, {Zheng}, {Zheng}, \& {Zhu}}]{SDSSDR17}
{Abdurro'uf}, {Accetta}, K., {Aerts}, C., {et~al.} 2022, \apjs, 259, 35,
  \dodoi{10.3847/1538-4365/ac4414}

\bibitem[{{Allende Prieto} {et~al.}(2006){Allende Prieto}, {Beers}, {Wilhelm},
  {Newberg}, {Rockosi}, {Yanny}, \& {Lee}}]{Prieto2006}
{Allende Prieto}, C., {Beers}, T.~C., {Wilhelm}, R., {et~al.} 2006, \apj, 636,
  804, \dodoi{10.1086/498131}

\bibitem[{{Anders} {et~al.}(2022){Anders}, {Khalatyan}, {Queiroz}, {Chiappini},
  {Ard{\`e}vol}, {Casamiquela}, {Figueras}, {Jim{\'e}nez-Arranz}, {Jordi},
  {Mongui{\'o}}, {Romero-G{\'o}mez}, {Altamirano}, {Antoja}, {Assaad},
  {Cantat-Gaudin}, {Castro-Ginard}, {Enke}, {Girardi}, {Guiglion}, {Khan},
  {Luri}, {Miglio}, {Minchev}, {Ramos}, {Santiago}, \&
  {Steinmetz}}]{Anders2022}
{Anders}, F., {Khalatyan}, A., {Queiroz}, A.~B.~A., {et~al.} 2022, \aap, 658,
  A91, \dodoi{10.1051/0004-6361/202142369}

\bibitem[{{Antoja} {et~al.}(2023){Antoja}, {Ramos}, {Garc{\'\i}a-Conde},
  {Bernet}, {Laporte}, \& {Katz}}]{Antoja2023}
{Antoja}, T., {Ramos}, P., {Garc{\'\i}a-Conde}, B., {et~al.} 2023, \aap, 673,
  A115, \dodoi{10.1051/0004-6361/202245518}

\bibitem[{{Antoja} {et~al.}(2018){Antoja}, {Helmi}, {Romero-G{\'o}mez}, {Katz},
  {Babusiaux}, {Drimmel}, {Evans}, {Figueras}, {Poggio}, {Reyl{\'e}}, {Robin},
  {Seabroke}, \& {Soubiran}}]{Antoja2018}
{Antoja}, T., {Helmi}, A., {Romero-G{\'o}mez}, M., {et~al.} 2018, \nat, 561,
  360, \dodoi{10.1038/s41586-018-0510-7}

\bibitem[{{Astropy Collaboration} {et~al.}(2013){Astropy Collaboration},
  {Robitaille}, {Tollerud}, {Greenfield}, {Droettboom}, {Bray}, {Aldcroft},
  {Davis}, {Ginsburg}, {Price-Whelan}, {Kerzendorf}, {Conley}, {Crighton},
  {Barbary}, {Muna}, {Ferguson}, {Grollier}, {Parikh}, {Nair}, {Unther},
  {Deil}, {Woillez}, {Conseil}, {Kramer}, {Turner}, {Singer}, {Fox}, {Weaver},
  {Zabalza}, {Edwards}, {Azalee Bostroem}, {Burke}, {Casey}, {Crawford},
  {Dencheva}, {Ely}, {Jenness}, {Labrie}, {Lim}, {Pierfederici}, {Pontzen},
  {Ptak}, {Refsdal}, {Servillat}, \& {Streicher}}]{astropy2013}
{Astropy Collaboration}, {Robitaille}, T.~P., {Tollerud}, E.~J., {et~al.} 2013,
  \aap, 558, A33, \dodoi{10.1051/0004-6361/201322068}

\bibitem[{{Astropy Collaboration} {et~al.}(2018){Astropy Collaboration},
  {Price-Whelan}, {Sip{\H{o}}cz}, {G{\"u}nther}, {Lim}, {Crawford}, {Conseil},
  {Shupe}, {Craig}, {Dencheva}, {Ginsburg}, {VanderPlas}, {Bradley},
  {P{\'e}rez-Su{\'a}rez}, {de Val-Borro}, {Aldcroft}, {Cruz}, {Robitaille},
  {Tollerud}, {Ardelean}, {Babej}, {Bach}, {Bachetti}, {Bakanov}, {Bamford},
  {Barentsen}, {Barmby}, {Baumbach}, {Berry}, {Biscani}, {Boquien}, {Bostroem},
  {Bouma}, {Brammer}, {Bray}, {Breytenbach}, {Buddelmeijer}, {Burke},
  {Calderone}, {Cano Rodr{\'\i}guez}, {Cara}, {Cardoso}, {Cheedella}, {Copin},
  {Corrales}, {Crichton}, {D'Avella}, {Deil}, {Depagne}, {Dietrich}, {Donath},
  {Droettboom}, {Earl}, {Erben}, {Fabbro}, {Ferreira}, {Finethy}, {Fox},
  {Garrison}, {Gibbons}, {Goldstein}, {Gommers}, {Greco}, {Greenfield},
  {Groener}, {Grollier}, {Hagen}, {Hirst}, {Homeier}, {Horton}, {Hosseinzadeh},
  {Hu}, {Hunkeler}, {Ivezi{\'c}}, {Jain}, {Jenness}, {Kanarek}, {Kendrew},
  {Kern}, {Kerzendorf}, {Khvalko}, {King}, {Kirkby}, {Kulkarni}, {Kumar},
  {Lee}, {Lenz}, {Littlefair}, {Ma}, {Macleod}, {Mastropietro}, {McCully},
  {Montagnac}, {Morris}, {Mueller}, {Mumford}, {Muna}, {Murphy}, {Nelson},
  {Nguyen}, {Ninan}, {N{\"o}the}, {Ogaz}, {Oh}, {Parejko}, {Parley}, {Pascual},
  {Patil}, {Patil}, {Plunkett}, {Prochaska}, {Rastogi}, {Reddy Janga},
  {Sabater}, {Sakurikar}, {Seifert}, {Sherbert}, {Sherwood-Taylor}, {Shih},
  {Sick}, {Silbiger}, {Singanamalla}, {Singer}, {Sladen}, {Sooley},
  {Sornarajah}, {Streicher}, {Teuben}, {Thomas}, {Tremblay}, {Turner},
  {Terr{\'o}n}, {van Kerkwijk}, {de la Vega}, {Watkins}, {Weaver}, {Whitmore},
  {Woillez}, {Zabalza}, \& {Astropy Contributors}}]{astropy2018}
{Astropy Collaboration}, {Price-Whelan}, A.~M., {Sip{\H{o}}cz}, B.~M., {et~al.}
  2018, \aj, 156, 123, \dodoi{10.3847/1538-3881/aabc4f}

\bibitem[{{Astropy Collaboration} {et~al.}(2022){Astropy Collaboration},
  {Price-Whelan}, {Lim}, {Earl}, {Starkman}, {Bradley}, {Shupe}, {Patil},
  {Corrales}, {Brasseur}, {N{\"o}the}, {Donath}, {Tollerud}, {Morris},
  {Ginsburg}, {Vaher}, {Weaver}, {Tocknell}, {Jamieson}, {van Kerkwijk},
  {Robitaille}, {Merry}, {Bachetti}, {G{\"u}nther}, {Aldcroft},
  {Alvarado-Montes}, {Archibald}, {B{\'o}di}, {Bapat}, {Barentsen},
  {Baz{\'a}n}, {Biswas}, {Boquien}, {Burke}, {Cara}, {Cara}, {Conroy},
  {Conseil}, {Craig}, {Cross}, {Cruz}, {D'Eugenio}, {Dencheva}, {Devillepoix},
  {Dietrich}, {Eigenbrot}, {Erben}, {Ferreira}, {Foreman-Mackey}, {Fox},
  {Freij}, {Garg}, {Geda}, {Glattly}, {Gondhalekar}, {Gordon}, {Grant},
  {Greenfield}, {Groener}, {Guest}, {Gurovich}, {Handberg}, {Hart},
  {Hatfield-Dodds}, {Homeier}, {Hosseinzadeh}, {Jenness}, {Jones}, {Joseph},
  {Kalmbach}, {Karamehmetoglu}, {Ka{\l}uszy{\'n}ski}, {Kelley}, {Kern},
  {Kerzendorf}, {Koch}, {Kulumani}, {Lee}, {Ly}, {Ma}, {MacBride}, {Maljaars},
  {Muna}, {Murphy}, {Norman}, {O'Steen}, {Oman}, {Pacifici}, {Pascual},
  {Pascual-Granado}, {Patil}, {Perren}, {Pickering}, {Rastogi}, {Roulston},
  {Ryan}, {Rykoff}, {Sabater}, {Sakurikar}, {Salgado}, {Sanghi}, {Saunders},
  {Savchenko}, {Schwardt}, {Seifert-Eckert}, {Shih}, {Jain}, {Shukla}, {Sick},
  {Simpson}, {Singanamalla}, {Singer}, {Singhal}, {Sinha}, {Sip{\H{o}}cz},
  {Spitler}, {Stansby}, {Streicher}, {{\v{S}}umak}, {Swinbank}, {Taranu},
  {Tewary}, {Tremblay}, {de Val-Borro}, {Van Kooten}, {Vasovi{\'c}}, {Verma},
  {de Miranda Cardoso}, {Williams}, {Wilson}, {Winkel}, {Wood-Vasey}, {Xue},
  {Yoachim}, {Zhang}, {Zonca}, \& {Astropy Project Contributors}}]{astropy2022}
{Astropy Collaboration}, {Price-Whelan}, A.~M., {Lim}, P.~L., {et~al.} 2022,
  \apj, 935, 167, \dodoi{10.3847/1538-4357/ac7c74}

\bibitem[{{Beaton} {et~al.}(2021){Beaton}, {Oelkers}, {Hayes}, {Covey},
  {Chojnowski}, {De Lee}, {Sobeck}, {Majewski}, {Cohen}, {Fernandez-Trincado},
  {Longa-Pena}, {O'Connell}, {Santana}, {Stringfellow}, {Zasowski}, {Aerts},
  {Anguiano}, {Bender}, {Canas}, {Cunha}, {Fleming}, {Frinchaboy}, {Feuillet},
  {Harding}, {Hasselquist}, {Holtzman}, {Johnson}, {Kollmeier}, {Kounkel},
  {Mahadevan}, {Price-Whelan}, {Rojas-Arriagada}, {Roman-Zuniga}, {Schlafly},
  {Schultheis}, {Shetrone}, {Simon}, {Stassun}, {Stutz}, {Tayar}, {Teske},
  {Tkachenko}, {Troup}, {Albareti}, {Bizyaev}, {Bovy}, {Burgasser}, {Comparat},
  {Downes}, {Geisler}, {Inno}, {Manchado}, {Ness}, {Pinsonneault}, {Prada},
  {Roman-Lopes}, {Simonian}, {Smith}, {Yan}, \& {Zamora}}]{Beaton2021}
{Beaton}, R.~L., {Oelkers}, R.~J., {Hayes}, C.~R., {et~al.} 2021, arXiv
  e-prints, arXiv:2108.11907.
\newblock \doarXiv{2108.11907}

\bibitem[{{Bennett} \& {Bovy}(2019)}]{Bennett2019}
{Bennett}, M., \& {Bovy}, J. 2019, \mnras, 482, 1417,
  \dodoi{10.1093/mnras/sty2813}

\bibitem[{{Bienaym{\'e}} {et~al.}(2014){Bienaym{\'e}}, {Famaey}, {Siebert},
  {Freeman}, {Gibson}, {Gilmore}, {Grebel}, {Bland-Hawthorn}, {Kordopatis},
  {Munari}, {Navarro}, {Parker}, {Reid}, {Seabroke}, {Siviero}, {Steinmetz},
  {Watson}, {Wyse}, \& {Zwitter}}]{Bienayme2014}
{Bienaym{\'e}}, O., {Famaey}, B., {Siebert}, A., {et~al.} 2014, \aap, 571, A92,
  \dodoi{10.1051/0004-6361/201424478}

\bibitem[{{Binney} \& {Sch{\"o}nrich}(2018)}]{Binney2018}
{Binney}, J., \& {Sch{\"o}nrich}, R. 2018, \mnras, 481, 1501,
  \dodoi{10.1093/mnras/sty2378}

\bibitem[{{Binney} \& {Tremaine}(2008)}]{Binney2008}
{Binney}, J., \& {Tremaine}, S. 2008, {Galactic Dynamics: Second Edition}

\bibitem[{{Binney} \& {Vasiliev}(2023)}]{Binney2023}
{Binney}, J., \& {Vasiliev}, E. 2023, arXiv e-prints, arXiv:2306.11602,
  \dodoi{10.48550/arXiv.2306.11602}

\bibitem[{{Bland-Hawthorn} \& {Gerhard}(2016)}]{Bland2016}
{Bland-Hawthorn}, J., \& {Gerhard}, O. 2016, \araa, 54, 529,
  \dodoi{10.1146/annurev-astro-081915-023441}

\bibitem[{{Bland-Hawthorn} \& {Tepper-Garc{\'\i}a}(2021)}]{Bland2021}
{Bland-Hawthorn}, J., \& {Tepper-Garc{\'\i}a}, T. 2021, \mnras, 504, 3168,
  \dodoi{10.1093/mnras/stab704}

\bibitem[{Blondel {et~al.}(2021)Blondel, Berthet, Cuturi, Frostig, Hoyer,
  Llinares-L{\'o}pez, Pedregosa, \& Vert}]{jaxopt:2021}
Blondel, M., Berthet, Q., Cuturi, M., {et~al.} 2021, arXiv preprint
  arXiv:2105.15183

\bibitem[{{Bovy} \& {Rix}(2013)}]{Bovy2013}
{Bovy}, J., \& {Rix}, H.-W. 2013, \apj, 779, 115,
  \dodoi{10.1088/0004-637X/779/2/115}

\bibitem[{{Bovy} {et~al.}(2016){Bovy}, {Rix}, {Schlafly}, {Nidever},
  {Holtzman}, {Shetrone}, \& {Beers}}]{Bovy2016}
{Bovy}, J., {Rix}, H.-W., {Schlafly}, E.~F., {et~al.} 2016, \apj, 823, 30,
  \dodoi{10.3847/0004-637X/823/1/30}

\bibitem[{{Bowen} \& {Vaughan}(1973)}]{BowenVaughan1973}
{Bowen}, I.~S., \& {Vaughan}, A.~H., J. 1973, \ao, 12, 1430,
  \dodoi{10.1364/AO.12.001430}

\bibitem[{Bradbury {et~al.}(2018)Bradbury, Frostig, Hawkins, Johnson, Leary,
  Maclaurin, Necula, Paszke, Vander{P}las, Wanderman-{M}ilne, \&
  Zhang}]{jax2018github}
Bradbury, J., Frostig, R., Hawkins, P., {et~al.} 2018, {JAX}: composable
  transformations of {P}ython+{N}um{P}y programs, 0.3.13.
\newblock \url{http://github.com/google/jax}

\bibitem[{{Cheng} {et~al.}(2023){Cheng}, {Anguiano}, {Majewski}, \&
  {Arras}}]{Cheng2023}
{Cheng}, X., {Anguiano}, B., {Majewski}, S.~R., \& {Arras}, P. 2023, arXiv
  e-prints, arXiv:2309.17405, \dodoi{10.48550/arXiv.2309.17405}

\bibitem[{{Creze} {et~al.}(1998){Creze}, {Chereul}, {Bienayme}, \&
  {Pichon}}]{Creze1998}
{Creze}, M., {Chereul}, E., {Bienayme}, O., \& {Pichon}, C. 1998, \aap, 329,
  920, \dodoi{10.48550/arXiv.astro-ph/9709022}

\bibitem[{{Cui} {et~al.}(2012){Cui}, {Zhao}, {Chu}, {Li}, {Li}, {Zhang}, {Su},
  {Yao}, {Wang}, {Xing}, {Li}, {Zhu}, {Wang}, {Gu}, {Luo}, {Xu}, {Zhang},
  {Liu}, {Zhang}, {Yang}, {Cao}, {Chen}, {Chen}, {Chen}, {Chen}, {Chu}, {Feng},
  {Gong}, {Hou}, {Hu}, {Hu}, {Hu}, {Jia}, {Jiang}, {Jiang}, {Jiang}, {Jin},
  {Li}, {Li}, {Li}, {Liu}, {Liu}, {Lu}, {Mao}, {Men}, {Qi}, {Qi}, {Shi},
  {Tang}, {Tao}, {Wang}, {Wang}, {Wang}, {Wang}, {Wang}, {Wang}, {Wang},
  {Wang}, {Wang}, {Wang}, {Wang}, {Wang}, {Xu}, {Xu}, {Yang}, {Yu}, {Yuan},
  {Yuan}, {Zhai}, {Zhang}, {Zhang}, {Zhang}, {Zhao}, {Zhou}, {Zhou}, {Zhu}, \&
  {Zou}}]{Cui2012}
{Cui}, X.-Q., {Zhao}, Y.-H., {Chu}, Y.-Q., {et~al.} 2012, Research in Astronomy
  and Astrophysics, 12, 1197, \dodoi{10.1088/1674-4527/12/9/003}

\bibitem[{{Cunha} {et~al.}(2017){Cunha}, {Smith}, {Hasselquist}, {Souto},
  {Shetrone}, {Allende Prieto}, {Bizyaev}, {Frinchaboy},
  {Garc{\'\i}a-Hern{\'a}ndez}, {Holtzman}, {Johnson}, {J{\H{o}}nsson},
  {Majewski}, {M{\'e}sz{\'a}ros}, {Nidever}, {Pinsonneault}, {Schiavon},
  {Sobeck}, {Skrutskie}, {Zamora}, {Zasowski}, \&
  {Fern{\'a}ndez-Trincado}}]{Cunha2017}
{Cunha}, K., {Smith}, V.~V., {Hasselquist}, S., {et~al.} 2017, \apj, 844, 145,
  \dodoi{10.3847/1538-4357/aa7beb}

\bibitem[{{Dalton} {et~al.}(2012){Dalton}, {Trager}, {Abrams}, {Carter},
  {Bonifacio}, {Aguerri}, {MacIntosh}, {Evans}, {Lewis}, {Navarro}, {Agocs},
  {Dee}, {Rousset}, {Tosh}, {Middleton}, {Pragt}, {Terrett}, {Brock}, {Benn},
  {Verheijen}, {Cano Infantes}, {Bevil}, {Steele}, {Mottram}, {Bates},
  {Gribbin}, {Rey}, {Rodriguez}, {Delgado}, {Guinouard}, {Walton}, {Irwin},
  {Jagourel}, {Stuik}, {Gerlofsma}, {Roelfsma}, {Skillen}, {Ridings},
  {Balcells}, {Daban}, {Gouvret}, {Venema}, \& {Girard}}]{Dalton2012}
{Dalton}, G., {Trager}, S.~C., {Abrams}, D.~C., {et~al.} 2012, in Society of
  Photo-Optical Instrumentation Engineers (SPIE) Conference Series, Vol. 8446,
  Ground-based and Airborne Instrumentation for Astronomy IV, ed. I.~S.
  {McLean}, S.~K. {Ramsay}, \& H.~{Takami}, 84460P, \dodoi{10.1117/12.925950}

\bibitem[{{de Jong} {et~al.}(2012){de Jong}, {Bellido-Tirado}, {Chiappini},
  {Depagne}, {Haynes}, {Johl}, {Schnurr}, {Schwope}, {Walcher}, {Dionies},
  {Haynes}, {Kelz}, {Kitaura}, {Lamer}, {Minchev}, {M{\"u}ller}, {Nuza},
  {Olaya}, {Piffl}, {Popow}, {Steinmetz}, {Ural}, {Williams}, {Winkler},
  {Wisotzki}, {Ansorge}, {Banerji}, {Gonzalez Solares}, {Irwin}, {Kennicutt},
  {King}, {McMahon}, {Koposov}, {Parry}, {Sun}, {Walton}, {Finger}, {Iwert},
  {Krumpe}, {Lizon}, {Vincenzo}, {Amans}, {Bonifacio}, {Cohen}, {Francois},
  {Jagourel}, {Mignot}, {Royer}, {Sartoretti}, {Bender}, {Grupp}, {Hess},
  {Lang-Bardl}, {Muschielok}, {B{\"o}hringer}, {Boller}, {Bongiorno}, {Brusa},
  {Dwelly}, {Merloni}, {Nandra}, {Salvato}, {Pragt}, {Navarro}, {Gerlofsma},
  {Roelfsema}, {Dalton}, {Middleton}, {Tosh}, {Boeche}, {Caffau}, {Christlieb},
  {Grebel}, {Hansen}, {Koch}, {Ludwig}, {Quirrenbach}, {Sbordone}, {Seifert},
  {Thimm}, {Trifonov}, {Helmi}, {Trager}, {Feltzing}, {Korn}, \&
  {Boland}}]{DeJong2012}
{de Jong}, R.~S., {Bellido-Tirado}, O., {Chiappini}, C., {et~al.} 2012, in
  Society of Photo-Optical Instrumentation Engineers (SPIE) Conference Series,
  Vol. 8446, Ground-based and Airborne Instrumentation for Astronomy IV, ed.
  I.~S. {McLean}, S.~K. {Ramsay}, \& H.~{Takami}, 84460T,
  \dodoi{10.1117/12.926239}

\bibitem[{{de Salas} \& {Widmark}(2021)}]{DeSalas2021}
{de Salas}, P.~F., \& {Widmark}, A. 2021, Reports on Progress in Physics, 84,
  104901, \dodoi{10.1088/1361-6633/ac24e7}

\bibitem[{{Eilers} {et~al.}(2019){Eilers}, {Hogg}, {Rix}, \&
  {Ness}}]{Eilers2019}
{Eilers}, A.-C., {Hogg}, D.~W., {Rix}, H.-W., \& {Ness}, M.~K. 2019, \apj, 871,
  120, \dodoi{10.3847/1538-4357/aaf648}

\bibitem[{{Eilers} {et~al.}(2022){Eilers}, {Hogg}, {Rix}, {Ness},
  {Price-Whelan}, {M{\'e}sz{\'a}ros}, \& {Nitschelm}}]{Eilers2022}
{Eilers}, A.-C., {Hogg}, D.~W., {Rix}, H.-W., {et~al.} 2022, \apj, 928, 23,
  \dodoi{10.3847/1538-4357/ac54ad}

\bibitem[{Fletcher(1987)}]{Flet87}
Fletcher, R. 1987, Practical Methods of Optimization, 2nd edn. (New York, NY,
  USA: John Wiley \& Sons)

\bibitem[{{Flynn} {et~al.}(2006){Flynn}, {Holmberg}, {Portinari}, {Fuchs}, \&
  {Jahrei{\ss}}}]{Flynn2006}
{Flynn}, C., {Holmberg}, J., {Portinari}, L., {Fuchs}, B., \& {Jahrei{\ss}}, H.
  2006, \mnras, 372, 1149, \dodoi{10.1111/j.1365-2966.2006.10911.x}

\bibitem[{{Frankel} {et~al.}(2023){Frankel}, {Bovy}, {Tremaine}, \&
  {Hogg}}]{Frankel2023}
{Frankel}, N., {Bovy}, J., {Tremaine}, S., \& {Hogg}, D.~W. 2023, \mnras, 521,
  5917, \dodoi{10.1093/mnras/stad908}

\bibitem[{{Frankel} {et~al.}(2018){Frankel}, {Rix}, {Ting}, {Ness}, \&
  {Hogg}}]{Frankel2018}
{Frankel}, N., {Rix}, H.-W., {Ting}, Y.-S., {Ness}, M., \& {Hogg}, D.~W. 2018,
  \apj, 865, 96, \dodoi{10.3847/1538-4357/aadba5}

\bibitem[{{Gaia Collaboration} {et~al.}(2016){Gaia Collaboration}, {Prusti},
  {de Bruijne}, {Brown}, {Vallenari}, {Babusiaux}, {Bailer-Jones}, {Bastian},
  {Biermann}, {Evans}, {Eyer}, {Jansen}, {Jordi}, {Klioner}, {Lammers},
  {Lindegren}, {Luri}, {Mignard}, {Milligan}, {Panem}, {Poinsignon},
  {Pourbaix}, {Randich}, {Sarri}, {Sartoretti}, {Siddiqui}, {Soubiran},
  {Valette}, {van Leeuwen}, {Walton}, {Aerts}, {Arenou}, {Cropper}, {Drimmel},
  {H{\o}g}, {Katz}, {Lattanzi}, {O'Mullane}, {Grebel}, {Holland}, {Huc},
  {Passot}, {Bramante}, {Cacciari}, {Casta{\~n}eda}, {Chaoul}, {Cheek}, {De
  Angeli}, {Fabricius}, {Guerra}, {Hern{\'a}ndez}, {Jean-Antoine-Piccolo},
  {Masana}, {Messineo}, {Mowlavi}, {Nienartowicz}, {Ord{\'o}{\~n}ez-Blanco},
  {Panuzzo}, {Portell}, {Richards}, {Riello}, {Seabroke}, {Tanga},
  {Th{\'e}venin}, {Torra}, {Els}, {Gracia-Abril}, {Comoretto},
  {Garcia-Reinaldos}, {Lock}, {Mercier}, {Altmann}, {Andrae}, {Astraatmadja},
  {Bellas-Velidis}, {Benson}, {Berthier}, {Blomme}, {Busso}, {Carry},
  {Cellino}, {Clementini}, {Cowell}, {Creevey}, {Cuypers}, {Davidson}, {De
  Ridder}, {de Torres}, {Delchambre}, {Dell'Oro}, {Ducourant}, {Fr{\'e}mat},
  {Garc{\'\i}a-Torres}, {Gosset}, {Halbwachs}, {Hambly}, {Harrison}, {Hauser},
  {Hestroffer}, {Hodgkin}, {Huckle}, {Hutton}, {Jasniewicz}, {Jordan},
  {Kontizas}, {Korn}, {Lanzafame}, {Manteiga}, {Moitinho}, {Muinonen},
  {Osinde}, {Pancino}, {Pauwels}, {Petit}, {Recio-Blanco}, {Robin}, {Sarro},
  {Siopis}, {Smith}, {Smith}, {Sozzetti}, {Thuillot}, {van Reeven}, {Viala},
  {Abbas}, {Abreu Aramburu}, {Accart}, {Aguado}, {Allan}, {Allasia},
  {Altavilla}, {{\'A}lvarez}, {Alves}, {Anderson}, {Andrei}, {Anglada Varela},
  {Antiche}, {Antoja}, {Ant{\'o}n}, {Arcay}, {Atzei}, {Ayache}, {Bach},
  {Baker}, {Balaguer-N{\'u}{\~n}ez}, {Barache}, {Barata}, {Barbier}, {Barblan},
  {Baroni}, {Barrado y Navascu{\'e}s}, {Barros}, {Barstow}, {Becciani},
  {Bellazzini}, {Bellei}, {Bello Garc{\'\i}a}, {Belokurov}, {Bendjoya},
  {Berihuete}, {Bianchi}, {Bienaym{\'e}}, {Billebaud}, {Blagorodnova},
  {Blanco-Cuaresma}, {Boch}, {Bombrun}, {Borrachero}, {Bouquillon}, {Bourda},
  {Bouy}, {Bragaglia}, {Breddels}, {Brouillet}, {Br{\"u}semeister},
  {Bucciarelli}, {Budnik}, {Burgess}, {Burgon}, {Burlacu}, {Busonero}, {Buzzi},
  {Caffau}, {Cambras}, {Campbell}, {Cancelliere}, {Cantat-Gaudin}, {Carlucci},
  {Carrasco}, {Castellani}, {Charlot}, {Charnas}, {Charvet}, {Chassat},
  {Chiavassa}, {Clotet}, {Cocozza}, {Collins}, {Collins}, {Costigan}, {Crifo},
  {Cross}, {Crosta}, {Crowley}, {Dafonte}, {Damerdji}, {Dapergolas}, {David},
  {David}, {De Cat}, {de Felice}, {de Laverny}, {De Luise}, {De March}, {de
  Martino}, {de Souza}, {Debosscher}, {del Pozo}, {Delbo}, {Delgado},
  {Delgado}, {di Marco}, {Di Matteo}, {Diakite}, {Distefano}, {Dolding}, {Dos
  Anjos}, {Drazinos}, {Dur{\'a}n}, {Dzigan}, {Ecale}, {Edvardsson}, {Enke},
  {Erdmann}, {Escolar}, {Espina}, {Evans}, {Eynard Bontemps}, {Fabre},
  {Fabrizio}, {Faigler}, {Falc{\~a}o}, {Farr{\`a}s Casas}, {Faye}, {Federici},
  {Fedorets}, {Fern{\'a}ndez-Hern{\'a}ndez}, {Fernique}, {Fienga}, {Figueras},
  {Filippi}, {Findeisen}, {Fonti}, {Fouesneau}, {Fraile}, {Fraser}, {Fuchs},
  {Furnell}, {Gai}, {Galleti}, {Galluccio}, {Garabato}, {Garc{\'\i}a-Sedano},
  {Gar{\'e}}, {Garofalo}, {Garralda}, {Gavras}, {Gerssen}, {Geyer}, {Gilmore},
  {Girona}, {Giuffrida}, {Gomes}, {Gonz{\'a}lez-Marcos},
  {Gonz{\'a}lez-N{\'u}{\~n}ez}, {Gonz{\'a}lez-Vidal}, {Granvik}, {Guerrier},
  {Guillout}, {Guiraud}, {G{\'u}rpide}, {Guti{\'e}rrez-S{\'a}nchez}, {Guy},
  {Haigron}, {Hatzidimitriou}, {Haywood}, {Heiter}, {Helmi}, {Hobbs},
  {Hofmann}, {Holl}, {Holland}, {Hunt}, {Hypki}, {Icardi}, {Irwin}, {Jevardat
  de Fombelle}, {Jofr{\'e}}, {Jonker}, {Jorissen}, {Julbe}, {Karampelas},
  {Kochoska}, {Kohley}, {Kolenberg}, {Kontizas}, {Koposov}, {Kordopatis},
  {Koubsky}, {Kowalczyk}, {Krone-Martins}, {Kudryashova}, {Kull}, {Bachchan},
  {Lacoste-Seris}, {Lanza}, {Lavigne}, {Le Poncin-Lafitte}, {Lebreton},
  {Lebzelter}, {Leccia}, {Leclerc}, {Lecoeur-Taibi}, {Lemaitre}, {Lenhardt},
  {Leroux}, {Liao}, {Licata}, {Lindstr{\o}m}, {Lister}, {Livanou}, {Lobel},
  {L{\"o}ffler}, {L{\'o}pez}, {Lopez-Lozano}, {Lorenz}, {Loureiro},
  {MacDonald}, {Magalh{\~a}es Fernandes}, {Managau}, {Mann}, {Mantelet},
  {Marchal}, {Marchant}, {Marconi}, {Marie}, {Marinoni}, {Marrese},
  {Marschalk{\'o}}, {Marshall}, {Mart{\'\i}n-Fleitas}, {Martino}, {Mary},
  {Matijevi{\v{c}}}, {Mazeh}, {McMillan}, {Messina}, {Mestre}, {Michalik},
  {Millar}, {Miranda}, {Molina}, {Molinaro}, {Molinaro}, {Moln{\'a}r},
  {Moniez}, {Montegriffo}, {Monteiro}, {Mor}, {Mora}, {Morbidelli}, {Morel},
  {Morgenthaler}, {Morley}, {Morris}, {Mulone}, {Muraveva}, {Musella},
  {Narbonne}, {Nelemans}, {Nicastro}, {Noval}, {Ord{\'e}novic},
  {Ordieres-Mer{\'e}}, {Osborne}, {Pagani}, {Pagano}, {Pailler}, {Palacin},
  {Palaversa}, {Parsons}, {Paulsen}, {Pecoraro}, {Pedrosa}, {Pentik{\"a}inen},
  {Pereira}, {Pichon}, {Piersimoni}, {Pineau}, {Plachy}, {Plum}, {Poujoulet},
  {Pr{\v{s}}a}, {Pulone}, {Ragaini}, {Rago}, {Rambaux}, {Ramos-Lerate},
  {Ranalli}, {Rauw}, {Read}, {Regibo}, {Renk}, {Reyl{\'e}}, {Ribeiro},
  {Rimoldini}, {Ripepi}, {Riva}, {Rixon}, {Roelens}, {Romero-G{\'o}mez},
  {Rowell}, {Royer}, {Rudolph}, {Ruiz-Dern}, {Sadowski}, {Sagrist{\`a}
  Sell{\'e}s}, {Sahlmann}, {Salgado}, {Salguero}, {Sarasso}, {Savietto},
  {Schnorhk}, {Schultheis}, {Sciacca}, {Segol}, {Segovia}, {Segransan},
  {Serpell}, {Shih}, {Smareglia}, {Smart}, {Smith}, {Solano}, {Solitro},
  {Sordo}, {Soria Nieto}, {Souchay}, {Spagna}, {Spoto}, {Stampa}, {Steele},
  {Steidelm{\"u}ller}, {Stephenson}, {Stoev}, {Suess}, {S{\"u}veges}, {Surdej},
  {Szabados}, {Szegedi-Elek}, {Tapiador}, {Taris}, {Tauran}, {Taylor},
  {Teixeira}, {Terrett}, {Tingley}, {Trager}, {Turon}, {Ulla}, {Utrilla},
  {Valentini}, {van Elteren}, {Van Hemelryck}, {van Leeuwen}, {Varadi},
  {Vecchiato}, {Veljanoski}, {Via}, {Vicente}, {Vogt}, {Voss}, {Votruba},
  {Voutsinas}, {Walmsley}, {Weiler}, {Weingrill}, {Werner}, {Wevers},
  {Whitehead}, {Wyrzykowski}, {Yoldas}, {{\v{Z}}erjal}, {Zucker}, {Zurbach},
  {Zwitter}, {Alecu}, {Allen}, {Allende Prieto}, {Amorim},
  {Anglada-Escud{\'e}}, {Arsenijevic}, {Azaz}, {Balm}, {Beck}, {Bernstein},
  {Bigot}, {Bijaoui}, {Blasco}, {Bonfigli}, {Bono}, {Boudreault}, {Bressan},
  {Brown}, {Brunet}, {Bunclark}, {Buonanno}, {Butkevich}, {Carret}, {Carrion},
  {Chemin}, {Ch{\'e}reau}, {Corcione}, {Darmigny}, {de Boer}, {de Teodoro}, {de
  Zeeuw}, {Delle Luche}, {Domingues}, {Dubath}, {Fodor}, {Fr{\'e}zouls},
  {Fries}, {Fustes}, {Fyfe}, {Gallardo}, {Gallegos}, {Gardiol}, {Gebran},
  {Gomboc}, {G{\'o}mez}, {Grux}, {Gueguen}, {Heyrovsky}, {Hoar}, {Iannicola},
  {Isasi Parache}, {Janotto}, {Joliet}, {Jonckheere}, {Keil}, {Kim},
  {Klagyivik}, {Klar}, {Knude}, {Kochukhov}, {Kolka}, {Kos}, {Kutka}, {Lainey},
  {LeBouquin}, {Liu}, {Loreggia}, {Makarov}, {Marseille}, {Martayan},
  {Martinez-Rubi}, {Massart}, {Meynadier}, {Mignot}, {Munari}, {Nguyen},
  {Nordlander}, {Ocvirk}, {O'Flaherty}, {Olias Sanz}, {Ortiz}, {Osorio},
  {Oszkiewicz}, {Ouzounis}, {Palmer}, {Park}, {Pasquato}, {Peltzer}, {Peralta},
  {P{\'e}turaud}, {Pieniluoma}, {Pigozzi}, {Poels}, {Prat}, {Prod'homme},
  {Raison}, {Rebordao}, {Risquez}, {Rocca-Volmerange}, {Rosen}, {Ruiz-Fuertes},
  {Russo}, {Sembay}, {Serraller Vizcaino}, {Short}, {Siebert}, {Silva},
  {Sinachopoulos}, {Slezak}, {Soffel}, {Sosnowska}, {Strai{\v{z}}ys}, {ter
  Linden}, {Terrell}, {Theil}, {Tiede}, {Troisi}, {Tsalmantza}, {Tur},
  {Vaccari}, {Vachier}, {Valles}, {Van Hamme}, {Veltz}, {Virtanen}, {Wallut},
  {Wichmann}, {Wilkinson}, {Ziaeepour}, \& {Zschocke}}]{Gaia2016}
{Gaia Collaboration}, {Prusti}, T., {de Bruijne}, J.~H.~J., {et~al.} 2016,
  \aap, 595, A1, \dodoi{10.1051/0004-6361/201629272}

\bibitem[{{Gaia Collaboration} {et~al.}(2022){Gaia Collaboration}, {Vallenari},
  {Brown}, {Prusti}, {de Bruijne}, {Arenou}, {Babusiaux}, {Biermann},
  {Creevey}, {Ducourant}, {Evans}, {Eyer}, {Guerra}, {Hutton}, {Jordi},
  {Klioner}, {Lammers}, {Lindegren}, {Luri}, {Mignard}, {Panem}, {Pourbaix},
  {Randich}, {Sartoretti}, {Soubiran}, {Tanga}, {Walton}, {Bailer-Jones},
  {Bastian}, {Drimmel}, {Jansen}, {Katz}, {Lattanzi}, {van Leeuwen}, {Bakker},
  {Cacciari}, {Casta{\~n}eda}, {De Angeli}, {Fabricius}, {Fouesneau},
  {Fr{\'e}mat}, {Galluccio}, {Guerrier}, {Heiter}, {Masana}, {Messineo},
  {Mowlavi}, {Nicolas}, {Nienartowicz}, {Pailler}, {Panuzzo}, {Riclet}, {Roux},
  {Seabroke}, {Sordo{\o}rcit}, {Th{\'e}venin}, {Gracia-Abril}, {Portell},
  {Teyssier}, {Altmann}, {Andrae}, {Audard}, {Bellas-Velidis}, {Benson},
  {Berthier}, {Blomme}, {Burgess}, {Busonero}, {Busso}, {C{\'a}novas}, {Carry},
  {Cellino}, {Cheek}, {Clementini}, {Damerdji}, {Davidson}, {de Teodoro},
  {Nu{\~n}ez Campos}, {Delchambre}, {Dell'Oro}, {Esquej},
  {Fern{\'a}ndez-Hern{\'a}ndez}, {Fraile}, {Garabato}, {Garc{\'\i}a-Lario},
  {Gosset}, {Haigron}, {Halbwachs}, {Hambly}, {Harrison}, {Hern{\'a}ndez},
  {Hestroffer}, {Hodgkin}, {Holl}, {Jan{\ss}en}, {Jevardat de Fombelle},
  {Jordan}, {Krone-Martins}, {Lanzafame}, {L{\"o}ffler}, {Marchal}, {Marrese},
  {Moitinho}, {Muinonen}, {Osborne}, {Pancino}, {Pauwels}, {Recio-Blanco},
  {Reyl{\'e}}, {Riello}, {Rimoldini}, {Roegiers}, {Rybizki}, {Sarro}, {Siopis},
  {Smith}, {Sozzetti}, {Utrilla}, {van Leeuwen}, {Abbas}, {{\'A}brah{\'a}m},
  {Abreu Aramburu}, {Aerts}, {Aguado}, {Ajaj}, {Aldea-Montero}, {Altavilla},
  {{\'A}lvarez}, {Alves}, {Anders}, {Anderson}, {Anglada Varela}, {Antoja},
  {Baines}, {Baker}, {Balaguer-N{\'u}{\~n}ez}, {Balbinot}, {Balog}, {Barache},
  {Barbato}, {Barros}, {Barstow}, {Bartolom{\'e}}, {Bassilana}, {Bauchet},
  {Becciani}, {Bellazzini}, {Berihuete}, {Bernet}, {Bertone}, {Bianchi},
  {Binnenfeld}, {Blanco-Cuaresma}, {Blazere}, {Boch}, {Bombrun}, {Bossini},
  {Bouquillon}, {Bragaglia}, {Bramante}, {Breedt}, {Bressan}, {Brouillet},
  {Brugaletta}, {Bucciarelli}, {Burlacu}, {Butkevich}, {Buzzi}, {Caffau},
  {Cancelliere}, {Cantat-Gaudin}, {Carballo}, {Carlucci}, {Carnerero},
  {Carrasco}, {Casamiquela}, {Castellani}, {Castro-Ginard}, {Chaoul},
  {Charlot}, {Chemin}, {Chiaramida}, {Chiavassa}, {Chornay}, {Comoretto},
  {Contursi}, {Cooper}, {Cornez}, {Cowell}, {Crifo}, {Cropper}, {Crosta},
  {Crowley}, {Dafonte}, {Dapergolas}, {David}, {David}, {de Laverny}, {De
  Luise}, {De March}, {De Ridder}, {de Souza}, {de Torres}, {del Peloso}, {del
  Pozo}, {Delbo}, {Delgado}, {Delisle}, {Demouchy}, {Dharmawardena}, {Di
  Matteo}, {Diakite}, {Diener}, {Distefano}, {Dolding}, {Edvardsson}, {Enke},
  {Fabre}, {Fabrizio}, {Faigler}, {Fedorets}, {Fernique}, {Fienga}, {Figueras},
  {Fournier}, {Fouron}, {Fragkoudi}, {Gai}, {Garcia-Gutierrez},
  {Garcia-Reinaldos}, {Garc{\'\i}a-Torres}, {Garofalo}, {Gavel}, {Gavras},
  {Gerlach}, {Geyer}, {Giacobbe}, {Gilmore}, {Girona}, {Giuffrida}, {Gomel},
  {Gomez}, {Gonz{\'a}lez-N{\'u}{\~n}ez}, {Gonz{\'a}lez-Santamar{\'\i}a},
  {Gonz{\'a}lez-Vidal}, {Granvik}, {Guillout}, {Guiraud},
  {Guti{\'e}rrez-S{\'a}nchez}, {Guy}, {Hatzidimitriou}, {Hauser}, {Haywood},
  {Helmer}, {Helmi}, {Sarmiento}, {Hidalgo}, {Hilger}, {H{\l}adczuk}, {Hobbs},
  {Holland}, {Huckle}, {Jardine}, {Jasniewicz}, {Jean-Antoine Piccolo},
  {Jim{\'e}nez-Arranz}, {Jorissen}, {Juaristi Campillo}, {Julbe}, {Karbevska},
  {Kervella}, {Khanna}, {Kontizas}, {Kordopatis}, {Korn}, {K{\'o}sp{\'a}l},
  {Kostrzewa-Rutkowska}, {Kruszy{\'n}ska}, {Kun}, {Laizeau}, {Lambert},
  {Lanza}, {Lasne}, {Le Campion}, {Lebreton}, {Lebzelter}, {Leccia}, {Leclerc},
  {Lecoeur-Taibi}, {Liao}, {Licata}, {Lindstr{\o}m}, {Lister}, {Livanou},
  {Lobel}, {Lorca}, {Loup}, {Madrero Pardo}, {Magdaleno Romeo}, {Managau},
  {Mann}, {Manteiga}, {Marchant}, {Marconi}, {Marcos}, {Marcos Santos},
  {Mar{\'\i}n Pina}, {Marinoni}, {Marocco}, {Marshall}, {Polo},
  {Mart{\'\i}n-Fleitas}, {Marton}, {Mary}, {Masip}, {Massari},
  {Mastrobuono-Battisti}, {Mazeh}, {McMillan}, {Messina}, {Michalik}, {Millar},
  {Mints}, {Molina}, {Molinaro}, {Moln{\'a}r}, {Monari}, {Mongui{\'o}},
  {Montegriffo}, {Montero}, {Mor}, {Mora}, {Morbidelli}, {Morel}, {Morris},
  {Muraveva}, {Murphy}, {Musella}, {Nagy}, {Noval}, {Oca{\~n}a}, {Ogden},
  {Ordenovic}, {Osinde}, {Pagani}, {Pagano}, {Palaversa}, {Palicio},
  {Pallas-Quintela}, {Panahi}, {Payne-Wardenaar}, {Pe{\~n}alosa Esteller},
  {Penttil{\"a}}, {Pichon}, {Piersimoni}, {Pineau}, {Plachy}, {Plum}, {Poggio},
  {Pr{\v{s}}a}, {Pulone}, {Racero}, {Ragaini}, {Rainer}, {Raiteri}, {Rambaux},
  {Ramos}, {Ramos-Lerate}, {Re Fiorentin}, {Regibo}, {Richards}, {Rios Diaz},
  {Ripepi}, {Riva}, {Rix}, {Rixon}, {Robichon}, {Robin}, {Robin}, {Roelens},
  {Rogues}, {Rohrbasser}, {Romero-G{\'o}mez}, {Rowell}, {Royer}, {Ruz Mieres},
  {Rybicki}, {Sadowski}, {S{\'a}ez N{\'u}{\~n}ez}, {Sagrist{\`a} Sell{\'e}s},
  {Sahlmann}, {Salguero}, {Samaras}, {Sanchez Gimenez}, {Sanna},
  {Santove{\~n}a}, {Sarasso}, {Schultheis}, {Sciacca}, {Segol}, {Segovia},
  {S{\'e}gransan}, {Semeux}, {Shahaf}, {Siddiqui}, {Siebert}, {Siltala},
  {Silvelo}, {Slezak}, {Slezak}, {Smart}, {Snaith}, {Solano}, {Solitro},
  {Souami}, {Souchay}, {Spagna}, {Spina}, {Spoto}, {Steele},
  {Steidelm{\"u}ller}, {Stephenson}, {S{\"u}veges}, {Surdej}, {Szabados},
  {Szegedi-Elek}, {Taris}, {Taylo}, {Teixeira}, {Tolomei}, {Tonello}, {Torra},
  {Torra}, {Torralba Elipe}, {Trabucchi}, {Tsounis}, {Turon}, {Ulla}, {Unger},
  {Vaillant}, {van Dillen}, {van Reeven}, {Vanel}, {Vecchiato}, {Viala},
  {Vicente}, {Voutsinas}, {Weiler}, {Wevers}, {Wyrzykowski}, {Yoldas}, {Yvard},
  {Zhao}, {Zorec}, {Zucker}, \& {Zwitter}}]{Gaia2022}
{Gaia Collaboration}, {Vallenari}, A., {Brown}, A.~G.~A., {et~al.} 2022, arXiv
  e-prints, arXiv:2208.00211, \dodoi{10.48550/arXiv.2208.00211}

\bibitem[{{Gaia Collaboration} {et~al.}(2023){Gaia Collaboration},
  {Recio-Blanco}, {Kordopatis}, {de Laverny}, {Palicio}, {Spagna}, {Spina},
  {Katz}, {Re Fiorentin}, {Poggio}, {McMillan}, {Vallenari}, {Lattanzi},
  {Seabroke}, {Casamiquela}, {Bragaglia}, {Antoja}, {Bailer-Jones},
  {Schultheis}, {Andrae}, {Fouesneau}, {Cropper}, {Cantat-Gaudin}, {Bijaoui},
  {Heiter}, {Brown}, {Prusti}, {de Bruijne}, {Arenou}, {Babusiaux}, {Biermann},
  {Creevey}, {Ducourant}, {Evans}, {Eyer}, {Guerra}, {Hutton}, {Jordi},
  {Klioner}, {Lammers}, {Lindegren}, {Luri}, {Mignard}, {Panem}, {Pourbaix},
  {Randich}, {Sartoretti}, {Soubiran}, {Tanga}, {Walton}, {Bastian}, {Drimmel},
  {Jansen}, {van Leeuwen}, {Bakker}, {Cacciari}, {Casta{\~n}eda}, {De Angeli},
  {Fabricius}, {Fr{\'e}mat}, {Galluccio}, {Guerrier}, {Masana}, {Messineo},
  {Mowlavi}, {Nicolas}, {Nienartowicz}, {Pailler}, {Panuzzo}, {Riclet}, {Roux},
  {Sordo}, {Th{\'e}venin}, {Gracia-Abril}, {Portell}, {Teyssier}, {Altmann},
  {Audard}, {Bellas-Velidis}, {Benson}, {Berthier}, {Blomme}, {Burgess},
  {Busonero}, {Busso}, {C{\'a}novas}, {Carry}, {Cellino}, {Cheek},
  {Clementini}, {Damerdji}, {Davidson}, {de Teodoro}, {Nu{\~n}ez Campos},
  {Delchambre}, {Dell'Oro}, {Esquej}, {Fern{\'a}ndez-Hern{\'a}ndez}, {Fraile},
  {Garabato}, {Garc{\'\i}a-Lario}, {Gosset}, {Haigron}, {Halbwachs}, {Hambly},
  {Harrison}, {Hern{\'a}ndez}, {Hestroffer}, {Hodgkin}, {Holl}, {Jan{\ss}en},
  {Jevardat de Fombelle}, {Jordan}, {Krone-Martins}, {Lanzafame},
  {L{\"o}ffler}, {Marchal}, {Marrese}, {Moitinho}, {Muinonen}, {Osborne},
  {Pancino}, {Pauwels}, {Reyl{\'e}}, {Riello}, {Rimoldini}, {Roegiers},
  {Rybizki}, {Sarro}, {Siopis}, {Smith}, {Sozzetti}, {Utrilla}, {van Leeuwen},
  {Abbas}, {{\'A}brah{\'a}m}, {Abreu Aramburu}, {Aerts}, {Aguado}, {Ajaj},
  {Aldea-Montero}, {Altavilla}, {{\'A}lvarez}, {Alves}, {Anders}, {Anderson},
  {Anglada Varela}, {Baines}, {Baker}, {Balaguer-N{\'u}{\~n}ez}, {Balbinot},
  {Balog}, {Barache}, {Barbato}, {Barros}, {Barstow}, {Bartolom{\'e}},
  {Bassilana}, {Bauchet}, {Becciani}, {Bellazzini}, {Berihuete}, {Bernet},
  {Bertone}, {Bianchi}, {Binnenfeld}, {Blanco-Cuaresma}, {Boch}, {Bombrun},
  {Bossini}, {Bouquillon}, {Bramante}, {Breedt}, {Bressan}, {Brouillet},
  {Brugaletta}, {Bucciarelli}, {Burlacu}, {Butkevich}, {Buzzi}, {Caffau},
  {Cancelliere}, {Carballo}, {Carlucci}, {Carnerero}, {Carrasco}, {Castellani},
  {Castro-Ginard}, {Chaoul}, {Charlot}, {Chemin}, {Chiaramida}, {Chiavassa},
  {Chornay}, {Comoretto}, {Contursi}, {Cooper}, {Cornez}, {Cowell}, {Crifo},
  {Crosta}, {Crowley}, {Dafonte}, {Dapergolas}, {David}, {De Luise}, {De
  March}, {De Ridder}, {de Souza}, {de Torres}, {del Peloso}, {del Pozo},
  {Delbo}, {Delgado}, {Delisle}, {Demouchy}, {Dharmawardena}, {Di Matteo},
  {Diakite}, {Diener}, {Distefano}, {Dolding}, {Edvardsson}, {Enke}, {Fabre},
  {Fabrizio}, {Faigler}, {Fedorets}, {Fernique}, {Figueras}, {Fournier},
  {Fouron}, {Fragkoudi}, {Gai}, {Garcia-Gutierrez}, {Garcia-Reinaldos},
  {Garc{\'\i}a-Torres}, {Garofalo}, {Gavel}, {Gavras}, {Gerlach}, {Geyer},
  {Giacobbe}, {Gilmore}, {Girona}, {Giuffrida}, {Gomel}, {Gomez},
  {Gonz{\'a}lez-N{\'u}{\~n}ez}, {Gonz{\'a}lez-Santamar{\'\i}a},
  {Gonz{\'a}lez-Vidal}, {Granvik}, {Guillout}, {Guiraud},
  {Guti{\'e}rrez-S{\'a}nchez}, {Guy}, {Hatzidimitriou}, {Hauser}, {Haywood},
  {Helmer}, {Helmi}, {Sarmiento}, {Hidalgo}, {H{\l}adczuk}, {Hobbs}, {Holland},
  {Huckle}, {Jardine}, {Jasniewicz}, {Jean-Antoine Piccolo},
  {Jim{\'e}nez-Arranz}, {Juaristi Campillo}, {Julbe}, {Karbevska}, {Kervella},
  {Khanna}, {Korn}, {K{\'o}sp{\'a}l}, {Kostrzewa-Rutkowska}, {Kruszy{\'n}ska},
  {Kun}, {Laizeau}, {Lambert}, {Lanza}, {Lasne}, {Le Campion}, {Lebreton},
  {Lebzelter}, {Leccia}, {Leclerc}, {Lecoeur-Taibi}, {Liao}, {Licata},
  {Lindstr{\o}m}, {Lister}, {Livanou}, {Lobel}, {Lorca}, {Loup}, {Madrero
  Pardo}, {Magdaleno Romeo}, {Managau}, {Mann}, {Manteiga}, {Marchant},
  {Marconi}, {Marcos}, {Marcos Santos}, {Mar{\'\i}n Pina}, {Marinoni},
  {Marocco}, {Marshall}, {Martin Polo}, {Mart{\'\i}n-Fleitas}, {Marton},
  {Mary}, {Masip}, {Massari}, {Mastrobuono-Battisti}, {Mazeh}, {Messina},
  {Michalik}, {Millar}, {Mints}, {Molina}, {Molinaro}, {Moln{\'a}r}, {Monari},
  {Mongui{\'o}}, {Montegriffo}, {Montero}, {Mor}, {Mora}, {Morbidelli},
  {Morel}, {Morris}, {Muraveva}, {Murphy}, {Musella}, {Nagy}, {Noval},
  {Oca{\~n}a}, {Ogden}, {Ordenovic}, {Osinde}, {Pagani}, {Pagano}, {Palaversa},
  {Pallas-Quintela}, {Panahi}, {Payne-Wardenaar}, {Pe{\~n}alosa Esteller},
  {Penttil{\"a}}, {Pichon}, {Piersimoni}, {Pineau}, {Plachy}, {Plum},
  {Pr{\v{s}}a}, {Pulone}, {Racero}, {Ragaini}, {Rainer}, {Raiteri}, {Ramos},
  {Ramos-Lerate}, {Regibo}, {Richards}, {Rios Diaz}, {Ripepi}, {Riva}, {Rix},
  {Rixon}, {Robichon}, {Robin}, {Robin}, {Roelens}, {Rogues}, {Rohrbasser},
  {Romero-G{\'o}mez}, {Rowell}, {Royer}, {Ruz Mieres}, {Rybicki}, {Sadowski},
  {S{\'a}ez N{\'u}{\~n}ez}, {Sagrist{\`a} Sell{\'e}s}, {Sahlmann}, {Salguero},
  {Samaras}, {Sanchez Gimenez}, {Sanna}, {Santove{\~n}a}, {Sarasso}, {Sciacca},
  {Segol}, {Segovia}, {S{\'e}gransan}, {Semeux}, {Shahaf}, {Siddiqui},
  {Siebert}, {Siltala}, {Silvelo}, {Slezak}, {Slezak}, {Smart}, {Snaith},
  {Solano}, {Solitro}, {Souami}, {Souchay}, {Spoto}, {Steele},
  {Steidelm{\"u}ller}, {Stephenson}, {S{\"u}veges}, {Surdej}, {Szabados},
  {Szegedi-Elek}, {Taris}, {Taylor}, {Teixeira}, {Tolomei}, {Tonello}, {Torra},
  {Torra}, {Torralba Elipe}, {Trabucchi}, {Tsounis}, {Turon}, {Ulla}, {Unger},
  {Vaillant}, {van Dillen}, {van Reeven}, {Vanel}, {Vecchiato}, {Viala},
  {Vicente}, {Voutsinas}, {Weiler}, {Wevers}, {Wyrzykowski}, {Yoldas}, {Yvard},
  {Zhao}, {Zorec}, {Zucker}, \& {Zwitter}}]{Gaia2023_spiral}
{Gaia Collaboration}, {Recio-Blanco}, A., {Kordopatis}, G., {et~al.} 2023,
  \aap, 674, A38, \dodoi{10.1051/0004-6361/202243511}

\bibitem[{{Garbari} {et~al.}(2012){Garbari}, {Liu}, {Read}, \&
  {Lake}}]{Garbari2012}
{Garbari}, S., {Liu}, C., {Read}, J.~I., \& {Lake}, G. 2012, \mnras, 425, 1445,
  \dodoi{10.1111/j.1365-2966.2012.21608.x}

\bibitem[{{Garc{\'\i}a P{\'e}rez} {et~al.}(2016){Garc{\'\i}a P{\'e}rez},
  {Allende Prieto}, {Holtzman}, {Shetrone}, {M{\'e}sz{\'a}ros}, {Bizyaev},
  {Carrera}, {Cunha}, {Garc{\'\i}a-Hern{\'a}ndez}, {Johnson}, {Majewski},
  {Nidever}, {Schiavon}, {Shane}, {Smith}, {Sobeck}, {Troup}, {Zamora},
  {Weinberg}, {Bovy}, {Eisenstein}, {Feuillet}, {Frinchaboy}, {Hayden},
  {Hearty}, {Nguyen}, {O'Connell}, {Pinsonneault}, {Wilson}, \&
  {Zasowski}}]{Perez2015}
{Garc{\'\i}a P{\'e}rez}, A.~E., {Allende Prieto}, C., {Holtzman}, J.~A.,
  {et~al.} 2016, \aj, 151, 144, \dodoi{10.3847/0004-6256/151/6/144}

\bibitem[{{Grand} {et~al.}(2022){Grand}, {Pakmor}, {Fragkoudi}, {G{\'o}mez},
  {Trick}, {Simpson}, {van de Voort}, \& {Bieri}}]{Grand2022}
{Grand}, R. J.~J., {Pakmor}, R., {Fragkoudi}, F., {et~al.} 2022, arXiv
  e-prints, arXiv:2211.08437, \dodoi{10.48550/arXiv.2211.08437}

\bibitem[{{GRAVITY Collaboration} {et~al.}(2022){GRAVITY Collaboration},
  {Abuter}, {Aimar}, {Amorim}, {Ball}, {Baub{\"o}ck}, {Berger}, {Bonnet},
  {Bourdarot}, {Brandner}, {Cardoso}, {Cl{\'e}net}, {Dallilar}, {Davies}, {de
  Zeeuw}, {Dexter}, {Drescher}, {Eisenhauer}, {F{\"o}rster Schreiber},
  {Foschi}, {Garcia}, {Gao}, {Gendron}, {Genzel}, {Gillessen}, {Habibi},
  {Haubois}, {Hei{\ss}el}, {Henning}, {Hippler}, {Horrobin}, {Jochum}, {Jocou},
  {Kaufer}, {Kervella}, {Lacour}, {Lapeyr{\`e}re}, {Le Bouquin}, {L{\'e}na},
  {Lutz}, {Ott}, {Paumard}, {Perraut}, {Perrin}, {Pfuhl}, {Rabien},
  {Shangguan}, {Shimizu}, {Scheithauer}, {Stadler}, {Stephens}, {Straub},
  {Straubmeier}, {Sturm}, {Tacconi}, {Tristram}, {Vincent}, {von Fellenberg},
  {Widmann}, {Wieprecht}, {Wiezorrek}, {Woillez}, {Yazici}, \&
  {Young}}]{Gravity2022}
{GRAVITY Collaboration}, {Abuter}, R., {Aimar}, N., {et~al.} 2022, \aap, 657,
  L12, \dodoi{10.1051/0004-6361/202142465}

\bibitem[{{Griffith} {et~al.}(2023){Griffith}, {Hogg}, {Dalcanton},
  {Hasselquist}, {Ratcilffe}, {Ness}, \& {Weinberg}}]{Griffith2023}
{Griffith}, E.~J., {Hogg}, D.~W., {Dalcanton}, J.~J., {et~al.} 2023, arXiv
  e-prints, arXiv:2307.05691, \dodoi{10.48550/arXiv.2307.05691}

\bibitem[{{Gunn} {et~al.}(2006){Gunn}, {Siegmund}, {Mannery}, {Owen}, {Hull},
  {Leger}, {Carey}, {Knapp}, {York}, {Boroski}, {Kent}, {Lupton}, {Rockosi},
  {Evans}, {Waddell}, {Anderson}, {Annis}, {Barentine}, {Bartoszek}, {Bastian},
  {Bracker}, {Brewington}, {Briegel}, {Brinkmann}, {Brown}, {Carr},
  {Czarapata}, {Drennan}, {Dombeck}, {Federwitz}, {Gillespie}, {Gonzales},
  {Hansen}, {Harvanek}, {Hayes}, {Jordan}, {Kinney}, {Klaene}, {Kleinman},
  {Kron}, {Kresinski}, {Lee}, {Limmongkol}, {Lindenmeyer}, {Long}, {Loomis},
  {McGehee}, {Mantsch}, {Neilsen}, {Neswold}, {Newman}, {Nitta}, {Peoples},
  {Pier}, {Prieto}, {Prosapio}, {Rivetta}, {Schneider}, {Snedden}, \&
  {Wang}}]{Gunn2006}
{Gunn}, J.~E., {Siegmund}, W.~A., {Mannery}, E.~J., {et~al.} 2006, \aj, 131,
  2332, \dodoi{10.1086/500975}

\bibitem[{Harris {et~al.}(2020)Harris, Millman, van~der Walt, Gommers,
  Virtanen, Cournapeau, Wieser, Taylor, Berg, Smith, Kern, Picus, Hoyer, van
  Kerkwijk, Brett, Haldane, del R{\'{i}}o, Wiebe, Peterson,
  G{\'{e}}rard-Marchant, Sheppard, Reddy, Weckesser, Abbasi, Gohlke, \&
  Oliphant}]{numpy}
Harris, C.~R., Millman, K.~J., van~der Walt, S.~J., {et~al.} 2020, Nature, 585,
  357, \dodoi{10.1038/s41586-020-2649-2}

\bibitem[{{Holmberg} \& {Flynn}(2000)}]{Holmberg2000}
{Holmberg}, J., \& {Flynn}, C. 2000, \mnras, 313, 209,
  \dodoi{10.1046/j.1365-8711.2000.02905.x}

\bibitem[{{Horta} {et~al.}(2022){Horta}, {Ness}, {Rybizki}, {Schiavon}, \&
  {Buder}}]{Horta2022}
{Horta}, D., {Ness}, M.~K., {Rybizki}, J., {Schiavon}, R.~P., \& {Buder}, S.
  2022, \mnras, 513, 5477, \dodoi{10.1093/mnras/stac953}

\bibitem[{{Horta} {et~al.}(2020){Horta}, {Schiavon}, {Mackereth}, {Beers},
  {Fern{\'a}ndez-Trincado}, {Frinchaboy}, {Garc{\'\i}a-Hern{\'a}ndez},
  {Geisler}, {Hasselquist}, {J{\"o}nsson}, {Lane}, {Majewski},
  {M{\'e}sz{\'a}ros}, {Bidin}, {Nataf}, {Roman-Lopes}, {Nitschelm},
  {Vargas-Gonz{\'a}lez}, \& {Zasowski}}]{Horta2020}
{Horta}, D., {Schiavon}, R.~P., {Mackereth}, J.~T., {et~al.} 2020, \mnras, 493,
  3363, \dodoi{10.1093/mnras/staa478}

\bibitem[{{Hunt} {et~al.}(2022){Hunt}, {Price-Whelan}, {Johnston}, \&
  {Darragh-Ford}}]{Hunt2022}
{Hunt}, J. A.~S., {Price-Whelan}, A.~M., {Johnston}, K.~V., \& {Darragh-Ford},
  E. 2022, \mnras, 516, L7, \dodoi{10.1093/mnrasl/slac082}

\bibitem[{{Hunt} {et~al.}(2021){Hunt}, {Stelea}, {Johnston}, {Gandhi},
  {Laporte}, \& {B{\'e}dorf}}]{Hunt2021}
{Hunt}, J. A.~S., {Stelea}, I.~A., {Johnston}, K.~V., {et~al.} 2021, \mnras,
  508, 1459, \dodoi{10.1093/mnras/stab2580}

\bibitem[{Hunter(2007)}]{Hunter:2007}
Hunter, J.~D. 2007, Computing in Science \& Engineering, 9, 90,
  \dodoi{10.1109/MCSE.2007.55}

\bibitem[{{Jeans}(1919)}]{Jeans1919}
{Jeans}, J.~H. 1919, {Problems of cosmogony and stellar dynamics}

\bibitem[{{J{\"o}nsson} {et~al.}(2020){J{\"o}nsson}, {Holtzman}, {Allende
  Prieto}, {Cunha}, {Garc{\'\i}a-Hern{\'a}ndez}, {Hasselquist}, {Masseron},
  {Osorio}, {Shetrone}, {Smith}, {Stringfellow}, {Bizyaev}, {Edvardsson},
  {Majewski}, {M{\'e}sz{\'a}ros}, {Souto}, {Zamora}, {Beaton}, {Bovy}, {Donor},
  {Pinsonneault}, {Poovelil}, \& {Sobeck}}]{Jonsson2020}
{J{\"o}nsson}, H., {Holtzman}, J.~A., {Allende Prieto}, C., {et~al.} 2020, \aj,
  160, 120, \dodoi{10.3847/1538-3881/aba592}

\bibitem[{{Khoperskov} {et~al.}(2019){Khoperskov}, {Di Matteo}, {Gerhard},
  {Katz}, {Haywood}, {Combes}, {Berczik}, \& {Gomez}}]{Khoperskov2019}
{Khoperskov}, S., {Di Matteo}, P., {Gerhard}, O., {et~al.} 2019, \aap, 622, L6,
  \dodoi{10.1051/0004-6361/201834707}

\bibitem[{{Kollmeier} {et~al.}(2017){Kollmeier}, {Zasowski}, {Rix}, {Johns},
  {Anderson}, {Drory}, {Johnson}, {Pogge}, {Bird}, {Blanc}, {Brownstein},
  {Crane}, {De Lee}, {Klaene}, {Kreckel}, {MacDonald}, {Merloni}, {Ness},
  {O'Brien}, {Sanchez-Gallego}, {Sayres}, {Shen}, {Thakar}, {Tkachenko},
  {Aerts}, {Blanton}, {Eisenstein}, {Holtzman}, {Maoz}, {Nandra}, {Rockosi},
  {Weinberg}, {Bovy}, {Casey}, {Chaname}, {Clerc}, {Conroy}, {Eracleous},
  {G{\"a}nsicke}, {Hekker}, {Horne}, {Kauffmann}, {McQuinn}, {Pellegrini},
  {Schinnerer}, {Schlafly}, {Schwope}, {Seibert}, {Teske}, \& {van
  Saders}}]{Kollmeier2017}
{Kollmeier}, J.~A., {Zasowski}, G., {Rix}, H.-W., {et~al.} 2017, arXiv
  e-prints, arXiv:1711.03234, \dodoi{10.48550/arXiv.1711.03234}

\bibitem[{{Kuijken} \& {Gilmore}(1989)}]{Kuijken1989}
{Kuijken}, K., \& {Gilmore}, G. 1989, \mnras, 239, 651,
  \dodoi{10.1093/mnras/239.2.651}

\bibitem[{{Kuijken} \& {Gilmore}(1991)}]{Kuijken1991}
---. 1991, \apjl, 367, L9, \dodoi{10.1086/185920}

\bibitem[{{Lane} {et~al.}(2022){Lane}, {Bovy}, \& {Mackereth}}]{Lane2022}
{Lane}, J. M.~M., {Bovy}, J., \& {Mackereth}, J.~T. 2022, \mnras, 510, 5119,
  \dodoi{10.1093/mnras/stab3755}

\bibitem[{{Laporte} {et~al.}(2019){Laporte}, {Minchev}, {Johnston}, \&
  {G{\'o}mez}}]{Laporte2019}
{Laporte}, C. F.~P., {Minchev}, I., {Johnston}, K.~V., \& {G{\'o}mez}, F.~A.
  2019, \mnras, 485, 3134, \dodoi{10.1093/mnras/stz583}

\bibitem[{{Li} \& {Widrow}(2021)}]{Li2021}
{Li}, H., \& {Widrow}, L.~M. 2021, \mnras, 503, 1586,
  \dodoi{10.1093/mnras/stab574}

\bibitem[{{Li} \& {Widrow}(2023)}]{Li2023}
---. 2023, \mnras, 520, 3329, \dodoi{10.1093/mnras/stad244}

\bibitem[{{Li} \& {Shen}(2020)}]{Li2020}
{Li}, Z.-Y., \& {Shen}, J. 2020, \apj, 890, 85,
  \dodoi{10.3847/1538-4357/ab6b21}

\bibitem[{{Mackereth} {et~al.}(2017){Mackereth}, {Bovy}, {Schiavon},
  {Zasowski}, {Cunha}, {Frinchaboy}, {Garc{\'\i}a Perez}, {Hayden}, {Holtzman},
  {Majewski}, {M{\'e}sz{\'a}ros}, {Nidever}, {Pinsonneault}, \&
  {Shetrone}}]{Mackereth2017}
{Mackereth}, J.~T., {Bovy}, J., {Schiavon}, R.~P., {et~al.} 2017, \mnras, 471,
  3057, \dodoi{10.1093/mnras/stx1774}

\bibitem[{{Majewski} {et~al.}(2017){Majewski}, {Schiavon}, {Frinchaboy},
  {Allende Prieto}, {Barkhouser}, {Bizyaev}, {Blank}, {Brunner}, {Burton},
  {Carrera}, {Chojnowski}, {Cunha}, {Epstein}, {Fitzgerald}, {Garc{\'\i}a
  P{\'e}rez}, {Hearty}, {Henderson}, {Holtzman}, {Johnson}, {Lam}, {Lawler},
  {Maseman}, {M{\'e}sz{\'a}ros}, {Nelson}, {Nguyen}, {Nidever}, {Pinsonneault},
  {Shetrone}, {Smee}, {Smith}, {Stolberg}, {Skrutskie}, {Walker}, {Wilson},
  {Zasowski}, {Anders}, {Basu}, {Beland}, {Blanton}, {Bovy}, {Brownstein},
  {Carlberg}, {Chaplin}, {Chiappini}, {Eisenstein}, {Elsworth}, {Feuillet},
  {Fleming}, {Galbraith-Frew}, {Garc{\'\i}a}, {Garc{\'\i}a-Hern{\'a}ndez},
  {Gillespie}, {Girardi}, {Gunn}, {Hasselquist}, {Hayden}, {Hekker}, {Ivans},
  {Kinemuchi}, {Klaene}, {Mahadevan}, {Mathur}, {Mosser}, {Muna}, {Munn},
  {Nichol}, {O'Connell}, {Parejko}, {Robin}, {Rocha-Pinto}, {Schultheis},
  {Serenelli}, {Shane}, {Silva Aguirre}, {Sobeck}, {Thompson}, {Troup},
  {Weinberg}, \& {Zamora}}]{Majewski2017}
{Majewski}, S.~R., {Schiavon}, R.~P., {Frinchaboy}, P.~M., {et~al.} 2017, \aj,
  154, 94, \dodoi{10.3847/1538-3881/aa784d}

\bibitem[{{Martell} {et~al.}(2017){Martell}, {Sharma}, {Buder}, {Duong},
  {Schlesinger}, {Simpson}, {Lind}, {Ness}, {Marshall}, {Asplund},
  {Bland-Hawthorn}, {Casey}, {De Silva}, {Freeman}, {Kos}, {Lin}, {Zucker},
  {Zwitter}, {Anguiano}, {Bacigalupo}, {Carollo}, {Casagrande}, {Da Costa},
  {Horner}, {Huber}, {Hyde}, {Kafle}, {Lewis}, {Nataf}, {Navin}, {Stello},
  {Tinney}, {Watson}, \& {Wittenmyer}}]{Martell2017}
{Martell}, S.~L., {Sharma}, S., {Buder}, S., {et~al.} 2017, \mnras, 465, 3203,
  \dodoi{10.1093/mnras/stw2835}

\bibitem[{{McKee} {et~al.}(2015){McKee}, {Parravano}, \&
  {Hollenbach}}]{McKee2015}
{McKee}, C.~F., {Parravano}, A., \& {Hollenbach}, D.~J. 2015, \apj, 814, 13,
  \dodoi{10.1088/0004-637X/814/1/13}

\bibitem[{{McMillan}(2011)}]{McMillan2011}
{McMillan}, P.~J. 2011, \mnras, 414, 2446,
  \dodoi{10.1111/j.1365-2966.2011.18564.x}

\bibitem[{{McMillan} \& {Binney}(2013)}]{McMillan2013}
{McMillan}, P.~J., \& {Binney}, J.~J. 2013, \mnras, 433, 1411,
  \dodoi{10.1093/mnras/stt814}

\bibitem[{{Ness} {et~al.}(2019){Ness}, {Johnston}, {Blancato}, {Rix}, {Beane},
  {Bird}, \& {Hawkins}}]{Ness2019}
{Ness}, M.~K., {Johnston}, K.~V., {Blancato}, K., {et~al.} 2019, \apj, 883,
  177, \dodoi{10.3847/1538-4357/ab3e3c}

\bibitem[{{Ness} {et~al.}(2022){Ness}, {Wheeler}, {McKinnon}, {Horta}, {Casey},
  {Cunningham}, \& {Price-Whelan}}]{Ness2022}
{Ness}, M.~K., {Wheeler}, A.~J., {McKinnon}, K., {et~al.} 2022, \apj, 926, 144,
  \dodoi{10.3847/1538-4357/ac4754}

\bibitem[{{Nidever} {et~al.}(2015){Nidever}, {Holtzman}, {Allende Prieto},
  {Beland}, {Bender}, {Bizyaev}, {Burton}, {Desphande}, {Fleming}, {Garc{\'\i}a
  P{\'e}rez}, {Hearty}, {Majewski}, {M{\'e}sz{\'a}ros}, {Muna}, {Nguyen},
  {Schiavon}, {Shetrone}, {Skrutskie}, {Sobeck}, \& {Wilson}}]{Nidever2015}
{Nidever}, D.~L., {Holtzman}, J.~A., {Allende Prieto}, C., {et~al.} 2015, \aj,
  150, 173, \dodoi{10.1088/0004-6256/150/6/173}

\bibitem[{{Oort}(1960)}]{Oort1960}
{Oort}, J.~H. 1960, \bain, 15, 45

\bibitem[{{Piffl} {et~al.}(2014){Piffl}, {Scannapieco}, {Binney}, {Steinmetz},
  {Scholz}, {Williams}, {de Jong}, {Kordopatis}, {Matijevi{\v{c}}},
  {Bienaym{\'e}}, {Bland-Hawthorn}, {Boeche}, {Freeman}, {Gibson}, {Gilmore},
  {Grebel}, {Helmi}, {Munari}, {Navarro}, {Parker}, {Reid}, {Seabroke},
  {Watson}, {Wyse}, \& {Zwitter}}]{Piffl2014}
{Piffl}, T., {Scannapieco}, C., {Binney}, J., {et~al.} 2014, \aap, 562, A91,
  \dodoi{10.1051/0004-6361/201322531}

\bibitem[{Price-Whelan(2018)}]{pyia}
Price-Whelan, A. 2018, \dodoi{10.5281/zenodo.1228136}

\bibitem[{{Price-Whelan}(2017{\natexlab{a}})}]{Price2017}
{Price-Whelan}, A.~M. 2017{\natexlab{a}}, The Journal of Open Source Software,
  2, 388, \dodoi{10.21105/joss.00388}

\bibitem[{{Price-Whelan}(2017{\natexlab{b}})}]{gala}
---. 2017{\natexlab{b}}, The Journal of Open Source Software, 2, 388,
  \dodoi{10.21105/joss.00388}

\bibitem[{{Price-Whelan} {et~al.}(2021){Price-Whelan}, {Hogg}, {Johnston},
  {Ness}, {Rix}, {Beaton}, {Brownstein}, {Garc{\'\i}a-Hern{\'a}ndez},
  {Hasselquist}, {Hayes}, {Lane}, {Shetrone}, {Sobeck}, \&
  {Zasowski}}]{Price2021}
{Price-Whelan}, A.~M., {Hogg}, D.~W., {Johnston}, K.~V., {et~al.} 2021, \apj,
  910, 17, \dodoi{10.3847/1538-4357/abe1b7}

\bibitem[{{Sanders} \& {Binney}(2015)}]{Sanders2015}
{Sanders}, J.~L., \& {Binney}, J. 2015, \mnras, 449, 3479,
  \dodoi{10.1093/mnras/stv578}

\bibitem[{{Santana} {et~al.}(2021){Santana}, {Beaton}, {Covey}, {O'Connell},
  {Longa-Pe{\~n}a}, {Cohen}, {Fern{\'a}ndez-Trincado}, {Hayes}, {Zasowski},
  {Sobeck}, {Majewski}, {Chojnowski}, {De Lee}, {Oelkers}, {Stringfellow},
  {Almeida}, {Anguiano}, {Donor}, {Frinchaboy}, {Hasselquist}, {Johnson},
  {Kollmeier}, {Nidever}, {Price-Whelan}, {Rojas-Arriagada}, {Schultheis},
  {Shetrone}, {Simon}, {Aerts}, {Borissova}, {Drout}, {Geisler}, {Law},
  {Medina}, {Minniti}, {Monachesi}, {Mu{\~n}oz}, {Poleski}, {Roman-Lopes},
  {Schlaufman}, {Stutz}, {Teske}, {Tkachenko}, {Van Saders}, {Weinberger}, \&
  {Zoccali}}]{Santana2021}
{Santana}, F.~A., {Beaton}, R.~L., {Covey}, K.~R., {et~al.} 2021, arXiv
  e-prints, arXiv:2108.11908.
\newblock \doarXiv{2108.11908}

\bibitem[{{Sch{\"o}nrich} {et~al.}(2010){Sch{\"o}nrich}, {Binney}, \&
  {Dehnen}}]{Schonrich2010}
{Sch{\"o}nrich}, R., {Binney}, J., \& {Dehnen}, W. 2010, \mnras, 403, 1829,
  \dodoi{10.1111/j.1365-2966.2010.16253.x}

\bibitem[{{Smith} {et~al.}(2021){Smith}, {Bizyaev}, {Cunha}, {Shetrone},
  {Souto}, {Allende Prieto}, {Masseron}, {M{\'e}sz{\'a}ros}, {J{\"o}nsson},
  {Hasselquist}, {Osorio}, {Garc{\'\i}a-Hern{\'a}ndez}, {Plez}, {Beaton},
  {Holtzman}, {Majewski}, {Stringfellow}, \& {Sobeck}}]{Smith2021}
{Smith}, V.~V., {Bizyaev}, D., {Cunha}, K., {et~al.} 2021, \aj, 161, 254,
  \dodoi{10.3847/1538-3881/abefdc}

\bibitem[{{Tremaine} {et~al.}(2023){Tremaine}, {Frankel}, \&
  {Bovy}}]{Tremaine2023}
{Tremaine}, S., {Frankel}, N., \& {Bovy}, J. 2023, \mnras, 521, 114,
  \dodoi{10.1093/mnras/stad577}

\bibitem[{{Vasiliev}(2019)}]{Vasiliev2019}
{Vasiliev}, E. 2019, \mnras, 482, 1525, \dodoi{10.1093/mnras/sty2672}

\bibitem[{Virtanen {et~al.}(2020)Virtanen, Gommers, Oliphant, Haberland, Reddy,
  Cournapeau, Burovski, Peterson, Weckesser, Bright, {van der Walt}, Brett,
  Wilson, Millman, Mayorov, Nelson, Jones, Kern, Larson, Carey, Polat, Feng,
  Moore, {VanderPlas}, Laxalde, Perktold, Cimrman, Henriksen, Quintero, Harris,
  Archibald, Ribeiro, Pedregosa, {van Mulbregt}, \& {SciPy 1.0
  Contributors}}]{scipy}
Virtanen, P., Gommers, R., Oliphant, T.~E., {et~al.} 2020, Nature Methods, 17,
  261, \dodoi{10.1038/s41592-019-0686-2}

\bibitem[{{Weinberg} {et~al.}(2022){Weinberg}, {Holtzman}, {Johnson}, {Hayes},
  {Hasselquist}, {Shetrone}, {Ting}, {Beaton}, {Beers}, {Bird}, {Bizyaev},
  {Blanton}, {Cunha}, {Fern{\'a}ndez-Trincado}, {Frinchaboy},
  {Garc{\'\i}a-Hern{\'a}ndez}, {Griffith}, {Johnson}, {J{\"o}nsson}, {Lane},
  {Leung}, {Mackereth}, {Majewski}, {M{\'e}sz{\'a}ros}, {Nitschelm}, {Pan},
  {Schiavon}, {Schneider}, {Schultheis}, {Smith}, {Sobeck}, {Stassun},
  {Stringfellow}, {Vincenzo}, {Wilson}, \& {Zasowski}}]{Weinberg2022}
{Weinberg}, D.~H., {Holtzman}, J.~A., {Johnson}, J.~A., {et~al.} 2022, \apjs,
  260, 32, \dodoi{10.3847/1538-4365/ac6028}

\bibitem[{{White} \& {Frenk}(1991)}]{White1991}
{White}, S. D.~M., \& {Frenk}, C.~S. 1991, \apj, 379, 52,
  \dodoi{10.1086/170483}

\bibitem[{{Widmark} {et~al.}(2022{\natexlab{a}}){Widmark}, {Hunt}, {Laporte},
  \& {Monari}}]{Widmark2022b}
{Widmark}, A., {Hunt}, J.~A.~S., {Laporte}, C.~F.~P., \& {Monari}, G.
  2022{\natexlab{a}}, \aap, 663, A16, \dodoi{10.1051/0004-6361/202243173}

\bibitem[{{Widmark} {et~al.}(2022{\natexlab{b}}){Widmark}, {Laporte}, \&
  {Monari}}]{Widmark2022}
{Widmark}, A., {Laporte}, C.~F.~P., \& {Monari}, G. 2022{\natexlab{b}}, \aap,
  663, A15, \dodoi{10.1051/0004-6361/202142819}

\bibitem[{{Widmark} {et~al.}(2022{\natexlab{c}}){Widmark}, {Laporte}, \&
  {Monari}}]{Widmark2022a}
---. 2022{\natexlab{c}}, \aap, 663, A15, \dodoi{10.1051/0004-6361/202142819}

\bibitem[{{Widrow} {et~al.}(2012){Widrow}, {Gardner}, {Yanny}, {Dodelson}, \&
  {Chen}}]{Widrow2012}
{Widrow}, L.~M., {Gardner}, S., {Yanny}, B., {Dodelson}, S., \& {Chen}, H.-Y.
  2012, \apjl, 750, L41, \dodoi{10.1088/2041-8205/750/2/L41}

\bibitem[{{Wilson} {et~al.}(2019){Wilson}, {Hearty}, {Skrutskie}, {Majewski},
  {Holtzman}, {Eisenstein}, {Gunn}, {Blank}, {Henderson}, {Smee}, {Nelson},
  {Nidever}, {Arns}, {Barkhouser}, {Barr}, {Beland}, {Bershady}, {Blanton},
  {Brunner}, {Burton}, {Carey}, {Carr}, {Colque}, {Crane}, {Damke}, {Davidson},
  {Dean}, {Di Mille}, {Don}, {Ebelke}, {Evans}, {Fitzgerald}, {Gillespie},
  {Hall}, {Harding}, {Harding}, {Hammond}, {Hancock}, {Harrison}, {Hope},
  {Horne}, {Karakla}, {Lam}, {Leger}, {MacDonald}, {Maseman}, {Matsunari},
  {Melton}, {Mitcheltree}, {O'Brien}, {O'Connell}, {Patten}, {Richardson},
  {Rieke}, {Rieke}, {Roman-Lopes}, {Schiavon}, {Sobeck}, {Stolberg}, {Stoll},
  {Tembe}, {Trujillo}, {Uomoto}, {Vernieri}, {Walker}, {Weinberg}, {Young},
  {Anthony-Brumfield}, {Bizyaev}, {Breslauer}, {De Lee}, {Downey}, {Halverson},
  {Huehnerhoff}, {Klaene}, {Leon}, {Long}, {Mahadevan}, {Malanushenko},
  {Nguyen}, {Owen}, {S{\'a}nchez-Gallego}, {Sayres}, {Shane}, {Shectman},
  {Shetrone}, {Skinner}, {Stauffer}, \& {Zhao}}]{Wilson2019}
{Wilson}, J.~C., {Hearty}, F.~R., {Skrutskie}, M.~F., {et~al.} 2019, \pasp,
  131, 055001, \dodoi{10.1088/1538-3873/ab0075}

\bibitem[{{Zasowski} {et~al.}(2013){Zasowski}, {Johnson}, {Frinchaboy},
  {Majewski}, {Nidever}, {Rocha Pinto}, {Girardi}, {Andrews}, {Chojnowski},
  {Cudworth}, {Jackson}, {Munn}, {Skrutskie}, {Beaton}, {Blake}, {Covey},
  {Deshpande}, {Epstein}, {Fabbian}, {Fleming}, {Garcia Hernandez}, {Herrero},
  {Mahadevan}, {M{\'e}sz{\'a}ros}, {Schultheis}, {Sellgren}, {Terrien}, {van
  Saders}, {Allende Prieto}, {Bizyaev}, {Burton}, {Cunha}, {da Costa},
  {Hasselquist}, {Hearty}, {Holtzman}, {Garc{\'\i}a P{\'e}rez}, {Maia},
  {O'Connell}, {O'Donnell}, {Pinsonneault}, {Santiago}, {Schiavon}, {Shetrone},
  {Smith}, \& {Wilson}}]{Zasowski2013}
{Zasowski}, G., {Johnson}, J.~A., {Frinchaboy}, P.~M., {et~al.} 2013, \aj, 146,
  81, \dodoi{10.1088/0004-6256/146/4/81}

\bibitem[{{Zasowski} {et~al.}(2017){Zasowski}, {Cohen}, {Chojnowski},
  {Santana}, {Oelkers}, {Andrews}, {Beaton}, {Bender}, {Bird}, {Bovy},
  {Carlberg}, {Covey}, {Cunha}, {Dell'Agli}, {Fleming}, {Frinchaboy},
  {Garc{\'\i}a-Hern{\'a}ndez}, {Harding}, {Holtzman}, {Johnson}, {Kollmeier},
  {Majewski}, {M{\'e}sz{\'a}ros}, {Munn}, {Mu{\~n}oz}, {Ness}, {Nidever},
  {Poleski}, {Rom{\'a}n-Z{\'u}{\~n}iga}, {Shetrone}, {Simon}, {Smith},
  {Sobeck}, {Stringfellow}, {Szigeti{\'a}ros}, {Tayar}, \&
  {Troup}}]{Zasowski2017}
{Zasowski}, G., {Cohen}, R.~E., {Chojnowski}, S.~D., {et~al.} 2017, \aj, 154,
  198, \dodoi{10.3847/1538-3881/aa8df9}

\bibitem[{{Zhang} {et~al.}(2013){Zhang}, {Rix}, {van de Ven}, {Bovy}, {Liu}, \&
  {Zhao}}]{Zhang2013}
{Zhang}, L., {Rix}, H.-W., {van de Ven}, G., {et~al.} 2013, \apj, 772, 108,
  \dodoi{10.1088/0004-637X/772/2/108}

\end{thebibliography}

\end{document}